\pdfoutput=1
\documentclass[pdflatex,sn-mathphys-ay]{sn-jnl}

\usepackage{graphicx}%
\usepackage{multirow}%
\usepackage{amsmath,amssymb,amsfonts}%
\numberwithin{equation}{section}
\usepackage{amsthm}%
\usepackage{mathrsfs}%
\usepackage[title]{appendix}%
\usepackage{xcolor}%
\usepackage{textcomp}%
\usepackage{manyfoot}%
\usepackage{booktabs}%
\usepackage{algorithm}%
\usepackage{algorithmicx}%
\usepackage{algpseudocode}%
\usepackage{listings}%
\usepackage{bm}
\usepackage{newtxtext}
\usepackage{natbib}
\usepackage{setspace}
\usepackage{ulem} 
\usepackage{accents} 
\usepackage{subfigure}
\theoremstyle{thmstyleone}%
\theoremstyle{thmstyletwo}%
\newtheorem{remark}{Remark}%

\theoremstyle{thmstylethree}%

\begin{document}

\title[Article Title]{Lagrangian Dynamical Theory of the Velocity Gradient Tensor: Real Schur Form, Schur Frame, Characteristic Vorticity Modes, and Commutative Vorticity Operators}


\author*[1]{\fnm{Tao} \sur{Chen}}\email{chentao2023@njust.edu.cn}

\affil*[1]{\orgdiv{School of Physics}, \orgname{Nanjing University of Science and Technology}, \orgaddress{ \city{Nanjing}, \postcode{210094}, \country{China}}}




\abstract{XXX}


\abstract{The present work proposes a novel Lagrangian dynamical theory for the velocity gradient tensor (VGT) $\bm{A}\equiv\bm{\nabla}\bm{u}$, formulated on the basis of its canonical real Schur form. Starting from the Navier--Stokes equations for compressible Newtonian fluids, we derive the general evolution equations for the six rotational invariants in the real Schur form, in which the angular velocity of the Schur frame along any Lagrangian trajectory is explicitly determined. These results then enable us to obtain the evolution equations for the derived VGT invariants, which also encompass the three principal VGT invariants and their Schur representations. Possibly for the first time, the evolution equations for the characteristic vorticity modes $(\bm{R}_{N},\bm{S}_{N})$ are derived in both intrinsic and component forms within the Schur frame. Beyond the conventional vortex stretching term in the Helmholtz vorticity transport equation, the theory reveals all stretching and tilting mechanisms of the distinct vorticity modes due to their interactions with the straining and shear strain-rate tensors. Moreover, in the present formulation, it unveils a new physical role of the enthalpy Hessian tensor (replaced by the pressure Hessian tensor in the incompressible limit) in the mutual transformation and redistribution of the two vorticity modes, as embodied in the longitudinal-transverse dynamical transfer term $\bm{\Upsilon}$ in the evolution equations. 
Physically, this term elucidates the mechanism of mutual transformation between axial vortices and surrounding shear layers coexisting within a swirling flow system, a process occurring in both two- and three-dimensional flows. Furthermore, the viscous source pair $(\bm{J}[\bm{R}_{N}],\bm{J}[\bm{S}_{N}])$ in the evolution equations is shown to arise from the spatial derivatives of the Schur frame, the dilatation, and the external force. As an illustrative example, we analyze the kinematics and dynamics of the vorticity modes for the Lamb--Oseen vortex, which validates the proposed theory. Using a similar approach, we discuss the Lagrangian dynamics of the commutative vorticity-operator pair $(\bm{\Psi}_{R},\bm{\Gamma}_{S})$. Finally, neglecting the nonlocal pressure Hessian contribution, we examine the restricted Euler dynamics for incompressible inviscid flow as a reduced case, which yields the Lagrangian invariance of the Schur frame along a pathline and the well--known Vieillefosse--Cantwell closed system of equations for the second and third principal VGT invariants. The proposed theory is expected to be useful for understanding the evolution dynamics of the VGT rotational invariants and intrinsic vorticity modes in a variety of vortical flows.}





\maketitle
\tableofcontents
\section{Introduction}\label{Introduction}
The fundamental features and properties of small-scale turbulence can be comprehensively understood through the velocity gradient tensor (VGT), denoted as $\bm{A}\equiv\bm{\nabla}\bm{u}$, and its invariants~\citep{Chong1990General,Meneveau2011Lagrangian,Johnson2024Multiscale}. The Lagrangian dynamics of the VGT
provides a unified framework that succinctly encapsulates the nonlinearity, nonlocality and multiscale nature of turbulent flows, thereby enabling the establishment of intrinsic connections among turbulence dynamics, statistics, and coherent flow structures~\citep{Sreenivasan1997,Johnson2024Multiscale}. Over the years, a variety of reduced-order models have been proposed, which simplify the evolution equation of the VGT to a small set of ordinary differential equations (ODEs), formulated either as low-dimensional dynamical systems or as stochastic differential equations in terms of the VGT invariants~\citep{Martin1998Dynamics,Meneveau2011Lagrangian,Lawson2015}. In particular, when the anisotropy of the pressure Hessian tensor and viscous diffusion effects are neglected along Lagrangian particle trajectories, one obtains the autonomous restricted Euler dynamics~\citep{Vieillefosse1982Local,Vieillefosse1984Internal,Cantwell1992}. This reduced description gives rise to a finite-time singularity, driven by gradient self-amplification mechanism ~\citep{Tsinober2009Informal}, and reproduces the characteristic asymmetric teardrop shape observed in the joint probability density function (PDF) in the $(\mathcal{Q},\mathcal{R})$-plane~\citep{Cantwell1993,Soria1994Study,Chu2013,Chacin2000Dynamics}.

To comprehensively extract information regarding straining, shear, and rotation in general vortical flows, various kinematic decompositions of the VGT have been formulated from distinct perspectives. The conventional approach is the symmetric-antisymmetric decomposition (SAD), which expresses the VGT as the sum of the strain-rate tensor $\bm{D}$ and the rotation-rate tensor $\bm{\varOmega}$, i.e., $\bm{A}=\bm{D}+\bm{\varOmega}$~\citep{Truesdell1954KinematicsVorticity,Batchelor1967IntroFD,Wu2006vorticity}. The dual vector of $\bm{\varOmega}$ gives the vorticity $\bm{\omega}\equiv\bm{\nabla}\times\bm{u}$, which is mathematically defined as the curl of the velocity field~\citep{Helmholtz1858}. Applying the SAD to fluid motion in a small vicinity of a point yields the classical Cauchy-Stokes theorem~\citep{Cauchy1841Memoire,Stokes1845InternalFriction}, which implies that vorticity is twice the angular velocity of a fluid volume element~\citep{Truesdell1954KinematicsVorticity}. Compared with primitive velocity-pressure visualizations, the vorticity paradigm often offers superior physical insights into vortical-flow phenomena and vortices \citep{Wu2006vorticity,YangYue2020}.
Vortices have long been celebrated as the sinews and muscles of turbulence~\citep{Moffatt1994}, although a precise and universal definition of a vortex remains an open issue of ongoing debate~\citep{LiuCQ2018}. From a phenomenological morphological perspective, the elementary vortex structures emerging under high-Reynolds-number conditions primarily manifest as (tubular) axial vortices and (sheet-like) shear layers, which often coexist within a single swirling vortex structure~\citep{Wu2006vorticity,Chen2026General}.Axial vortices can form through the winding of shear layers, a process referred to as the Klein--Kaden--Betz mechanism, arising from the strong coupling between longitudinal and transverse flow processes. ~\citep{Klein1910wirbeln,Kaden1931aufwicklung,Betz1950wirbel,MaoFeng2023}. However, vorticity alone cannot distinguish between pure shearing motions and the rigid-rotation (swirling) motion of a vortex~\citep{LiuCQ2025}, nor can the strain-rate tensor differentiate irrotational normal straining from pure shear~\citep{Kolar2004}. Furthermore, the existing Eulerian vortex criteria tend to exclude shear layers and are often sensitive to threshold selection in extracting axial vortices, which prevents the full capture of the KKB mechanism~\citep{Chen2026General}.

These limitations have motivated the proposal of the triple decomposition of motion (TDM) for $\bm{A}$ as an optimization problem for an orthogonal transformation matrix or the basic reference frame (BRF)~\citep{Kolar2004,Kolar2007VortexID}. The TDM is formulated as $\bm{A}=\bm{A}_{RES}+\bm{A}_{SH}=\bm{A}_{EL}+\bm{A}_{RR}+\bm{A}_{SH}$, where $\bm{A}_{EL}$ describes the irrotational straining motion (``EL'' for elongation), $\bm{A}_{RR}$ represents the rigid-rotation motion, and $\bm{A}_{SH}$ accounts for the effective pure shearing; the residual tensor $\bm{A}_{RES}$ is the sum of $\bm{A}_{EL}$ and $\bm{A}_{RR}$. The resulting vorticity decomposition conceptually separates the shear vorticity $\bm{\omega}_{SH}$ (i.e., the spin mode) from the residual vorticity $\bm{\omega}_{RES}$ (i.e, the rigid-rotation mode), although the explicit expressions for these two vorticity modes were not provided therein. The residual vorticity has been used to quantify the swirling strength of a vortex with new implementation algorithms~\citep{Kolar2007VortexID,Kolar2011Triple,Nagata2020,Hayashi2021,Boukharfane2021s}, and the shear vorticity for intraventricular blood flow~\citep{Kronborg2022Computational}, while the TDM has also inspired a new energy stability analysis related to the lifetimes of vortex structures across different scales~\citep{Hoffman2021}. Subsequently, by applying the real Schur form to the VGT matrix~\citep{Schur1909,Murnaghan1931}, the first version of the normal-nilpotent decomposition (NND-I) was proposed in several studies with different derivation approaches: $\bm{A}=\bm{N}+\bm{S}$, where $\bm{N}$ is a normal tensor and $\bm{S}$ is a nilpotent tensor~\citep{LiZhen2010,LiZhen2014,LiuCQ2018,Tian2018,GaoLiu2019}. The NND-I leads to the Liutex-shear decomposition (LSD) for vorticity, $\bm{\omega}=\bm{R}_{N}^{+}+\bm{S}_{N}^{+}$, in which the rigid-rotation mode $\bm{R}_{N}^{+}$ extracts the vortex (referred to as the Liutex/Rortex), and $\bm{S}_{N}^{+}$ denotes the spin/shear mode~\citep{LiuCQ2018,LiuCQ2020}. Explicit expressions for the Liutex vector were subsequently reported in \citet{Wang2019Liutex} and \citet{XuWQ2019}. [As described below, the plus-sign superscript for physical quantities was introduced by~\citet{Chen2026Kinematic} to distinguish the two distinct forms of IVD.] It is noted that the matrices of $(\bm{N},\bm{S})$ in the NND-I adopt different forms depending on the sign of $\Delta$ (the discriminant of the VGT), and $\bm{R}_{N}^{+}$ emerges only in the region where $\Delta>0$.~\citet{Kronborg2023} demonstrated that the standardized real Schur form was a solution to the optimization problem originally proposed
by~\citet{Kolar2004,Kolar2007VortexID}, and put forward a new simplified algorithm for computing the TDM.

Recently,~\citet{LiZhen2024} clarified several points of confusion pertaining to the Schur forms and the NND variants, the conceptual
distinctions between NND and TDM, as well as the intrinsic gap between complex and real NNDs. The TDM of~\citet{Kolar2007VortexID} was interpreted as a mixture of two distinct NNDs. Under certain conditions of uniqueness, the second version of NND, denoted as NND-II, was proposed. NND-II maintains a unified matrix representation across the entire flow domain, although the standardized real Schur form remains valid only for the region where $\Delta>0$. Compared with NND-I, the vorticity decomposition arising from NND-II extends the definition of the rigid-rotation mode to the whole domain~\citep{LiZhen2024,Chen2025Kinematic}. The concept of rotation-resistance for non-rotational points $(\Delta<0)$ was introduced by~\citet{Ma2024uniform} in a uniform decomposition of the VGT. Based on the real Schur form of the VGT,~\citet{Chen2026General} proposed a pair of invariant vorticity decompositions (IVDs) to account for different orientations of the spin mode. Specifically, IVD-I is identical to LSD, while IVD-II with a negative axial spin component takes the form $\bm{\omega}=\bm{R}_{N}^{-}+\bm{S}_{N}^{-}$, where $\bm{R}_{N}^{-}$ denotes the corresponding rigid-rotation mode. The Schur-form-based decompositions of both the VGT and vorticity have enabled novel analyses of turbulence statistics and coherent structures, as demonstrated in some recent studies~\citep{Keylock2018,Nagata2020,Das2020,Watanabe2020,Yu2021,Boukharfane2021,Jiang2022,YinW2023,Arun2024velocity,Arun2024,Keylock2025,BilbaoLudena2025}.
The real Schur form is regarded as an intrinsic local property of arbitrary flows, and the Lie invariances of the associated vorticity 2-forms have been analyzed~\citep{Zhu2021Thermodynamic}. Apart from these characteristic algebraic decompositions, several direction--dependent vorticity decompositions (DVDs) have been developed on the basis of oriented line and surface elements, utilizing both material and field descriptions in conjunction with differential geometry~\citep{Chen2025Kinematic,Chen2026General,Zhang2026Bifurcation}. Subsequently,~\citet{Chen2026Kinematic} introduced an operator--form vorticity decomposition (OVD), in which the two distinct vorticity modes and the Schur frame (bases) are directly linked to the eigenvalues and eigenvectors of a pair of commutative vorticity (vortex) operators. These two vorticity operators provide a complete rational description of the fundamental forms of vortex structures. It is noted that vorticity decomposition with respect to the surface principal frame was studied by~\citet{Xie2020} based on a generalized nonholonomic basis theory.

The aforementioned studies mainly focus on the kinematic decompositions of VGT and vorticity in instantaneous flow fields, their advantages with respect to the classical vortex identification criteria, as well as some novel turbulence statistics based on Schur forms and NNDs. Nevertheless, to the best of the author's knowledge, a corresponding Lagrangian dynamical theory has not yet been established. It remains unclear how Lagrangian dynamical evolution correlates several key physical quantities, including the rotational VGT invariants in the real Schur form, the Schur frames, the characteristic vorticity modes, and the commuting vorticity operators. The main goal of the present paper is to answer these questions.

The structure of this article is as follows. In \textsection\ref{RSR}, we introduce the real Schur form of the VGT, which serves as the foundation for subsequent analyses. In \textsection\ref{Invariant vorticity decomposition}, two invariant vorticity decompositions (IVDs) are presented which are unified in an operator-form vorticity decomposition (OVD) in \textsection \ref{Operator-form vorticity decomposition}. In \textsection\ref{UU1}, the material derivatives of the real Schur form, the Schur frame, and the associated rotational invariants are systematically derived. Detailed discussions on the degenerate cases are presented, including zero discriminant, negative discriminant, zero vorticity, $\gamma=0$, and $\gamma\rightarrow 0$ (where $\gamma$ denotes the axial spin component). This sets the stage for the Lagrangian dynamics of the VGT and its rotational invariants (including the derived invariants) in \textsection\ref{UU2}. In \textsection\ref{UU3}, we split the vorticity transport equation into two separate evolution equations governing the Lagrangian dynamics of the characteristic rigid-rotation and spin modes. Then, the kinematics and dynamics of the vorticity modes within the Lamb--Oseen vortex are investigated using the proposed theory. Section~\ref{UU4} develops the Lagrangian dynamics for the commutative vorticity operators. As a reduced case, an analogous real-Schur formalism is constructed for Lagrangian dynamics under the restricted Euler approximation. Conclusions and discussions are presented in \textsection\ref{Conclusions}.

\section{Real Schur form of the velocity gradient tensor}\label{RSR}
\begin{figure}[t]
	\centering 
	\includegraphics[width=1.0\columnwidth,trim={3cm 6cm 9cm 8cm},clip]{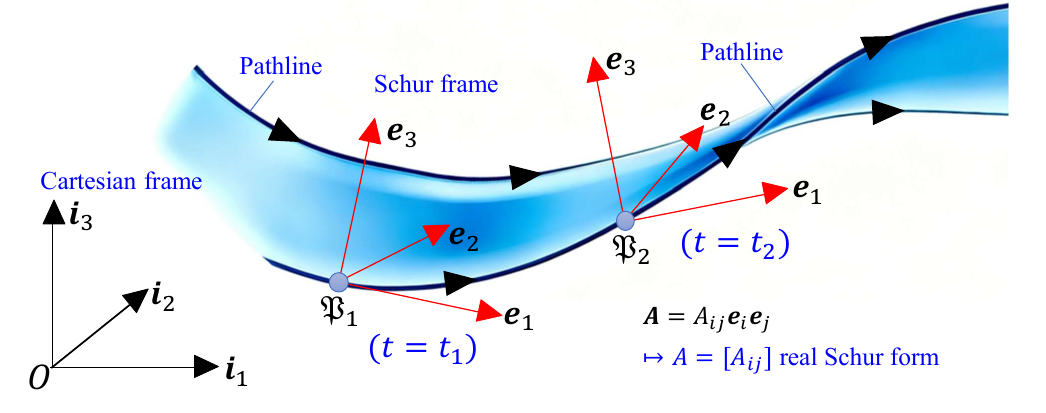}
	\caption{Schematic of the Cartesian frame $(\bm{i}_{1},\bm{i}_{2},\bm{i}_{3})$ and the Schur frame $(\bm{e}_{1},\bm{e}_{2},\bm{e}_{3})$ along a pathline. The VGT matrix takes a real Schur from in the Schur frame for $\Delta>0$. The points $\mathfrak{P}_{1}$ and $\mathfrak{P}_{2}$ correspond to the time instants $t_{1}$ and $t_{2}$, respectively.} 
	\label{Schur_frame_along_pathline}
\end{figure}

From the characteristic algebraic perspective, the vortex region $\mathscr{V}\subset\mathbb{R}^3$ should satisfy the condition $\Delta_{\rm 3D}>0$, where $\Delta_{\rm 3D}$ denotes the discriminant of the characteristic polynomial of the VGT $\bm{A}\equiv\bm{\nabla}\bm{u}$~\citep{Chong1990General}. At a point $\mathfrak{P}\in\mathscr{V}$, the VGT $\bm{A}$ admits one real eigenvalue $\lambda_{r}$ and a pair of complex conjugate eigenvalues $\lambda_{1,2} = \lambda_{\mathrm{cr}} \pm{\rm i} \lambda_{\mathrm{ci}}$, with $\chi\equiv\lambda_{\mathrm{cr}}$ and $\lambda_{\mathrm{ci}}$ representing their real and imaginary parts, respectively. In a sufficiently small neighborhood of $\mathfrak{P}$, the local streamline pattern necessarily exhibits either a closed orbit or a spiral configuration~\citep{Chong1990General,Chu2013}. Imposing the requirement $\omega_{3}\equiv\bm{\omega}\bm{\cdot}\bm{e}_{3}>0$ on the axial vorticity component, the real eigenvector $\bm{e}_{3}$ uniquely determines the local axis of rotation at $\mathfrak{P}$, where the eigenpair $(\lambda_{r},\bm{e}_{3})$ satisfies $\bm{e}_{3}\bm{\cdot}\bm{A}=\lambda_{r}\bm{e}_{3}$~\citep{zhou1999mechanisms,LiuCQ2018,GaoLiu2019}. The plane perpendicular to this axis is denoted by $\mathscr{P}$, which forms a two-dimensional (2D) subspace of the Euclidean space $\mathbb{R}^3$ such that $\mathbb{R}^3\equiv\mathscr{P}\oplus\left\{\zeta\bm{e}_{3}\lvert \zeta\in\mathbb{R}\right\}$. Under the global Cartesian coordinate system $(X_{1},X_{2},X_{3})$ with the orthonormal basis $\left\{\bm{i}_{1},\bm{i}_{2},\bm{i}_{3}\right\}$, the VGT $\bm{A}$ is expressed as $\bm{A}=C_{ij}\bm{i}_{i}\bm{i}_{j}$ with $C_{ij}\equiv\bm{i}_{i}\bm{\cdot}\bm{A}\bm{\cdot}\bm{i}_{j}$ being the $(i,j)$-th VGT component. As illustrated in figure~\ref{Schur_frame_along_pathline}, in the orthonormal Schur frame $\left\{\bm{e}_{1},\bm{e}_{2},\bm{e}_{3}\right\}$ where $\mathscr{P}={\rm Span}\left\{\bm{e}_{1},\bm{e}_{2}\right\}$, the VGT $\bm{A}$ is expressed as $\bm{A}=A_{ij}\bm{e}_{i}\bm{e}_{j}$ where $A_{ij}\equiv\bm{e}_{i}\bm{\cdot}\bm{A}\bm{\cdot}\bm{e}_{j}=\partial_{i}u_{j}$ ($i$ denotes the row index and $j$ the column index). The corresponding VGT matrix takes the standardized real Schur form as~\citep{Murnaghan1931,LiZhen2014,LiuCQ2018}
\begin{eqnarray}\label{Schur}
	\bm{A}\mapsto A\equiv\left[A_{ij}\right]=\begin{bmatrix}
		\chi & \psi + \gamma & -\beta \\
		-\psi & \chi & \alpha \\
		0 & 0 & \lambda_r
	\end{bmatrix}.
\end{eqnarray}
The notations herein and their relations to the published literature are discussed in Appendix~\ref{Notations herein and connections with published literature}.
We refer to~\eqref{Schur} as the Schur representation of the VGT, which is realized as a block upper-triangular matrix in the real-number field. In the context of general compressible flows, three independent scalars are needed to define the orientation of a reference frame, leaving six degrees of freedom for the VGT matrix~\citep{Meneveau2011Lagrangian}. The real Schur form introduces a new set of rotational invariants, $(\chi,\lambda_{r},\psi,\alpha,\beta,\gamma)$, expressed in the Schur frame. Specifically, the diagonal elements $(\chi,\lambda_{r})$ quantify the rates of stretching or contraction experienced by the material line elements instantaneously parallel to the principal axes. Meanwhile, the off-diagonal entry $\psi$ denotes the characteristic angular velocity of the rigid rotation about $\bm{e}_{3}$. The remaining invariants $(\alpha,\beta,\gamma)$ collectively describe the angular deformation of the three mutually perpendicular faces of an infinitesimal cubic fluid element centered at $\mathfrak{P}$. An equivalent choice of invariant set is $(\chi,\lambda_{r},\alpha,\beta,\lambda_{\mathrm{ci}},\omega_{3})$, which comprises the complete information of eigenvalues and vorticity components.

\begin{subequations}
	It is worth noting that the transformation relations between the Cartesian and Schur bases are given by
	\begin{equation}\label{basis_transform}
		\bm{e}_{i}=\bm{i}_{k}Q_{ki}~~\text{or}~~\bm{i}_{k}={Q}_{ki}\bm{e}_{i}.
	\end{equation}
	where $Q\equiv[Q_{ij}]$ is a real orthogonal matrix satisfying $Q^{\rm T}Q=QQ^{\rm T}=I$, with ${I}\equiv[\delta_{ij}]$ denoting the identity matrix. Since the VGT remains invariant under this basis change, it follows that
	\begin{equation}\label{vvv1}
		QAQ^{\rm T}=C~~\text{or}~~A=Q^{\rm T}CQ.
	\end{equation}
\end{subequations}

Following the Schur form in~\eqref{Schur}, the normal-nilpotent decomposition (NND) of the VGT reads~\citep{LiZhen2014,LiuCQ2018,GaoLiu2019}
\begin{subequations}\label{ANS}
	\begin{equation}
		\bm{A}=\bm{N}+\bm{S},
	\end{equation}
	\begin{equation}
		\bm{N}\mapsto N = 
		\begin{bmatrix}
			\chi & \psi & 0 \\
			-\psi & \chi & 0 \\
			0 & 0 & \lambda_r
		\end{bmatrix}, \qquad
		\bm{S}\mapsto S = 
		\begin{bmatrix}
			0 & \gamma & -\beta \\
			0 & 0 & \alpha \\
			0 & 0 & 0
		\end{bmatrix},
	\end{equation}
\end{subequations}
where $\bm{N}$ and $\bm{S}$ are the normal and nilpotent tensors, respectively. Accordingly, the orthonormal \textit{Schur frame} $(\bm{e}_{1},\bm{e}_{2},\bm{e}_{3})$ is referred to as \textit{the NND basis} or \textit{the NND triad}. Physically, $\bm{N}$ encapsulates the contributions from axial stretching and in-plane rigid rotation, whereas $\bm{S}$ represents shear deformation. When the normal straining components are separated from $\bm{N}$ in the Schur frame, the NND~\eqref{ANS} reduces to a triple decomposition akin to those in~\citet{Kolar2004,Kolar2007VortexID} and~\citet{Kronborg2023}. In the following discussion, all rotational invariants and Schur basis vectors are assumed to vary smoothly with both position and time over the spatial region and time range under consideration. Remarks on the degenerate cases (which include $\Delta=0$, $\Delta<0$, $\bm{\omega}=\bm{0}$, $\gamma=0$, and $\gamma\rightarrow 0$) are documented in \S\ref{Remarks on degenerate cases}.

\section{Invariant vorticity decomposition}\label{Invariant vorticity decomposition}
The VGT $\bm{A}$ is decomposed as the symmetric strain--rate tensor $\bm{D}$ and the skew--symmetric rotation-rate tensor $\bm{\varOmega}$~\citep{Truesdell1954KinematicsVorticity,Wu2006vorticity}:
\begin{subequations}
	\begin{gather}
		\bm{A} = \bm{D} + \bm{\varOmega},\label{decomp1} \\
		\bm{D} \equiv \frac{1}{2}\left(\bm{A} + \bm{A}^{\mathrm{T}}\right), \quad 
		\bm{\varOmega} \equiv \frac{1}{2}\left(\bm{A} - \bm{A}^{\mathrm{T}}\right),\label{decomp2}
	\end{gather}
\end{subequations}
where ``T'' denotes the transpose.
Using~\eqref{Schur}, the matrices of $\bm{D}\equiv D_{ij}\bm{e}_{i}\bm{e}_{j}$ and $\bm{\varOmega}\equiv\varOmega_{ij}\bm{e}_{i}\bm{e}_{j}$ are, respectively, given by
\begin{subequations}\label{DOmega}
	\begin{gather}
		\bm{D}\mapsto D =
		\begin{bmatrix}
			\chi & \dfrac12\gamma & -\dfrac12\beta \\[4pt]
			\dfrac12\gamma & \chi & \dfrac12\alpha \\[4pt]
			-\dfrac12\beta & \dfrac12\alpha & \lambda_r
		\end{bmatrix}, \label{eq:D_matrix} \\[8pt]
		\bm{\varOmega}\mapsto \varOmega =
		\begin{bmatrix}
			0 & \psi+\dfrac12\gamma & -\dfrac12\beta \\[4pt]
			-\left(\psi+\dfrac12\gamma\right) & 0 & \dfrac12\alpha \\[4pt]
			\dfrac12\beta & -\dfrac12\alpha & 0
		\end{bmatrix}, \label{eq:Omega_matrix}
	\end{gather}
\end{subequations}
where $D_{ij}\equiv\bm{e}_{i}\bm{\cdot}\bm{D}\bm{\cdot}\bm{e}_{j}$ and $\varOmega_{ij}\equiv\bm{e}_{i}\bm{\cdot}\bm{\varOmega}\bm{\cdot}\bm{e}_{j}$ are the corresponding components. It can be observed that, apart from the diagonal straining components, $\bm{D}$ in~\eqref{eq:D_matrix} also comprises shear components, being consistent with the earlier observation of~\citet{Kolar2007VortexID}.
The dual vector of $\bm{\varOmega}$ in~\eqref{eq:Omega_matrix} yields the invariant vorticity decomposition (IVD)~\citep{Chen2026General}:
\begin{subequations}\label{IVD_general}
	\begin{gather}
		\bm{\omega} = \bm{R}_N + \bm{S}_N, \label{IVD_general1} \\
		\bm{R}_{N}\equiv2\psi\bm{e}_{3},\quad \bm{S}_{N}\equiv\alpha \bm{e}_1 + \beta \bm{e}_2 + \gamma \bm{e}_3. \label{IVD_general2}
	\end{gather}
\end{subequations}
Here, $\bm{R}_N$ is termed the characteristic rigid-rotation mode, while $\bm{S}_{N}$ is termed the characteristic spin mode (or the characteristic shear mode in viscous flows). Note that the IVD occurs exclusively for the axial vorticity component
\begin{equation}
	\omega_{3}=2\psi+\gamma,
\end{equation}
which depends only on the velocity field projected on the plane $\mathscr{P}$. The restricted 2D VGT on $\mathscr{P}$ is $\bm{A}_{\rm 2D}=A_{ij}\bm{e}_{i}\bm{e}_{j}=\bm{D}_{\rm 2D}+\bm{\varOmega}_{\rm 2D}~(i,j=1,2)$ whose discriminant is~\citep{Chen2025Kinematic}
\begin{eqnarray}\label{delta}
\Delta=4\lambda_{\rm ci}^{2}=4\psi(\psi+\gamma)=\omega_{3}^2-\gamma^2.
\end{eqnarray}
Therefore, $\Delta>0$ implies $\omega_{3}>\lvert\gamma\rvert$ and $\lambda_{\rm ci}=(1/2)\sqrt{\Delta}$. Then, the 3D discriminant of $\bm{A}$ is expressed as
\begin{eqnarray}\label{delta3D}
\Delta_{\rm 3D}=\frac{1}{27}\lambda_{\rm ci}^{2}\left[\lambda_{\rm ci}^{2}+(\chi-\lambda_{r})^2\right]^{2}=\frac{1}{108}\Delta\left[\frac{1}{4}\Delta+(\chi-\lambda_{r})^2\right]^{2}.
\end{eqnarray}
Our definition of $\Delta_{\rm 3D}$ differs from that of~\citet{Ma2024uniform} by a minus sign, and this sign convention has no impact on the underlying physics.

Depending on the sign of the axial spin component $\gamma$, two different IVDs (IVD-I and IVD-II) have been proposed by~\citet{LiuCQ2018} and~\citet{Chen2026General}. These are briefly introduced below.
\begin{itemize}
	\item \textbf{Case A.} IVD-I with positive spin $(\gamma^+>0)$ 
	\begin{subequations}\label{IVD-I}
	\begin{equation}\label{IVD-Ia}
		\omega_3 = 2\psi^+ + \gamma^+,
	\end{equation}
	\begin{equation}\label{IVD-Ib}
		2\psi^+ = \omega_3 - \sqrt{\omega_3^2 - \Delta},\quad
		\gamma^+ = \sqrt{\omega_3^2 - \Delta}.
	\end{equation}
	\end{subequations}
	In this case, the characteristic rigid-rotation mode $\bm{R}_{N}^{+}\equiv2\psi^{+}\bm{e}_{3}$ coincides precisely the Liutex (or Rortex) vector in the region where $\Delta>0$, which have been widely employed for vortex identification in complex flows~\citep{LiuCQ2018,LiuCQ2025}. The characteristic spin mode is $\bm{S}_{N}^{+}=\bm{\omega}-\bm{R}_{N}^{+}$. Equation~\eqref{IVD-Ia} is referred to as the Liutex--shear decomposition (LSD), whereas ~\eqref{IVD-Ib} was previously reported by~\citet{Wang2019Liutex} and~\citet{XuWQ2019}. Physically, this synergistic effect manifests as the formation of axial vortices through the wrapping
	of shear layers, accompanied by a transfer from the spin mode to the rigid-rotation mode, which ultimately
	leads to vortex intensification. In fluids with	small viscosity, the rolling-up of thin shear layers is the only known mechanism for the rapid formation of axial vortices~\citep{Klein1910wirbeln,Kaden1931aufwicklung,Betz1950wirbel,Wu2006vorticity}. Thus, we refer to the pair $(2\psi^+,\gamma^+)$ as the Klein--Kaden--Betz (KKB) configuration, in honor of these pioneering scientists in the field of vorticity and vortex dynamics.
	\item  \textbf{Case B.} IVD-II with negative spin $(\gamma^-<0)$ 
		\begin{subequations}\label{IVD-II}
		\begin{equation}\label{IVD-IIa}
			\omega_3 = 2\psi^{-} + \gamma^{-},
		\end{equation}
		\begin{equation}\label{IVD-IIb}
			2\psi^- = \omega_3 + \sqrt{\omega_3^2 - 4\lambda_{\mathrm{ci}}^2},\quad
			\gamma^- = -\sqrt{\omega_3^2 - \Delta}.
		\end{equation}
	\end{subequations}
	Equation~\eqref{IVD-II} was proposed in~\citet{Chen2026General} and~\citet{Chen2026Kinematic}. Correspondingly, the characteristic rigid-rotation mode is $\bm{R}_{N}^{-}\equiv2\psi^{-}\bm{e}_{3}$, which can also be used to represent vortex. The characteristic spin mode is $\bm{S}_{N}^{-}=\bm{\omega}-\bm{R}_{N}^{-}$. In contrast to IVD-I, we refer to the pair $(2\psi^{-},\gamma^{-})$ as the anti-KKB configuration~\citep{Chen2026General}. This mechanism occurs when the axial vortex possesses sufficient swirling strength to maintain its coherence, such that opposing spin effects cannot significantly disrupt the primary vortex structure. During the formation of an axial vortex, the anti-KKB mechanism could be activated to suppress the unbounded growth of swirling strength in the inner core region, while the KKB mechanism may continue to govern the roll-up of the outer shear layers.
\end{itemize}
Four useful quantities can be derived from~\eqref{IVD-I} and~\eqref{IVD-II}:
\begin{equation}\label{four_ids}
	\psi^+ = \psi^- + \gamma^-,\quad
	\psi^- = \psi^+ + \gamma^+,\quad
	\gamma^- = -\gamma^+,\quad
	\psi^+ + \psi^- = \omega_3.
\end{equation}
From~\eqref{delta} and~\eqref{four_ids}, $\Delta$ and $\lambda_{\mathrm{ci}}$ admit the concise representations as
\begin{equation}
	\Delta=4\psi^{+}\psi^{-},\quad \lambda_{\mathrm{ci}}=\sqrt{\psi^{+}\psi^{-}}.
\end{equation}
For IVD-I and IVD-II, the restricted Schur frames on $\mathscr{P}$ are respectively denoted by $(\bm{e}_{1}^{+},\bm{e}_{2}^{+})$ and $(\bm{e}_{1}^{-},\bm{e}_{2}^{-})$. The restricted VGT $\bm{A}_{\rm 2D}$ on $\mathscr{P}$ is expressed as (see figure~\ref{Schur_frame_NND_OVD})
\begin{subequations}\label{eq3p89}
	\begin{itemize}
		\item \textbf{Case A.} IVD-I with the Schur frame $(\bm{e}_{1}^{+},\bm{e}_{2}^{+})$.
		\begin{equation}\label{eq3p8}
			\bm{A}_{\rm 2D}\mapsto	\begin{bmatrix}
				\chi & \psi^{+}+\gamma^{+} \\
				-\psi^+ & \chi
			\end{bmatrix}=
			\begin{bmatrix}
				\chi & \psi^{-}\\
				-\psi^+ & \chi
			\end{bmatrix};
		\end{equation}
		\item \textbf{Case B.} IVD-II with the Schur frame $(\bm{e}_{1}^{-},\bm{e}_{2}^{-})$.
		\begin{equation}\label{eq3p9}
		\bm{A}_{\rm 2D}\mapsto
			\begin{bmatrix}
				\chi & \psi^{-}+\gamma^{-} \\
				-\psi^- & \chi
			\end{bmatrix}=
			\begin{bmatrix}
				\chi & \psi^{+}\\
				-\psi^- & \chi
			\end{bmatrix}.
		\end{equation}
	\end{itemize}
\end{subequations}
There remains ongoing debate over the matrix representation of the VGT in the region where $\Delta\leq 0$ from different perspectives~\citep{LiZhen2014,LiZhen2024,Ma2024uniform,Chen2025Kinematic,Chen2026Kinematic}. During the construction of the NND-I, different forms of matrices are adopted for $\Delta>0$ and $\Delta\leq 0$, where the real Schur form~\eqref{Schur} is used for $\Delta>0$ and a strictly upper triangular matrix for $\Delta\leq 0$~\citep{LiZhen2014}. The matrix in~\eqref{Schur} is consistent with the works focusing on the conventional vortex region where $\Delta>0$~\citep{Wang2019Liutex,LiuCQ2025}. It was shown by~\citet{LiZhen2024} that~\eqref{Schur} can be consistently applied to the VGT throughout the whole domain, although not as a real Schur form for $\Delta\leq0$. Under this choice, the vorticity decompositions in~\eqref{IVD-I} and~\eqref{IVD-II}, and the matrices in~\eqref{eq3p89}, remain valid for $\Delta\leq{0}$. However, the characteristic vorticity modes exhibit opposite signs for $\Delta\leq 0$: $\psi^{+}<0$ and $\gamma^{+}>0$, while $\psi^{-}>0$ and $\gamma^{-}<0$~\citep{Chen2026Kinematic}.

\section{Operator-form vorticity decomposition}\label{Operator-form vorticity decomposition}
As is well recognized, commutativity
and symmetry constitute the fundamental concepts in modern physics.
On $\mathscr{P}$,~\citet{Chen2026Kinematic} proposed an operator-form vorticity decomposition (OVD) as
\begin{subequations}\label{OVD2}
	\begin{gather}
		\omega_3 \bm{\sigma}_0 = \bm{\Psi}_R + \bm{\Gamma}_S, \label{OVD} \\
		[\bm{\Psi}_R,\bm{\Gamma}_S]\equiv\bm{\Psi}_R\circ\bm{\Gamma}_S-\bm{\Gamma}_S\circ\bm{\Psi}_R = \mathbf{O}. \label{commutator}
	\end{gather}
\end{subequations}
Here, the Pauli tensor $\bm{\sigma}_{0}\equiv\bm{e}_{1}\bm{e}_{1}+\bm{e}_{2}\bm{e}_{2}=\bm{I}-\bm{e}_{3}\bm{e}_{3}$ is equal to the identity tensor on $\mathscr{P}$; $(\bm{\Psi}_R,\bm{\Gamma}_S)$ denotes a pair of commutative vorticity operators, where $\bm{\Psi}_{R}$ is termed the rigid-rotation vorticity operator and $\bm{\Gamma}_{S}$ is termed the spin vorticity operator. Given an arbitrary orthonormal basis $\left\{\overline{\bm{e}}_{1},\overline{\bm{e}}_{2}\right\}$ on $\mathscr{P}$, we have $\bm{A}_{\rm 2D}=\overline{A}_{ij}\overline{\bm{e}}_{i}\overline{\bm{e}}_{j}$ and $\bm{D}_{\rm 2D}=\overline{D}_{ij}\overline{\bm{e}}_{i}\overline{\bm{e}}_{j}~(i,j=1,2)$. Then, the matrices of $(\bm{\Psi}_R,\bm{\Gamma}_S)$ are
\begin{subequations}
\begin{equation}
	\bm{\Psi}_R \mapsto
	\begin{bmatrix}
		2\overline{A}_{12} & -(\overline{A}_{11}-\overline{A}_{22}) \\
		-(\overline{A}_{11}-\overline{A}_{22}) & -2\overline{A}_{21}
	\end{bmatrix}, 
\end{equation}
\begin{equation}
	\bm{\Gamma}_S \mapsto
	\begin{bmatrix}
		-2\overline{D}_{12} & \overline{D}_{11}-\overline{D}_{22} \\
		\overline{D}_{11}-\overline{D}_{22} & 2\overline{D}_{21}
	\end{bmatrix}.
\end{equation}	
\end{subequations}

\begin{figure}[t]
	\centering 
	\includegraphics[width=0.9\columnwidth,trim={6cm 3cm 6cm 4cm},clip]{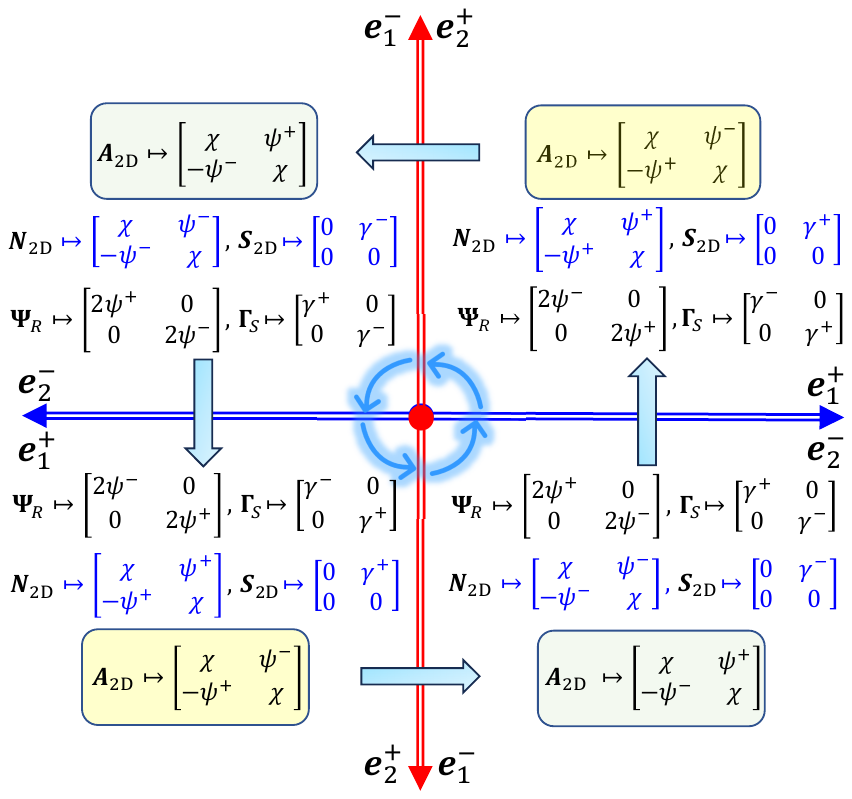}
	\caption{The Schur frames $(\bm{e}_{1}^{\pm},\bm{e}_{2}^{\pm},\bm{e}_{3})$, the associated real Schur forms of $\bm{A}_{\rm 2D}$, the NND matrices of $(\bm{N}_{\rm 2D},\bm{S}_{\rm 2D})$, and the OVD matrices of $(\bm{\Psi}_{R},\bm{\Gamma}_{S})$ on the rotation-axis-normal plane $\mathscr{P}$, where the rotation axis $\bm{e}_{3}$ obeys $\bm{e}_{3}=\bm{e}_{1}^{\pm}\times\bm{e}_{2}^{\pm}$.}
	\label{Schur_frame_NND_OVD}
\end{figure}

The information encoded by the commutator fully characterizes two fundamental morphological features of vortices: tubular/axial vortices and winding shear layers with different spin orientations. It was shown by~\citet{Chen2026Kinematic} that their real eigenvalues are precisely the characteristic vorticity modes $(2\psi^{\pm},\gamma^{\pm})$ appearing in the IVDs (\eqref{IVD-I} and~\eqref{IVD-II}), while their complete common eigenvectors $(\bm{e}_{1}^{\pm},\bm{e}_{2}^{\pm})$, together with $\bm{e}_{3}$, just constitute the Schur frame $\left\{\bm{e}_{1}^{\pm},\bm{e}_{2}^{\pm},\bm{e}_{3}\right\}$ used for the NND (\S\ref{RSR}). Therefore,~\eqref{OVD2} serves as the unifying principle underlying the previously distinct IVDs, which underscores the essential roles of both the IVD vorticity
modes and the Schur frames in formulating a complete theory of kinematic vorticity decomposition. Using~\eqref{eq3p8} and~\eqref{eq3p9}, the Schur representations of $(\bm{\Psi}_R,\bm{\Gamma}_S)$ are given as follows (figure~\ref{Schur_frame_NND_OVD}).
\begin{itemize}
	\item Case A. Under the Schur frame $(\bm{e}_{1}^{+},\bm{e}_{2}^{+})$
	\begin{subequations}
		\begin{equation}
		\bm{\Psi}_R \mapsto
		\begin{bmatrix}
			2\psi^{-} & 0 \\
			0 & 2\psi^{+}
		\end{bmatrix}, \qquad
		\bm{\Gamma}_S \mapsto
		\begin{bmatrix}
			\gamma^{-} & 0 \\
			0 & \gamma^{+}
		\end{bmatrix}.
	\end{equation}
	Therefore, we obtain
	\begin{equation}
		\bm{\Psi}_{R}=2\psi^{-}\bm{e}_{1}^{+}\bm{e}_{1}^{+}+2\psi^{+}\bm{e}_{2}^{+}\bm{e}_{2}^{+},
	\end{equation}
	\begin{equation}
		\bm{\Gamma}_{S}=\gamma^{-}\bm{e}_{1}^{+}\bm{e}_{1}^{+}+\gamma^{+}\bm{e}_{2}^{+}\bm{e}_{2}^{+}.
	\end{equation}	
	\end{subequations}
	\item Case B. Under the Schur frame $(\bm{e}_{1}^{-},\bm{e}_{2}^{-})$
	\begin{subequations}
			\begin{equation}
			\bm{\Psi}_R \mapsto
			\begin{bmatrix}
				2\psi^+ & 0 \\
				0 & 2\psi^- 
			\end{bmatrix}, \qquad
			\bm{\Gamma}_S \mapsto 
			\begin{bmatrix}
				\gamma^+ & 0 \\
				0 & \gamma^- 
			\end{bmatrix}.
		\end{equation}
		Therefore, we obtain
		\begin{equation}
			\bm{\Psi}_{R}=2\psi^{+}\bm{e}_{1}^{-}\bm{e}_{1}^{-}+2\psi^{-}\bm{e}_{2}^{-}\bm{e}_{2}^{-},
		\end{equation}
		\begin{equation}
			\bm{\Gamma}_{S}=\gamma^{+}\bm{e}_{1}^{-}\bm{e}_{1}^{-}+\gamma^{-}\bm{e}_{2}^{-}\bm{e}_{2}^{-}.
		\end{equation}	
	\end{subequations}
\end{itemize}
Since $\gamma^{+}=-\gamma^{-}$, it follows that at a fixed spatial point, $\bm{\Gamma}_{S}$ is proportional to the spin angular momentum operator in the Pauli representation of an electron in quantum mechanics, with its eigenvectors resembling two-component spinors~\citep{Chen2026Kinematic}. More importantly, we emphasize that the intensification of vortical structures can be achieved not only through the vortex stretching mechanism (only active in 3D flows), but also via the vorticity mode conversion mechanism, which acts in both 2D and 3D flows.

\section{Material derivatives of the real Schur form, the Schur frame, and the rotational invariants of the velocity gradient tensor}\label{UU1}
\subsection{Material derivative of the real Schur form}
Applying the material derivative operator to both sides of~\eqref{vvv1}, we obtain
\begin{equation}\label{ss1}
\dot{A}=\dot{Q}^{\rm T}CQ+Q^{\rm T}\dot{C}Q+Q^{T}C\dot{Q}.
\end{equation}
where the material derivative, denoted by an over-dot above the physical quantity, reads $\dot{(\cdot)}\equiv D(\cdot)/Dt$.
Invoking~\eqref{vvv1} a second time, we arrive at
\begin{equation}\label{ss2}
	\dot{A}=\dot{Q}^{\rm T}QA+AQ^{\rm T}\dot{Q}+Q^{\rm T}\dot{C}Q.
\end{equation}
By introducing the following matrices associated with the real orthogonal matrix
\begin{equation}\label{ss3}
	W\equiv\dot{Q}^{\rm T}Q=-{Q}^{\rm T}\dot{Q}~~\text{and}~~F\equiv Q^{\rm T}\dot{C}Q,
\end{equation}
~\eqref{ss2} can be concisely expressed as
\begin{subequations}
	\begin{equation}\label{ss4}
		\dot{A}=\left[W,A\right]+F,
	\end{equation}
where the commutator is defined as
	\begin{equation}\label{ss4a}
		\left[W,A\right]\equiv WA-AW.
	\end{equation}
\end{subequations}
Interestingly,~\eqref{ss3} is formally analogous to the Heisenberg equation of motion in the Heisenberg picture of quantum mechanics~\citep{Born1926}. The additional term $F=[F_{ij}]$ acts as a general source term. Therefore, ~\eqref{ss4} can be interpreted as a Heisenberg-type-like evolution equation under a real orthogonal transformation (rather than a unitary one), supplemented by a source contribution.

\subsection{Material derivative of the Schur frame}
Combining $W$ in~\eqref{ss3} with the Schur frame yields a second-order skew-symmetric tensor
	\begin{equation}\label{ss5}
	\bm{W}\equiv W_{ij}\bm{e}_{i}\bm{e}_{j},~W_{ij}=-W_{ji}.
	\end{equation}
	Along a pathline, it follows from~\eqref{basis_transform} and~\eqref{ss5} that
	\begin{eqnarray}\label{ss6}
	\dot{\bm{e}}_{i}&=&\bm{i}_{k}\dot{Q}_{ki}=Q_{kj}\dot{Q}_{ki}\bm{e}_{j}=(Q^{\rm T}\dot{Q})_{ji}\bm{e}_{j}=W_{ij}\bm{e}_{j}\nonumber\\
	&=&\bm{e}_{i}\bm{\cdot}\bm{W}=(\star\bm{W})\bm{\times}\bm{e}_{i},
	\end{eqnarray}
	where $\star$ denotes the Hodge star operator~\citep{Hodge1941}, and $\star\bm{W}$ is the dual vector of $\bm{W}$. Physically, $\star\bm{W}$ represents the angular velocity of the Schur frame along a pathline. It is evident that
	\begin{equation}\label{ss7}
		\bm{W}=\bm{e}_{i}\otimes\dot{\bm{e}}_{i}=-\dot{\bm{e}}_{i}\otimes\bm{e}_{i},~~W_{ij}=\dot{\bm{e}}_{i}\bm{\cdot}\bm{e}_{j}=-\bm{e}_{i}\bm{\cdot}\dot{\bm{e}}_{j}.
	\end{equation}
	From~\eqref{ss5}, we have
	\begin{eqnarray}\label{ss8}
	\star\bm{W}&=&\frac{1}{2}W_{ij}\star\left(\bm{e}_{i}\bm{e}_{j}-\bm{e}_{j}\bm{e}_{i}\right)\nonumber\\
	&=&\frac{1}{2}W_{ij}\star(\bm{e}_{i}\wedge\bm{e}_{j})\nonumber\\
	&=&\frac{1}{2}W_{ij}\bm{e}_{i}\times\bm{e}_{j}\nonumber\\
	&=&\frac{1}{2}e_{kij}W_{ij}\bm{e}_{k},
	\end{eqnarray}
	where $\wedge$ is the wedge product operator, and $e_{ijk}$ is the Levi-Civita permutation symbol~\citep{Chern2000Lectures}.
	On the other hand, using~\eqref{ss6}, we obtain
	\begin{eqnarray}\label{ss9}
	W_{ij}&=&\bm{e}_{i}\bm{\cdot}\bm{W}\bm{\cdot}\bm{e}_{j}\nonumber\\
	&=&(\star\bm{W})\times\bm{e}_{i}\bm{\cdot}\bm{e}_{j}\nonumber\\
	&=&(\star\bm{W})\bm{\cdot}(\bm{e}_{i}\times\bm{e}_{j})\nonumber\\
	&=&e_{ijk}(\star\bm{W})_{k}.
	\end{eqnarray}
	For brevity, we write the matrices of $(\star\bm{W},\bm{W})$ as
	\begin{eqnarray}\label{ss10}
	\star\bm{W}\mapsto\begin{bmatrix}
	    W_{1} \\
		W_{2} \\
		W_{3}
	\end{bmatrix},~~
	\bm{W}\mapsto{W}=\begin{bmatrix}
	0 & W_3 & -W_2 \\
	-W_3 & 0 & W_1 \\
	W_2 & -W_1 & 0
	\end{bmatrix}.
	\end{eqnarray}
	Substituting~\eqref{ss10} into~\eqref{ss6} yields the material derivatives for the Schur basis vectors:
\begin{subequations}\label{ss11}
	\begin{align}
 \dfrac{D\boldsymbol e_1}{Dt} &= W_3 \boldsymbol e_2 - W_2 \boldsymbol e_3, \label{ss11a} \\
 \dfrac{D\boldsymbol e_2}{Dt} &= -W_3 \boldsymbol e_1 + W_1 \boldsymbol e_3, \label{ss11b} \\
 \dfrac{D\boldsymbol e_3}{Dt} &= W_2 \boldsymbol e_1 - W_1 \boldsymbol e_2. \label{ss11c}
	\end{align}
\end{subequations}

\subsection{Evaluation of the commutator}
Using~\eqref{Schur} and~\eqref{ss10}, we evaluate the commutator as
\begin{equation}\label{ss16}
	[W,A] = 
	\begin{bmatrix}
		G_{11} & G_{12} & G_{13} \\
		G_{21} & G_{22} & G_{23} \\
		G_{31} & G_{32} & G_{33}
	\end{bmatrix},
\end{equation}
where the matrix elements are given by
\begin{subequations}\label{eq:Phi_system}
	\begin{align}
		G_{11} &= \gamma W_3 + \beta W_2, \label{eq:G11} \\
		G_{12} &= -\beta W_1, \label{eq:G12} \\
		G_{13} &= \alpha W_3 + (\chi-\lambda_r) W_2 - (\psi+\gamma) W_1, \label{eq:G13} \\
		G_{21} &= -\alpha W_2, \label{eq:G21} \\
		G_{22} &= -\gamma W_3 + \alpha W_1, \label{eq:G22} \\
		G_{23} &= \beta W_3 - \psi W_2 + (\lambda_r-\chi) W_1, \label{eq:G23} \\
		G_{31} &= (\chi-\lambda_r) W_2 + \psi W_1, \label{eq:G31} \\
		G_{32} &= (\psi+\gamma) W_2 + (\lambda_r-\chi) W_1, \label{eq:G32} \\
		G_{33} &= -\beta W_2 - \alpha W_1. \label{eq:G33}
	\end{align}
\end{subequations}

\subsection{Material derivatives of the rotational invariants of the velocity gradient tensor}
Substituting~\eqref{ss16} and~\eqref{eq:Phi_system} into~\eqref{ss4} yields the following system of equations:
\begin{subequations}\label{ss17}
	\begin{align}
		\dfrac{D\alpha}{Dt} &= W_3\beta + W_1(\lambda_r - \chi) - W_2\psi + F_{23}, \label{eq:alphas} \\
		\dfrac{D\beta}{Dt} &= -W_3\alpha + W_2(\lambda_r - \chi) + W_1(\psi + \gamma) - F_{13}, \label{eq:betas} \\
		\dfrac{D\gamma}{Dt} &= -W_1\beta - W_2\alpha + F_{12} + F_{21}, \label{eq:gammas} \\
		\dfrac{D\psi}{Dt} &= W_2\alpha - F_{21}, \label{eq:psis} \\
		\dfrac{D\lambda_r}{Dt} &= -W_2\beta - W_1\alpha + F_{33}, \label{eq:lambda_rs} \\
		\dfrac{D\chi}{Dt} &= W_3\gamma + W_2\beta + F_{11}, \label{eq:chi1s} \\
		\dfrac{D\chi}{Dt} &= -W_3\gamma + W_1\alpha + F_{22}, \label{eq:chi2s} \\
		F_{31} &= -W_2(\chi - \lambda_r) - W_1\psi, \label{eq:F31s} \\
		F_{32} &= -W_2(\psi+\gamma) - W_1(\lambda_r - \chi). \label{eq:F32s}
	\end{align}
\end{subequations}
By employing~\eqref{eq:F31s} and~\eqref{eq:F32s},~\eqref{eq:alphas} and~\eqref{eq:betas} can be equivalently recast as
\begin{subequations}\label{eqforalphabeta}
	\begin{align}
		\frac{D\alpha}{Dt} &= W_3 \beta - W_2(2\psi + \gamma) + F_{23} - F_{32}, \label{eq:alpha_new} \\
		\frac{D\beta}{Dt} &= -W_3\alpha + W_1(2\psi + \gamma) + F_{31} - F_{13}. \label{eq:beta_new}
	\end{align}
\end{subequations}
In terms of the spin components $(\alpha,\beta)$, the average of~\eqref{eq:chi1s} and~\eqref{eq:chi2s} yields
\begin{eqnarray}
	\frac{D\chi}{Dt}=\frac{1}{2}\left(W_{2}\beta+W_{1}\alpha\right)+\frac{1}{2}\left(F_{11}+F_{22}\right),
\end{eqnarray}
while their difference generates 
\begin{eqnarray}\label{W3a}
	\gamma W_{3}=\frac{1}{2}\left(\alpha W_{1}-\beta W_{2}+F_{22}-F_{11}\right).
\end{eqnarray}
Introduce the following factor
\begin{equation}\label{Xifactor}
	\Xi\equiv(\chi-\lambda_{r})^2+\psi(\psi+\gamma)=(\chi-\lambda_{r})^2+\frac{1}{4}\Delta.
\end{equation}
Clearly, $\Xi>0$ always holds in the region where $\Delta>0$ (even for $\chi=\lambda_{r}$), and~\eqref{delta3D} implies
\begin{eqnarray}\label{ratio}
	\Xi=6\sqrt{3}\sqrt{\dfrac{\Delta_{\rm 3D}}{\Delta}}.
\end{eqnarray}
Equation~\eqref{ratio} indicates that $\Xi$ is proportional to the square root of the ratio ${\Delta_{\rm 3D}}/{\Delta}$.

When $\Xi\neq 0$ and $\gamma\neq 0$, solving~\eqref{eq:F31s},~\eqref{eq:F32s}, and~\eqref{W3a} yields
\begin{subequations}\label{Schur_frame_angular1}
	\begin{align}
		W_1 &= -(\psi+\gamma)\frac{F_{31}}{\Xi}+(\chi-\lambda_{r})\frac{F_{32}}{\Xi}, \label{eq:W1} \\[10pt]
		W_2 &= -(\chi-\lambda_{r})\frac{F_{31}}{\Xi}-\psi\frac{F_{32}}{\Xi}, \label{eq:W2} \\[10pt]
		W_3 &= \frac{1}{2} \left[
		\left( \frac{\beta}{\gamma}(\chi - \lambda_r) - \frac{\alpha}{\gamma}(\psi + \gamma) \right)\frac{F_{31}}{\Xi}
		+ \left( \frac{\alpha}{\gamma}(\chi - \lambda_r) + \frac{\beta}{\gamma}\psi \right)\frac{F_{32}}{\Xi}
		+ \frac{F_{22} - F_{11}}{\gamma}
		\right]. \label{eq:W3}
	\end{align}
\end{subequations}
\subsection{Remarks on degenerate cases}\label{Remarks on degenerate cases}
This subsection discusses several degenerate cases, which do not affect the main derivations presented in this paper. Readers may skip this part and proceed directly to the subsequent sections.
\subsubsection{Cases with zero discriminant}
For $\Delta=4\psi(\psi+\gamma)=0$ at the point $\mathfrak{P}$ (assuming $\gamma\neq 0$), we have
\begin{itemize}
	\item \textbf{Case A.} IVD-I with $2\psi=2\psi^+=0$ and $\gamma=\gamma^+=\omega_{3}>0$. The corresponding Schur frame is $(\bm{e}_{1},\bm{e}_{2})=(\bm{e}_{1}^{+},\bm{e}_{2}^{+})$.\\
	If $\chi\neq\lambda_{r}$, then $\Xi=(\chi-\lambda_{r})^2>0$, and hence~\eqref{Schur_frame_angular1} reduces to
	\begin{subequations}
		\begin{align}
			W_1 &= -\frac{\gamma F_{31}}{(\chi-\lambda_r)^2}
			+\frac{F_{32}}{\chi-\lambda_r}, \label{eq:W1_psi0} \\[10pt]
			W_2 &= -\frac{F_{31}}{\chi-\lambda_r}, \label{eq:W2_psi0} \\[10pt]
			W_3 &= -\frac{\alpha F_{31}}{2(\chi-\lambda_r)^2}
			+\frac{\beta F_{31}+\alpha F_{32}}{2\gamma(\chi-\lambda_r)}
			+\frac{F_{22}-F_{11}}{2\gamma}. \label{eq:W3_psi0}
		\end{align}
	\end{subequations}
	If $\chi=\lambda_{r}$, then~\eqref{eq:F31s} and~\eqref{eq:F32s} imply that $F_{31}=0$ and $W_{2}=-F_{32}/\gamma$. In general, $W_{1}$ and $W_{3}$ cannot be determined at the point $\mathfrak{P}$.
	\item \textbf{Case B.} IVD-II with $2\psi=2\psi^-=2\omega_{3}>0$ and $\gamma=\gamma^{-}=-\omega_{3}<0$. The corresponding Schur frame is $(\bm{e}_{1},\bm{e}_{2})=(\bm{e}_{1}^{-},\bm{e}_{2}^{-})$.\\
If $\chi\neq\lambda_{r}$, then $\Xi=(\chi-\lambda_{r})^2>0$, and hence~\eqref{Schur_frame_angular1} reduces to
\begin{subequations}
	\begin{align}
		W_1 &= \frac{F_{32}}{\chi-\lambda_r}, \label{eq:W1_psimg} \\[8pt]
		W_2 &= -\frac{F_{31}}{\chi-\lambda_r}+\frac{\gamma F_{32}}{(\chi-\lambda_r)^2}, \label{eq:W2_psimg} \\[8pt]
		W_3 &= \frac{\alpha F_{32}+\beta F_{31}}{2\gamma(\chi-\lambda_r)}
		-\frac{\beta F_{32}}{2(\chi-\lambda_r)^2}
		+\frac{F_{22}-F_{11}}{2\gamma}. \label{eq:W3_psimg}
	\end{align}
\end{subequations} 
	If $\chi=\lambda_{r}$, then~\eqref{eq:F31s} and~\eqref{eq:F32s} imply that $F_{32}=0$ and $W_{1}=F_{31}/\gamma$. In general, $W_{2}$ and $W_{3}$ cannot be determined at the point $\mathfrak{P}$.
\end{itemize}
Nevertheless, all such degenerate cases satisfying $\chi=\lambda_{r}$ lie on the dividing surface implicitly defined by $\Delta=0$. These may be interpreted as limiting cases approached from the side $\Delta \gtrless 0$. For example, we consider unidirectional simple shear flow. In the Cartesian coordinate system $(x,y)$, the velocity field is expressed as $\bm{u}=ky\bm{i}_{x}~(k>0)$, where $(\bm{i}_{x},\bm{i}_{y},\bm{i}_{z})$ denotes the orthonormal Cartesian frame with $\bm{i}_{z}=\bm{i}_{x}\times\bm{i}_{y}$. The vorticity is $\omega_{z}=-\omega_3=-k<0$, so that the rotation axis is determined as $\bm{e}_{3}=-\bm{i}_{z}$. In the reordered frame $(\bm{i}_{y},\bm{i}_{x},-\bm{i}_{z})$, the VGT $\bm{A}\equiv\bm{\nabla}\bm{u}$ is expressed as
\begin{subequations}
\begin{equation}
\bm{A}\mapsto\begin{bmatrix}
	 A_{yy} & A_{yx} \\ A_{xy} & A_{xx}
	  \end{bmatrix}
	 	= \begin{bmatrix} \partial_y u_y & \partial_y u_x \\ \partial_x u_y & \partial_x u_x \end{bmatrix}
= \begin{bmatrix} 0 & k \\ 0 & 0
\end{bmatrix}.
\end{equation}
The vorticity operators $(\bm{\Psi}_{R},\bm{\Gamma}_{S})$ are evaluated as
\begin{equation}
	\bm{\Psi}_R \mapsto \begin{bmatrix} 2k & 0 \\ 0 & 0 \end{bmatrix} = \begin{bmatrix} 2\psi^- & 0 \\ 0 & 2\psi^+ \end{bmatrix}, \quad \bm{\Gamma}_S \mapsto \begin{bmatrix} -k  & 0 \\ 0 & k  \end{bmatrix} = \begin{bmatrix} \gamma^- & 0 \\ 0 & \gamma^+ \end{bmatrix}.
\end{equation}
\end{subequations}
Therefore, the Schur frames are determined as $	(\bm{e}_{1}^{+},\bm{e}_{2}^{+})= (\bm{i}_{y},\bm{i}_{x})~\text{or}~(-\bm{i}_{y},-\bm{i}_{x})$, and $(\bm{e}_{1}^{-},\bm{e}_{2}^{-}) = (\bm{i}_{x},-\bm{i}_{y})~\text{or}~(-\bm{i}_{x},\bm{i}_{y})$. Consequently, we have $\Delta=0$ and $\chi=\lambda_{r}=0$. Consider a perturbed velocity field $\bm{u}_{\varepsilon}=ky\bm{i}_{x}-\varepsilon x\bm{i}_{y}$, where $0<\varepsilon\ll 1$ and $k=\mathcal{O}(1)$. Since the vorticity is $\omega_{z}=-k-\varepsilon<0$, the rotation axis remains $\bm{e}_{3}=-\bm{e}_{z}$. In the reordered frame $(\bm{i}_{y},\bm{i}_{x},-\bm{i}_{z})$, the VGT $\bm{A}_{\varepsilon}\equiv\bm{\nabla}\bm{u}_{\varepsilon}$ is expressed as
\begin{subequations}
\begin{equation}
	\boldsymbol{A}_{\varepsilon}\mapsto\begin{bmatrix} 0 & k \\ -\varepsilon & 0 \end{bmatrix}.
\end{equation}
Accordingly, the vorticity operators $(\bm{\Psi}_{R\varepsilon},\bm{\Gamma}_{S\varepsilon})$ are evaluated as
\begin{equation}
	\bm{\Psi}_{R\varepsilon} \mapsto \begin{bmatrix} 2k & 0 \\ 0 & 2\varepsilon \end{bmatrix} 
	= \begin{bmatrix} 2\psi_{\varepsilon}^-& 0 \\ 0 & 2\psi_{\varepsilon}^+ \end{bmatrix}, \quad \bm{\Gamma}_{S\varepsilon} \mapsto \begin{bmatrix} -k+\varepsilon  & 0 \\ 0 & k-\varepsilon  \end{bmatrix} 
	= \begin{bmatrix} \gamma_{\varepsilon}^- & 0 \\ 0 & \gamma_{\varepsilon}^+ \end{bmatrix}.
\end{equation}
\end{subequations}
Therefore, the Schur frames $(\bm{e}_{1}^{\pm},\bm{e}_{2}^{\pm})$ for $\bm{u}_{\varepsilon}$ are the same as those of the unperturbed base flow $\bm{u}$. The discriminant becomes $\Delta_{\varepsilon}=4\psi_{\varepsilon}^{+}\psi_{\varepsilon}^{-}=4k\varepsilon>0$, and $\chi_{\varepsilon}=\lambda_{r\varepsilon}=0$. As $\varepsilon\rightarrow0^{+}$, $\Delta_{\varepsilon}\rightarrow 0^{+}$; the perturbed flow converges continuously to the unperturbed base flow. It is emphasized that, based on the foregoing discussion, $\bm{W}$ cannot be uniquely determined for the simple shear flow $\bm{u}$ when only the VGT at a single point is known. In contrast, one can obtain $\bm{W}=\bm{0}$ for all perturbed flows $\bm{u}_{\varepsilon}$ based on the VGT at a single point.

\subsubsection{Cases with negative discriminant}
Consider $\Delta<0$ at the point $\mathfrak{P}$. When $\gamma\neq 0$ and $\Xi\neq 0$, $\bm{W}$ can be uniquely determined from~\eqref{Schur_frame_angular1}, and the discussion is essentially analogous to that for $\Delta>0$. When $\gamma\neq0$ and $\Xi=0$, $\bm{W}$ cannot be uniquely determined at $\mathfrak{P}$. It follows from~\eqref{Xifactor} that $\Delta=-4(\chi-\lambda_{r})^2<0$, or equivalently $\psi^{+}\psi^{-}=-(\chi-\lambda_{r})^2<0$. This in fact implies an algebraic constraint for $F_{31}/F_{32}$.
\begin{itemize}
\item \textbf{Case A.} IVD-I with $2\psi=2\psi^+<0$ and $\gamma=\gamma^+>0$. The corresponding Schur frame is $(\bm{e}_{1},\bm{e}_{2})=(\bm{e}_{1}^{+},\bm{e}_{2}^{+})$.\\
In this case,~\eqref{eq:F31s} and~\eqref{eq:F32s} become
\begin{subequations}
\begin{align}
	F_{31} &= -W_2(\chi - \lambda_r) - W_1\psi^+, \label{eq:F31ss} \\
F_{32} &= -W_{2}\psi^{-} +W_1(\chi-\lambda_r). \label{eq:F32ss}
\end{align}
\end{subequations}
Hence, the ratio of $F_{31}/F_{32}$ is given by
\begin{equation}
	\frac{F_{31}}{F_{32}} = \operatorname{sgn}(\chi - \lambda_r) \sqrt{\left|\frac{\psi^+}{\psi^-}\right|}.
\end{equation}
where $\operatorname{sgn}$ denotes the sign function.
\item \textbf{Case B.} IVD-II with $2\psi=2\psi^->0$ and $\gamma=\gamma^-<0$. The corresponding Schur frame is $(\bm{e}_{1},\bm{e}_{2})=(\bm{e}_{1}^{-},\bm{e}_{2}^{-})$.\\
In this case,~\eqref{eq:F31s} and~\eqref{eq:F32s} become
\begin{subequations}
	\begin{align}
		F_{31} &= -W_2(\chi - \lambda_r) - W_1\psi^-, \label{eq:F31sss} \\
		F_{32} &= -W_{2}\psi^{+} +W_1(\chi-\lambda_r). \label{eq:F32sss}
	\end{align}
\end{subequations}
Hence, the ratio of $F_{31}/F_{32}$ is given by
\begin{equation}
	\frac{F_{31}}{F_{32}} = -\operatorname{sgn}(\chi - \lambda_r) \sqrt{\left|\frac{\psi^-}{\psi^+}\right|}.
\end{equation}
\end{itemize}

\subsubsection{Cases with zero vorticity}
When $\bm{\omega}=\bm{0}$ holds at $\mathfrak{P}$, $\bm{\varOmega}=\bm{O}$ and $\bm{A}=\bm{D}$. In this case, the requirement $\omega_{3}>0$ cannot be used to determine the rotation axis $\bm{e}_{3}$, which can be selected as one of the real eigenvectors based on an additional physical requirement in a certain problem. The relationship between the Schur basis vectors and the eigenvectors of $\bm{A}$ was discussed in~\citet{Chen2026Kinematic}. Under the Schur frame $(\bm{e}_{1},\bm{e}_{2},\bm{e}_{3})$, supposing $\gamma\neq 0$, it holds that $\alpha=\beta=0$ and $\gamma=-2\psi$, so that~\eqref{Schur} becomes
\begin{eqnarray}\label{Schur11}
	\bm{A}\mapsto A
	=\begin{bmatrix}
		\chi & -\psi & 0 \\
		-\psi & \chi & 0 \\
		0 & 0 & \lambda_r
	\end{bmatrix}.
\end{eqnarray}
The discriminant of $\bm{A}$ is $\Delta=-\gamma^2=-4\psi^2< 0$, and thus $\Xi=(\chi-\lambda_{r})^2-\psi^2$. For $\gamma\neq 0$ (i.e., $\psi\neq 0$) and $\Xi\neq 0$ (i.e., $\chi-\lambda_{r}\neq\pm\psi$), $\bm{W}$ can be uniquely determined from~\eqref{Schur_frame_angular1}. For $\gamma\neq 0$ and $\Xi=0$, we have $\gamma=-2\psi=-2(\chi-\lambda_{r})$ or $\gamma=-2\psi=2(\chi-\lambda_{r})$; in this case, $\bm{W}$ cannot be uniquely determined at $\mathfrak{P}$. Clearly, $\gamma\neq 0$ requires $\chi\neq\lambda_{r}$. Without loss of generality, we assume $\chi>\lambda_{r}$. Consequently, we have $2\psi^+=-2(\chi-\lambda_{r})$, $2\psi^-=2(\chi-\lambda_{r})$, $\gamma^+=2(\chi-\lambda_r)$, and $\gamma^-=-2(\chi-\lambda_r)$. 
\begin{itemize}
	\item \textbf{Case A.} IVD-I with $2\psi=2\psi^+<0$ and $\gamma=\gamma^+>0$. The corresponding Schur frame is $(\bm{e}_{1},\bm{e}_{2})=(\bm{e}_{1}^{+},\bm{e}_{2}^{+})$. In this case, $F_{31}=F_{32}$.
	\item \textbf{Case B.} IVD-II with $2\psi=2\psi^->0$ and $\gamma=\gamma^-<0$. The corresponding Schur frame is $(\bm{e}_{1},\bm{e}_{2})=(\bm{e}_{1}^{-},\bm{e}_{2}^{-})$. In this case, $F_{31}=-F_{32}$.
\end{itemize}

\subsubsection{Cases with $\gamma=0$ and $\gamma\rightarrow 0$}
For $\gamma=0$ at the point $\mathfrak{P}$, $\omega_{3}>0$ implies $\psi>0$. The fluid motion parallel to the plane $\mathscr{P}$ can be regarded as the superposition of isotropic expansion (with relative stretching rate $\chi$) and rigid rotation (with angular speed $\omega_{3}/2$). By using~\eqref{IVD-I} and~\eqref{IVD-II}, the vorticity modes are obtained as $(2\psi^{\pm},\gamma^{\pm})=(\omega_{3},0)$. Since $\Xi=(\chi-\lambda_r)^2+\psi^2>0$, $(W_{1},W_{2})$ can be uniquely determined from~\eqref{eq:F31s} and~\eqref{eq:F32s}. The vorticity operators reduce to $\bm{\Psi}_{R}=\omega_{3}(\mathbf{I}-\bm{e}_{3}\bm{e}_{3})$ and $\bm{\Gamma}_{S}=\mathbf{O}$. Therefore, any two right-handed
orthonormal vectors are their common eigenvectors, which form the Schur frame $(\bm{e}_{1}^{\pm},\bm{e}_{2}^{\pm})$. For $\gamma\rightarrow 0$ but $\gamma\neq 0$, we have $\Delta>0$ and $\Xi>0$. The flow behavior in the neighborhood of $\mathfrak{P}$ can be revealed by the motion of the Schur frame.
Assuming $f=\alpha W_{1}-\beta W_{2}+F_{22}-F_{11}=\mathcal{O}(\gamma^\eta)$, where $\eta>0$ is a positive constant,~\eqref{eq:W3} implies $W_{3}=f/(2\gamma)=\mathcal{O}(\gamma^{\eta-1})$. Here $f$ is a complex function that incorporates the effects of pressure and viscosity.
If $\eta\geq1$, then $W_3$ is either a higher-order small quantity or of the same order as $\gamma$, giving a finite contribution.
If $\eta<1$, then $W_{3}\rightarrow\infty$ (sufficiently large) as $\gamma\rightarrow 0$. Accordingly,~\eqref{ss11} gives ${D\bm{e}_{1}}/{Dt}\approx W_{3}\bm{e}_{2}$, ${D\bm{e}_{2}}/{Dt}\approx- W_{3}\bm{e}_{1}$, and $D\bm{e}_{3}/Dt\approx{0}$, which indicates a very rapid variation of the Schur basis vectors on $\mathscr{P}$. Physically, this could be associated with a strong swirling system consisting of an axial vortex and winding shear layers.
The condition $\gamma\rightarrow 0$ typically corresponds to the collapse of shear layers into the vortex core due to the intense entrainment, potentially leading to finite-time singularities.

\section{Lagrangian dynamics of the velocity gradient tensor and its rotational invariants}\label{UU2}
\subsection{Evolution equation for the velocity gradient tensor}
\begin{subequations}
For compressible viscous flow, the Navier-Stokes (NS) equations can be written as
\begin{equation}\label{NS_full}
\frac{\partial \bm{u}}{\partial t} + \bm{u} \cdot \bm{\nabla} \bm{u} = -\frac{1}{\rho}\bm{\nabla}{p}+\bm{\eta}^{\prime},
\end{equation}
where $\rho$ is the density and $p$ is the thermodynamic pressure. The source term $\bm{\eta}^{\prime}$ on the right-hand side is generally expressed as~\citep{Wu2006vorticity}
\begin{eqnarray}\label{eta_source_full}
	\bm{\eta}^\prime\equiv\frac{1}{\rho}\bm{\nabla}\left(\mu_{\vartheta}\vartheta\right)-\nu\bm{\nabla}\times\bm{\omega}+\frac{2}{\rho}\bm{\nabla}\mu\bm{\cdot}\bm{D}+\bm{f},
\end{eqnarray}
with $\bm{f}$ representing the external force per unit mass.
Note that the pressure gradient can be related to the gradients of the specific enthalpy $h=e+P$ ($P\equiv p/\rho$) and the specific entropy $s$ through the thermodynamic relation $-\bm{\nabla}p/\rho=-\bm{\nabla}h+T\bm{\nabla}s$. The longitudinal viscosity is defined as $\mu_{\vartheta}\equiv\mu_{b}+(4/3)\mu$, with $\mu_{b}$ and $\mu$ denoting the bulk and dynamic viscosities, respectively; the corresponding kinematic viscosities are denoted by $\nu_{\vartheta}$, $\nu_{b}$ and $\nu$.
Without affecting the primary physics, the linear diffusion approximation introduced by~\citet{Lighthill1956Viscosity} was adopted in which all viscosities are treated as constants. The term $T\bm{\nabla}s$ can be absorbed into $\bm{f}$ and therefore does not appear explicitly in the equations. It should be kept in mind, however, that the curl of \(T\bm{\nabla} s\) does not vanish in general; rather, it produces the torque \(\bm{\nabla} T\times\bm{\nabla} s\). This baroclinic--type contribution is negligible only when \(\bm{\nabla} T\) and \(\bm{\nabla} s\) are parallel. In strongly compressible flows, however, the two gradients are typically misaligned, so that the torque can play a non-negligible role in the vorticity and enstrophy balance and should therefore be retained. Finally, we arrive at the following compact form
\begin{equation}\label{NS}
\frac{\partial \bm{u}}{\partial t} + \bm{u} \cdot \bm{\nabla} \bm{u} = -\bm{\nabla}{h}+\bm{\eta},
\end{equation}
where the source term becomes
\begin{eqnarray}\label{expression_eta}
\bm{\eta}&\equiv&\nu_{\vartheta}\bm{\nabla}\vartheta- \nu \bm{\nabla} \times \bm{\omega} + \bm{f}\nonumber\\
&=&\left(\nu_{b}+\frac{1}{3}\nu\right)\bm{\nabla}\vartheta+\nu\nabla^2\bm{u}+\bm{f}.
\end{eqnarray}
\end{subequations}
Evaluating the spatial gradient of~\eqref{NS} yields the evolution equation for the VGT $\bm{A}$:
\begin{subequations}
\begin{equation}\label{VGT_eq1}
\frac{D\bm{A}}{Dt}=-\bm{A}^{2}-\bm{\nabla\nabla}{h}+\bm{\Sigma}.
\end{equation}
On the right-hand side, the first term $-\bm{A}^2$ originates from the divergence of the convection term $\bm{u} \cdot \bm{\nabla} \bm{u}$, which accounts for the self-amplification (or attenuation) mechanism of the VGT~\citep{Meneveau2011Lagrangian,Johnson2024Multiscale}. The symmetric tensor $\bm{\nabla\nabla}{h}=\partial_{i}\partial_{j}{h}\,\bm{e}_{i}\bm{e}_{j}$ is referred to as the enthalpy Hessian tensor, whose components in the Schur frame are denoted by $\partial_{i}\partial_{j}{h}=\bm{e}_{i}\bm{\cdot}\bm{\nabla\nabla}h\bm{\cdot}\bm{e}_{j}$. The source term $\bm{\Sigma}=\Sigma_{ij}\bm{e}_{i}\bm{e}_{j}$ collects the gradient effects associated with dilatation, viscous diffusion, and external forcing, i.e.,
\begin{equation}\label{source_term_Sigma}
\bm{\Sigma}\equiv\bm{\nabla\eta}=\left(\nu_{b}+\frac{1}{3}\nu\right)\bm{\nabla\nabla}\vartheta+\nu\nabla^2\bm{A}+\bm{\nabla}\bm{f}.
\end{equation}
\end{subequations}
In the incompressible limit, the enthalpy Hessian tensor $\bm{\nabla\nabla}h$ should be replaced by the pressure Hessian tensor $\bm{\nabla\nabla}P$, and~\eqref{VGT_eq1} and~\eqref{source_term_Sigma} reduce to
\begin{equation}\label{VGT_eq2}
	\frac{D\bm{A}}{Dt}=-\bm{A}^{2}-\bm{\nabla\nabla}{P}+\nu\nabla^{2}\bm{A}+\bm{\nabla}\bm{f}.
\end{equation}

\subsection{Schur representation of the material derivative of the velocity gradient tensor}
By virtue of~\eqref{basis_transform} and~\eqref{ss3}, we obtain
\begin{eqnarray}\label{ss12}
	F_{ij}&=&Q_{pi}\dot{C}_{pq}Q_{qj}=\left(\bm{e}_{i}\bm{\cdot}\bm{i}_{p}\right)\dot{C}_{pq}(\bm{i}_{q}\bm{\cdot}\bm{e}_{j})\nonumber\\
	&=&\bm{e}_{i}\bm{\cdot}(\dot{C}_{pq}\bm{i}_{p}\bm{i}_{q})\bm{\cdot}\bm{e}_{j}=\bm{e}_{i}\bm{\cdot}\dot{\bm{A}}\bm{\cdot}\bm{e}_{j}.
\end{eqnarray}
Therefore, $F_{ij}~(i,j=1,2,3)$ just denote the matrix entries of $\dot{\bm{A}}$ (rather than $\dot{A}$):
\begin{eqnarray}
	\dot{\bm{A}}=F_{ij}\bm{e}_{i}\bm{e}_{j}\mapsto F.
\end{eqnarray}
Because of $\bm{e}_{i}\bm{\cdot}\left(-\bm{A}^{2}\right)\bm{\cdot}\bm{e}_{j}=-A_{im}A_{mj}$, the matrix of $-\bm{A}^{2}$ is given by
\begin{eqnarray}\label{v1}
-A^2 = 
\begin{bmatrix}
	-\chi^2 + \psi(\psi+\gamma) & -2\chi(\psi+\gamma) & \beta(\chi+\lambda_{r}) - \alpha(\psi+\gamma) \\
	2\chi\psi & -\chi^2 + \psi(\psi+\gamma) & -\beta\psi - \alpha(\chi+\lambda_r) \\
	0 & 0 & -\lambda_r^2
\end{bmatrix}.
\end{eqnarray}
From~\eqref{VGT_eq1} and~\eqref{v1}, one obtains the Schur representation for the material derivative of $\bm{A}$:
\begin{eqnarray}\label{v2}
	F=
	\begin{bmatrix}
		F_{11} & F_{12} & F_{13} \\
		F_{21} & F_{22} & F_{23} \\
		F_{31} & F_{32} & F_{33}
	\end{bmatrix}=-A^2-[\partial_{i}\partial_{j}h]+[\Sigma_{ij}],
\end{eqnarray}
where the matrix elements are respectively given by
\begin{subequations}\label{v3}
	\begin{align}
		F_{11} &= -\chi^2 +\psi(\psi+\gamma) - \partial_{1}\partial_{1}h + \Sigma_{11}, \label{eq:F11} \\
		F_{12} &= -2\chi(\psi+\gamma) - \partial_{1}\partial_{2}h + \Sigma_{12}, \label{eq:F12} \\
		F_{13} &= \beta(\chi+\lambda_{r}) - \alpha(\psi+\gamma) - \partial_{1}\partial_{3} h + \Sigma_{13}, \label{eq:F13} \\
		F_{21} &= 2\chi\psi - \partial_2\partial_1 h + \Sigma_{21}, \label{eq:F21} \\
		F_{22} &= -\chi^2 +\psi(\psi+\gamma)- \partial_2\partial_2 h + \Sigma_{22}, \label{eq:F22} \\
		F_{23} &= -\beta\psi - \alpha(\chi+\lambda_{r}) - \partial_2\partial_3 h + \Sigma_{23}, \label{eq:F23} \\
		F_{31} &= -\partial_3\partial_1 h + \Sigma_{31}, \label{eq:F31_new} \\
		F_{32} &= -\partial_3\partial_2 h + \Sigma_{32}, \label{eq:F32_new} \\
		F_{33} &= -\lambda_r^2 - \partial_3\partial_3 h + \Sigma_{33}. \label{eq:F33}
	\end{align}
\end{subequations}

\subsection{Evolution equations for the rotational invariants of the velocity gradient tensor}
Substituting~\eqref{v3} into~\eqref{ss17}~--~\eqref{Schur_frame_angular1}, we obtain the evolution equations for the rotational VGT invariants:
\begin{subequations}\label{new1}
	\begin{align}
		\dfrac{D\alpha}{Dt} &= - W_2(2\psi + \gamma) + W_3\beta - \beta\psi - \alpha(\chi + \lambda_r) + \Sigma_{23} - \Sigma_{32}, \label{eq:new_alpha} \\
		\dfrac{D\beta}{Dt} &= W_1(2\psi + \gamma) - W_3\alpha - \beta(\chi + \lambda_r) + \alpha(\psi + \gamma) + \Sigma_{31} - \Sigma_{13}, \label{eq:new_beta} \\
		\dfrac{D\gamma}{Dt} &= -W_1\beta - W_2\alpha - 2\chi\gamma - 2\partial_1\partial_2 h + \Sigma_{12} + \Sigma_{21}, \label{eq:new_gamma} \\
		\dfrac{D\psi}{Dt} &= W_2\alpha - 2\chi\psi + \partial_{1}\partial_{2} h - \Sigma_{21}, \label{eq:new_psi} \\
		\dfrac{D\lambda_r}{Dt} &= -W_2\beta - W_1\alpha - \lambda_r^2 - \partial_3\partial_3 h + \Sigma_{33}, \label{eq:new_lambda_r} \\
\dfrac{D\chi}{Dt}&=\frac{1}{2}(W_{1}\alpha+W_{2}\beta)- \chi^2 + \psi(\psi + \gamma)-\frac{1}{2}\left(\partial_1\partial_1 h + \partial_2\partial_2 h\right)+\frac{1}{2}\left( \Sigma_{11} + \Sigma_{22}\right).
	\end{align}
\end{subequations}
where the angular velocity components are expressed as
\begin{subequations}\label{Schur_frame_angular2}
	\begin{align}
		W_1 &= -(\psi+\gamma)\frac{-\partial_3\partial_1{h}+\Sigma_{31}}{\Xi}
		+(\chi-\lambda_r)\frac{-\partial_3\partial_2{h}+\Sigma_{32}}{\Xi}, \label{eq:W1_new} \\
		W_2 &= -(\chi-\lambda_r)\frac{-\partial_3\partial_1{h}+\Sigma_{31}}{\Xi}
		-\psi\frac{-\partial_3\partial_2{h}+\Sigma_{32}}{\Xi}, \label{eq:W2_new} \\
		\begin{split}
			W_3 &= \dfrac{1}{2\gamma}\Big[W_1\alpha - W_2\beta + \partial_1\partial_1 h - \partial_2\partial_2 h + \Sigma_{22} - \Sigma_{11}\Big] \\
			&= \frac{1}{2}\left[
			\left(\frac{\beta}{\gamma}(\chi-\lambda_r)-\frac{\alpha}{\gamma}(\psi+\gamma)\right)\frac{-\partial_3\partial_1 h+\Sigma_{31}}{\Xi} \right. \\
			&\quad\left.+\left(\frac{\alpha}{\gamma}(\chi-\lambda_r)+\frac{\beta}{\gamma}\psi\right)\frac{-\partial_3\partial_2h+\Sigma_{32}}{\Xi} \right. \\
			&\quad\left.+\frac{1}{\gamma}(\partial_1\partial_1h-\partial_2\partial_2h+\Sigma_{22}-\Sigma_{11})
			\right].
		\end{split} \label{eq:W3_new}
	\end{align}
\end{subequations}
It is worth noting that both $F_{31}$ in~\eqref{eq:F31_new} and $F_{32}$ in~\eqref{eq:F32_new} are governed entirely by the off-diagonal components of the enthalpy Hessian tensor $\bm{\nabla\nabla}h$ and the source term $\bm{\Sigma}$. Therefore, from~\eqref{Schur_frame_angular1},~\eqref{v3} and~\eqref{Schur_frame_angular2}, it is seen that $W_{1}$ and $W_{2}$ depend only on the off-diagonal elements of $-\bm{\nabla\nabla}h+\bm{\Sigma}$, whereas $W_{3}$ is additionally influenced by $\partial_1\partial_1h-\partial_2\partial_2h+\Sigma_{22}-\Sigma_{11}=(-\partial_2\partial_2h+\Sigma_{22})-(-\partial_{1}\partial_{1}h+\Sigma_{11})$ due to the difference between the diagonal elements defined on the rotation-axis-normal plane $\mathscr{P}$. In regions where the pressure Hessian and the source term are not dominant, the above equations reduce to the following simplified form:
\begin{subequations}\label{x611}
	\begin{align}
		\frac{D\alpha}{Dt}            & \approx -\beta\psi-\alpha(\chi+\lambda_r), \label{x611a} \\
		\frac{D\beta}{Dt}             & \approx -\beta(\chi+\lambda_r)+\alpha(\psi+\gamma), \label{x611b} \\
		\frac{D\gamma}{Dt}            & \approx -2\chi\gamma, \label{x611c} \\
		\frac{D\psi}{Dt}              & = -2\chi\psi, \label{x611d} \\
		\frac{D\lambda_r}{Dt}         & \approx -\lambda_r^2, \label{x611e} \\
		\frac{D\chi}{Dt}              & \approx -\chi^2+\psi(\psi+\gamma)=-\chi^2 + \frac{1}{4}\Delta. \label{x611f}
	\end{align}
\end{subequations}
In~\eqref{x611a} and~\eqref{x611b}, the temporal variation rates of $\alpha$ and $\beta$ arise from their coupling with the normal straining components $(\chi,\lambda_{r})$ and the vorticity-mode invariants $(\psi,\gamma)$. For positive $\chi$, $\gamma$~\eqref{x611c} and $\psi$~\eqref{x611d} preserve their signs and are either both damped or both amplified, depending on the sign of the time average of $\chi$. If $\chi>0$ for a sufficiently long time, then $\gamma,\psi\rightarrow 0$; if $\chi<0$, they may grow exponentially. The axial stretching rate $\lambda_{r}$ in~\eqref{x611e} admits a solution $\lambda_{r}=\lambda_{r}(t_0)/[1+(t-t_{0})\lambda_{r}(t_0)]$, provided $1+(t-t_{0})\lambda_{r}(t_0)\neq 0$. Thus, if $\lambda_{r}(t_0)>0$, then $\lambda_{r}$ decays algebraically and tends to zero as $t\rightarrow+\infty$. If $\lambda_{r}(t_0)<0$, it blows up at the finite time $t=t_{0}-1/\lambda_{r}(t_0)$. Equation~\eqref{x611f} presents a Riccati-type equation for $\chi$, with a time-dependent forcing term $\Delta/4$. For $\Delta>0$, the sign of $D\chi/Dt$ depends on the relative magnitude of $\chi$ and $\sqrt{\Delta}/2$.
\subsection{Schur representation of the principal invariants of the velocity gradient tensor}
The characteristic equation for $\bm{A}$ is given by
\begin{equation}
	\lambda^3+\mathcal{P}\lambda^2+\mathcal{Q}\lambda+\mathcal{R}=0,
\end{equation}
where the first, second and third principal invariants of $\bm{A}$ are respectively defined as~\citep{Chong1990General}
\begin{subequations}
\begin{align}
	\mathcal{P} &\equiv -\operatorname{tr}(\bm{A}) = -\vartheta, \label{eq:P} \\
	\mathcal{Q} &\equiv \frac{1}{2}\left[\vartheta^2 - \operatorname{tr}(\bm{A}^2)\right], \label{eq:Q} \\
	\mathcal{R} &\equiv -\det(\bm{A}) 
	= \frac{1}{3}\left[\vartheta^3 - 3\vartheta \mathcal{Q} - \operatorname{tr}(\bm{A}^3)\right]. \label{eq:R}
\end{align}
\end{subequations}
Here,
${\rm tr}$ denotes the trace operator, and the dilatation  $\vartheta\equiv\bm{\nabla}\bm{\cdot}\bm{u}$ characterizes the fluid compressibility. Note that the Cayley-Hamilton theorem has been applied in the derivation of~\eqref{eq:R}.

For incompressible flow, the physical interpretations of $(\mathcal{Q},\mathcal{R})$ are as follows. The second principal invariant can be decomposed as $\mathcal{Q}=(\lVert\bm{\varOmega}\rVert^2-\lVert\bm{D}\rVert^2)/2$, where $\lVert\bm{\varOmega}\rVert^2\equiv\bm{\varOmega}\bm{:}\bm{\varOmega}=-{\rm tr}(\bm{\varOmega}^2)=\omega^2/2$ denotes the enstrophy, and $\lVert\bm{D}\rVert^2\equiv\bm{D:D}={\rm tr}(\bm{D}^2)$ is the squared magnitude of the rate-of-strain tensor. Thus, $\mathcal{Q}$ quantifies the local balance between rotational and straining motions~\citep{Jeong1995Vortex}. The $\mathcal{Q}$-criterion, which identifies vortical structures by the condition $\mathcal{Q}>0$, has been widely used in turbulence research, although it does not strictly guarantee the existence of a local pressure minimum within the identified region~\citep{Hunt1988Eddies,Jeong1995Vortex}; see also \S\ref{secV1}. Similarly, the third principal invariant is decomposed as $\mathcal{R}=-(1/3){\rm tr}(\bm{D}^3)-(1/4)\bm{\omega}\bm{\cdot}\bm{D}\bm{\cdot}\bm{\omega}$, which represents the balance between enstrophy production (associated with vortex stretching) and dissipation production (associated with strain self-amplification)~\citep{Johnson2024Multiscale,Chen2024boundary}. The gradient self-amplification mechanism tends to generate intermittent extreme events in turbulent flows.

By using~\eqref{v1}, the trace of $\bm{A}^2$ is evaluated as
\begin{eqnarray}\label{g4b}
	{\rm tr}(\bm{A}^2)
	&=&2\chi^2+\lambda_{r}^{2}-\frac{1}{2}\Delta\nonumber\\
	&=&2\chi^2+\lambda_{r}^{2}-2\lambda_{\rm ci}^2.
\end{eqnarray}
Similarly, the trace of $\bm{A}^3$ is evaluated as 
\begin{eqnarray}\label{e6p13}
{\rm tr}(\bm{A}^3)&=&2\chi^3+\lambda_{r}^3-\frac{3}{2}\chi\Delta\nonumber\\
&=&2\chi^3+\lambda_{r}^3-6\chi\lambda_{\mathrm{ci}}^2.
\end{eqnarray}
Substituting~\eqref{g4b} and~\eqref{e6p13} into~\eqref{eq:Q} and~\eqref{eq:R}, we obtain the Schur representations for the principal invariants in compressible flow:
\begin{subequations}
	\begin{align}
		\mathcal{P}&=-(2\chi+\lambda_{r}),\label{eq:P1}\\
		\mathcal{Q} &= \frac{1}{4}\Delta+\chi^2+2\chi\lambda_{r}, \label{eq:Q1} \\
		\mathcal{R} &=-\lambda_{r}\left(\frac{1}{4}\Delta+\chi^2\right)=-\lambda_{r}\left(\mathcal{Q}-2\chi\lambda_{r}\right). \label{eq:R1}
	\end{align}
\end{subequations}

For incompressible flow, we have $\mathcal{P}=0$ (i.e., $\lambda_{r}=-2\chi$), and therefore
\begin{subequations}
	\begin{align}
		\mathcal{Q} &=-\frac{1}{2}{\rm tr}(\bm{A}^2)= \frac{1}{4}\Delta-3\chi^2=\psi^2+\psi\gamma-3\chi^2, \label{eq:Q11} \\
		\mathcal{R} &=-\frac{1}{3}{\rm tr}(\bm{A}^3)= 2\chi\left(\frac{1}{4}\Delta+\chi^2\right) \notag \\
		&= 2\chi\left(\mathcal{Q}+4\chi^2\right) = 2\chi\mathcal{Q}+8\chi^3\notag\\
		&=2\psi\chi\psi+2\psi\chi\gamma+2\chi^3. \label{eq:R11}
	\end{align}
\end{subequations}
Equation~\eqref{eq:Q11} implies that $\mathcal{Q}>0\Rightarrow\Delta>0$, which aligns with the earlier finding of~\citet{Chakraborty2005} that ``$\mathcal{Q}>0$ criterion is more restrictive than $\Delta>0$ criterion''. Furthermore, it provides a triple decomposition of $\mathcal{Q}$ in incompressible flow: a positive contribution from the characteristic rigid-rotation mode $(\psi^2)$, a negative contribution from the axial straining $(-3\chi^2)$, and a cross term $(\psi\gamma)$ with sign flexibility. In~\eqref{eq:R11}, the terms $2\psi\chi\psi$ and $2\psi\chi\gamma$ arise from enstrophy production, whereas $2\chi^3$ originates from strain self-amplification. Indeed, by making use of~\eqref{eq:D_matrix} and~\eqref{IVD_general}, we obtain
\begin{subequations}
	\begin{align}
		-\frac{1}{3} \operatorname{tr}(\bm{D}^3) &= 2\chi^3 - \frac12 \chi\gamma^2 + \frac14 \chi(\alpha^2 + \beta^2) + \frac14 \alpha\beta\gamma, \label{a618} \\[4pt]
		-\frac{1}{4} \bm{\omega}\bm{\cdot}\bm{D}\bm{\cdot}\bm{\omega} &= \frac{1}{2} \chi \omega_{3}^2 -\frac14\chi(\alpha^2+\beta^2) - \frac14 \alpha\beta\gamma, \label{a619}
	\end{align}
\end{subequations}
where the first term on the right-hand side of~\eqref{a619} is partitioned into
\begin{equation}
\frac{1}{2} \chi \omega_{3}^2=2\psi\chi\psi+2\psi\chi\gamma+\frac{1}{2}\chi\gamma^2.
\end{equation}
When summing~\eqref{a618} and~\eqref{a619}, the last two terms in~\eqref{a618} cancel with those in~\eqref{a619}. Then, applying the last equality in~\eqref{delta}, we obtain
\begin{eqnarray}
	-\frac{1}{3} \operatorname{tr}(\bm{D}^3)-\frac{1}{4} \bm{\omega}\bm{\cdot}\bm{D}\bm{\cdot}\bm{\omega}&=&2\chi^3+\frac{1}{2}\chi\left(\omega_{3}^2-\gamma^2\right)\nonumber\\
	&=&2\chi^3+\frac{1}{2}\chi\Delta\nonumber\\
	&=&\mathcal{R}.
\end{eqnarray}

\subsection{Evolution equations for the derived invariants of the velocity gradient tensor}\label{secV1}
By applying~\eqref{new1}, we obtain the evolution equations for the derived VGT invariants below.
\begin{itemize}
	\item The discriminant of VGT $\Delta=4\psi(\psi+\gamma)$
	\begin{subequations}
	\begin{eqnarray}\label{g1}
\frac{D\Delta}{Dt} 
= 
4(\psi+\gamma) W_2 \alpha 
- 4\psi W_1 \beta 
- 4\chi \Delta 
+ 4\gamma \partial_1 \partial_2 h
+ 4\psi \Sigma_{12} 
- 4(\psi+\gamma) \Sigma_{21}.
	\end{eqnarray}
When the pressure Hessian and the source term are neglected,~\eqref{g1} simplifies to
\begin{eqnarray}
\frac{D\Delta}{Dt}\approx-4\chi\Delta.
\end{eqnarray}
\end{subequations}
	\item The imaginary part of the complex-conjugate eigenpair $\lambda_{\rm ci}=\frac{1}{2}\sqrt{\Delta}=\sqrt{\psi(\psi+\gamma)}$
		\begin{subequations}
	\begin{eqnarray}\label{g2}
\frac{D\lambda_{\text{ci}}}{Dt} = 
-2\chi\lambda_{\text{ci}} 
+ \frac{1}{2\lambda_{\text{ci}}}
\left[
(\psi+\gamma)W_2\alpha - \psi W_1\beta + \gamma \partial_1\partial_2 h+ \psi\Sigma_{12} - (\psi+\gamma)\Sigma_{21}
\right].
	\end{eqnarray}
When the pressure Hessian and the source term are neglected,~\eqref{g2} simplifies to
	\begin{eqnarray}\label{g2a}
		\frac{D\lambda_{\rm ci}}{Dt}\approx-2\chi\lambda_{\rm ci}.
	\end{eqnarray}
	\end{subequations}
	\item The axial vorticity component $\omega_{3}=2\psi+\gamma$
	\begin{subequations}
	\begin{eqnarray}\label{g3}
	\frac{D\omega_3}{Dt} = -W_1\beta + W_2\alpha - 2\chi\omega_3 + \Sigma_{12} - \Sigma_{21}.
	\end{eqnarray}
When the pressure Hessian and the source term are neglected,~\eqref{g3} simplifies to
	\begin{eqnarray}\label{g3a}
		\frac{D\omega_{3}}{Dt}\approx-2\chi\omega_{3}.
	\end{eqnarray}
	\end{subequations}
	\item The characteristic rigid-rotation mode $\phi\equiv\psi+\gamma$
	\begin{subequations}
	\begin{eqnarray}\label{DphiDt}
	\frac{D\phi}{Dt} = -W_1\beta - 2\chi\phi - \partial_1\partial_2 h+ \Sigma_{12}.
	\end{eqnarray}
	It is noted that, in the Schur frame $(\bm{e}_{1},\bm{e}_{2})=(\bm{e}_{1}^{+},\bm{e}_{2}^{+})$, we have $(\psi,\gamma)=(\psi^+,\gamma^+)$, which implies $\phi=\psi^{-}$. In the Schur frame, $(\bm{e}_{1},\bm{e}_{2})=(\bm{e}_{1}^{-},\bm{e}_{2}^{-})$, $(\psi,\gamma)=(\psi^-,\gamma^-)$ and thus $\phi=\psi^{+}$. When the pressure Hessian and the source term are neglected,~\eqref{DphiDt} simplifies to
		\begin{eqnarray}\label{DphiDt1}
		\frac{D\phi}{Dt} \approx - 2\chi\phi .
	\end{eqnarray}
	\end{subequations}
		\item The dilatation $\vartheta=-\mathcal{P}=2\chi+\lambda_{r}$
	\begin{subequations}
		\begin{eqnarray}\label{g4}
			\frac{D\vartheta}{Dt}&=& -\vartheta^2 + 4\chi\vartheta - 6\chi^2 + \frac{1}{2}\Delta- \nabla^2 h+ \mathrm{tr}(\bm{\Sigma})\nonumber\\
			&=&-\left(2\chi^2+\lambda_{r}^2-\frac{1}{2}\Delta\right)-\nabla^2 h+{\rm tr}(\bm{\Sigma}).
		\end{eqnarray}
		This equation may also be directly derived by taking the trace of~\eqref{VGT_eq1}, which yields
		\begin{eqnarray}\label{g4a}
			\frac{D\vartheta}{Dt}=-{\rm tr}(\bm{A}^2)-\nabla^2 h+{\rm tr}(\bm{\Sigma}).
		\end{eqnarray}
		This implies the Poisson equation for the enthalpy
		\begin{eqnarray}\label{Poisson1}
		\nabla^2 h=-\frac{D\vartheta}{Dt}-{\rm tr}(\bm{A}^2)+{\rm tr}(\bm{\Sigma}).
		\end{eqnarray}
		Substituting~\eqref{g4b} into~\eqref{g4a} recovers~\eqref{g4}. For incompressible flow with $\bm{\Sigma}=\bm{0}$ ($h$ should be replaced by $P$),~\eqref{Poisson1} implies $\nabla^2 P=-{\rm tr}(\bm{A}^2)=2\mathcal Q$, where $2\mathcal{Q}$ in~\eqref{eq:Q11} acts as the source term. The solution of the Poisson equation indicates the non-local dependence of the pressure field upon the spatial distribution of $\mathcal{Q}$~\citep{Lawson2015,Johnson2024Multiscale}.
	\end{subequations}
	\item The second principal invariant $\mathcal Q$
	\begin{subequations}
	\begin{equation}\label{eq:DQ}
		\begin{aligned}
			\frac{D\mathcal{Q}}{Dt} &= (\lambda_r - \chi)(W_1\alpha+W_2\beta) + (\psi+\gamma)W_2\alpha - \psi W_1\beta \\
			&\quad - (\lambda_r + 2\chi)\mathcal{Q} - 3\mathcal{R} \\
			&\quad -(\chi+\lambda_r)(\partial_1^2h+\partial_2^2h) + \gamma\partial_1\partial_2h - 2\chi\partial_3^2h \\
			&\quad + (\chi+\lambda_r)(\Sigma_{11}+\Sigma_{22}) + \psi\Sigma_{12} - (\psi+\gamma)\Sigma_{21} + 2\chi\Sigma_{33},
		\end{aligned}
	\end{equation}
	For incompressible flow $(\lambda_{r}=-2\chi)$, retaining only the isotropic pressure-Hessian contribution and neglecting the source term,~\eqref{eq:DQ} can be approximated as
	\begin{eqnarray}\label{eq7p16}
	\frac{D\mathcal{Q}}{Dt}&\approx&-3\mathcal{R}-(\chi+\lambda_r)\left(\frac{2}{3}\nabla^{2}P\right) - 2\chi\left(\frac{1}{3}\nabla^{2}P\right)\nonumber\\
	&=&-3\mathcal{R}+\chi\left(\frac{2}{3}\nabla^{2}P\right) - 2\chi\left(\frac{1}{3}\nabla^{2}P\right)\nonumber\\
	&=&-3\mathcal{R}.
	\end{eqnarray}
	\end{subequations}
	\item The third principal invariant $\mathcal{R}$
	\begin{subequations}
	\begin{equation}
		\label{eq:R_evolution}
		\begin{aligned}
			\frac{DR}{Dt} 
			&= \left( \xi - \lambda_r\chi \right)\left( W_1\alpha + W_2\beta \right) 
			- \lambda_r\left( \psi+\gamma \right) W_2\alpha 
			+ \lambda_r\psi W_1\beta \\
			&\quad-(\lambda_{r}+2\chi)\mathcal{R} \\
			&\quad + \lambda_r\chi\left( \partial_1\partial_1 h + \partial_2\partial_2 h \right) 
			- \lambda_r\gamma \partial_1\partial_2 h 
			+ \xi\partial_3\partial_3 h \\
			&\quad - \lambda_r\chi\left( \Sigma_{11}+\Sigma_{22} \right) 
			- \lambda_r\psi \Sigma_{12} 
			+ \lambda_r\left( \psi+\gamma \right)\Sigma_{21} 
			- \xi\Sigma_{33},
		\end{aligned}
	\end{equation}
	where the factor $\xi$ is defined as $\xi\equiv\chi^2+\Delta/4=\chi^2+\psi(\psi+\gamma)$. 	For incompressible flow $(\lambda_{r}=-2\chi)$, retaining only the isotropic pressure-Hessian contribution and neglecting the source term,~\eqref{eq:R_evolution} can be approximated as
	\begin{eqnarray}\label{eq7p18}
	\frac{D\mathcal{R}}{Dt}&\approx&\lambda_r\chi\left( \frac{2}{3}\nabla^{2}P \right) 
	+ \xi\left( \frac{1}{3}\nabla^{2}P \right) \nonumber\\
	&=&\frac{1}{3}\nabla^{2}P(\xi-4\chi^2)\nonumber\\
	&=&\frac{1}{3}\nabla^{2}P\left(\frac{1}{4}\Delta-3\chi^2\right)\nonumber\\
	&=&\frac{2}{3}\mathcal{Q}^2.
	\end{eqnarray}
	\end{subequations}
	Here,~\eqref{eq7p16} and~\eqref{eq7p18} just recover the restricted Euler equations for the principal invariant pair $(\mathcal{Q},\mathcal{R})$~\citep{Vieillefosse1982Local,Vieillefosse1984Internal}, which feature a time invariant given by the discriminant $27\mathcal{R}^2/4+\mathcal{Q}^3=\text{constant}$. In the limit $\mathcal{R}\rightarrow\infty$, $\mathcal{Q}$ approaches the Vieillefosse tail, $(\mathcal{Q}=-3\mathcal{R}^{2/3}/2^{2/3})$~\citep{Cantwell1992}. The joint probability density function (PDF) of $(\mathcal{Q},\mathcal{R})$ exhibits a universal asymmetric teardrop shape, with enhanced probability density along the Vieillefosse
	tail, across a wide range of flow configurations and Reynolds numbers~\citep{Cantwell1993,Johnson2024Multiscale}. The pressure and viscous terms, which are neglected in the restricted Euler system, are responsible for regularizing the finite-time singularity that would otherwise arise~\citep{Meneveau2011Lagrangian}.
\end{itemize}

\section{Lagrangian dynamics of the characteristic vorticity modes}\label{UU3}
\subsection{Evolution equation for the vorticity}
The antisymmetric part of $\bm{A}^2$ is $\mathscr{A}(\bm{A}^2)=\bm{D}\bm{\cdot}\bm{\varOmega}+\bm{\varOmega}\bm{\cdot}\bm{D}$, where $\mathscr{A}$ denotes the antisymmetrization operator.
Applying $\mathscr{A}$ to both sides of~\eqref{VGT_eq1} and multiplying the result by $2$, we obtain
\begin{subequations}
\begin{equation}\label{cc1}
	2\frac{D\bm{\varOmega}}{Dt}=-2\left(\bm{D}\bm{\cdot}\bm{\varOmega}+\bm{\varOmega}\bm{\cdot}\bm{D}\right)+2\mathscr{A}(\bm{\Sigma}),
\end{equation}
where the last term is given by
\begin{equation}
	2\mathscr{A}(\bm{\Sigma})\equiv\bm{\Sigma}-\bm{\Sigma}^{\rm T}=2\nu\nabla^2\bm{\varOmega}+2\mathscr{A}(\bm{\nabla}\bm{f}).
\end{equation} 
\end{subequations}
Upon applying the Hodge star operator to every term of~\eqref{cc1}, we get
\begin{subequations}\label{eq7p3}
	\begin{gather}
		\star\left(2\frac{D\bm{\varOmega}}{Dt}\right)=\frac{D\bm{\omega}}{Dt}, \label{eq7p3a} \\
		\star\left(-2\left(\bm{D}\bm{\cdot}\bm{\varOmega}+\bm{\varOmega}\bm{\cdot}\bm{D}\right)\right)=-\vartheta\bm{\omega}+\bm{\omega}\bm{\cdot}\bm{D}, \label{eq7p3b} \\
		\star\left(2\mathscr{A}(\bm{\Sigma})\right)=\nu\nabla^2\bm{\omega}+\bm{\nabla}\times\bm{f}. \label{eq7p3c}
	\end{gather}
\end{subequations}
A detailed proof of~\eqref{eq7p3b} is provided in Appendix~\ref{app1}. Therefore, the transport equation for the vorticity is derived as
\begin{eqnarray}\label{vv}
	\frac{D\bm{\omega}}{Dt}&=&-\vartheta\bm{\omega}+\bm{\omega}\bm{\cdot}\bm{A}+\star\left(2\mathscr{A}(\bm{\Sigma})\right)\nonumber\\
	&=&-\vartheta\bm{\omega}+\bm{\omega}\bm{\cdot}\bm{A}+\nu\nabla^2\bm{\omega}+\bm{\nabla}\times\bm{f}.
\end{eqnarray}
The left-hand side of~\eqref{vv} represents the material derivative of vorticity. On the right-hand side of~\eqref{vv}, the term $-\vartheta\bm{\omega}$ represents the coupling between dilatation and vorticity; $\bm{\omega}\bm{\cdot}\bm{A}=\bm{\omega}\bm{\cdot}\bm{D}$ represents the stretching and tilting of vorticity lines by the strain-rate tensor, which is usually referred to as the vortex stretching term; the last term $\star\left(2\mathscr{A}(\bm{\Sigma})\right)$ arises from the viscous diffusion of vorticity and the curl of external forcing. In the Schur frame $(\bm{e}_{1},\bm{e}_{2},\bm{e}_{3})$, these terms can be expanded as follows.
\begin{itemize}
	\item The first term  $D\bm{\omega}/Dt$
	\begin{subequations}\label{DW12}
		\begin{gather}
			\frac{D\bm{\omega}}{Dt}=\frac{D\omega_{i}}{Dt}\bm{e}_{i}+\omega_{i}\frac{D\bm{e}_{i}}{Dt}
			=\left(\frac{D\omega_{i}}{Dt}+\omega_{j}W_{ji}\right)\bm{e}_{i}, \label{DW1} \\[6pt]
			\frac{D\bm{\omega}}{Dt}\mapsto
			\begin{bmatrix}
				\dfrac{D\alpha}{Dt} \\[8pt]
				\dfrac{D\beta}{Dt} \\[8pt]
				\dfrac{D\omega_3}{Dt}
			\end{bmatrix}
			+
			\begin{bmatrix}
				-W_3\beta + W_2\omega_3 \\
				W_3\alpha - W_1\omega_3 \\
				-W_2\alpha + W_1\beta
			\end{bmatrix}. \label{DW2}
		\end{gather}
	\end{subequations}
Equations~\eqref{DW1} and~\eqref{DW2} provide the decomposition of the material derivative of the vorticity into one contribution associated with the intrinsic changes in the Schur frame and the other contribution due to frame rotation along a pathline.
\item  The second term $-\vartheta\bm{\omega}$
\begin{eqnarray}\label{The second term}
	-\vartheta\bm{\omega}=-\vartheta\omega_{i}\bm{e}_{i}\mapsto-\vartheta
	\begin{bmatrix}
		\alpha \\
		\beta \\
		\omega_3
	\end{bmatrix}
	=\begin{bmatrix}
		-(2\chi + \lambda_r)\alpha \\
		-(2\chi + \lambda_r)\beta \\
		-(2\chi + \lambda_r)\omega_3
	\end{bmatrix}.
\end{eqnarray}
\item The third term $\bm{\omega}\bm{\cdot}\bm{A}$
\begin{eqnarray}\label{The third term}
	\bm{\omega}\bm{\cdot}\bm{A}=A^{\rm T}_{ij}\omega_{j}\bm{e}_{i}\mapsto
	\begin{bmatrix}
		\chi & -\psi & 0 \\
		\psi + \gamma & \chi & 0 \\
		-\beta & \alpha & \lambda_r
	\end{bmatrix}
	\begin{bmatrix}
		\alpha \\
		\beta \\
		\omega_3
	\end{bmatrix}
	=	\begin{bmatrix}
		\chi\alpha-\psi\beta \\
		(\psi+\gamma)\alpha+\chi\beta \\
		\lambda_{r}\omega_3
	\end{bmatrix}.
\end{eqnarray}
\item The fourth term $\star\left(2\mathscr{A}(\bm{\Sigma})\right)$
\begin{eqnarray}\label{The fourth term}
	\star\left(2\mathscr{A}(\bm{\Sigma})\right)\mapsto
	\begin{bmatrix}
		\Sigma_{23} - \Sigma_{32} \\
		\Sigma_{31} - \Sigma_{13} \\
		\Sigma_{12} - \Sigma_{21}
	\end{bmatrix}
	=	\begin{bmatrix}
		\left(\nu\nabla^2\bm{\omega}+\bm{\nabla}\times\bm{f}\right)\bm{\cdot}\bm{e}_{1} \\
		\left(\nu\nabla^2\bm{\omega}+\bm{\nabla}\times\bm{f}\right)\bm{\cdot}\bm{e}_{2} \\
		\left(\nu\nabla^2\bm{\omega}+\bm{\nabla}\times\bm{f}\right)\bm{\cdot}\bm{e}_{3}
	\end{bmatrix}.
\end{eqnarray}
\end{itemize}
Substituting~\eqref{DW12}~--~\eqref{The fourth term} into~\eqref{vv} consistently yields~\eqref{eq:new_alpha} for $D\alpha/Dt$, ~\eqref{eq:new_beta} for $D\beta/Dt$ and~\eqref{g3} for $D\omega_{3}/Dt$. In addition, the Schur representation for $D\bm{\omega}/Dt$ is derived as 
\begin{eqnarray}\label{eq780}
	\frac{D\bm{\omega}}{Dt}
	&=& 
	\Big[ -\beta\psi - \alpha(\chi+\lambda_r) + \Sigma_{23} - \Sigma_{32} \Big] \bm{e}_1 \nonumber \\
	&& +
	\Big[ -\beta(\chi+\lambda_r) + \alpha(\psi+\gamma) + \Sigma_{31} - \Sigma_{13} \Big] \bm{e}_2 \nonumber \\
	&& +
	\Big[-2\chi\omega_{3} + \Sigma_{12} - \Sigma_{21} \Big] \bm{e}_3.
\end{eqnarray}
\begin{remark}
	 In fact, the vorticity transport equation is conventionally as follows. To begin with, the convective term appearing in \eqref{NS} satisfies the Lamb identity:
	 \begin{subequations}
	 	\begin{equation}\label{Lamb}
	 	\bm{u}\bm{\cdot}\bm{\nabla}\bm{u}=\bm{\omega}\times\bm{u}+\bm{\nabla}\left(\frac{1}{2}u^2\right).
	 \end{equation}
	 Taking the curl of~\eqref{Lamb} gives
	 \begin{equation}\label{eq712}
	 	\bm{\nabla}\times(\bm{u}\bm{\cdot}\bm{\nabla}\bm{u})=\bm{u}\bm{\cdot}\bm{\nabla}\bm{\omega}-\bm{\omega}\bm{\cdot}\bm{\nabla}\bm{u}+\vartheta\bm{\omega}.
	 \end{equation}
	 Taking the curl of~\eqref{expression_eta}, we obtain
	 \begin{equation}\label{eq711}
	 	\bm{\nabla}\times\bm{\eta}=\nu\nabla^2\bm{\omega}+\bm{\nabla}\times\bm{f}
	 \end{equation}
	 Taking the curl of~\eqref{NS} and invoking~\eqref{eq712} and~\eqref{eq711} also yields~\eqref{vv}.	
	 \end{subequations}
\end{remark}

\subsection{Evolution equations for the characteristic rigid-rotation and spin modes}\label{Sec7p2}
For 3D compressible flow, by using~\eqref{IVD_general},~\eqref{ss11} and~\eqref{new1}, the evolution equations for the characteristic vorticity modes $(\bm{R}_{N},\bm{S}_{N})$ are derived as 
\begin{subequations}\label{eq78ab}
	\begin{align}
		\frac{D\bm{R}_N}{Dt}
		&= 2\psi W_2 \, \bm{e}_1 
		- 2\psi W_1 \, \bm{e}_2 
		+ \left( 2\alpha W_2 - 4\chi\psi + 2\partial_2\partial_1 h - 2\Sigma_{21} \right) \bm{e}_3, \label{eq78a} \\[6pt]
		\frac{D\bm{S}_{N}}{Dt} 
		&= 
		\Big[ -2\psi W_2 - \beta\psi - \alpha(\chi+\lambda_r) + \Sigma_{23} - \Sigma_{32} \Big] \bm{e}_1 \notag \\
		&\quad + 
		\Big[ 2\psi W_1 - \beta(\chi+\lambda_r) + \alpha(\psi+\gamma) + \Sigma_{31} - \Sigma_{13} \Big] \bm{e}_2 \notag \\
		&\quad + 
		\Big[ -2\alpha W_2 - 2\chi\gamma 
		- 2\partial_1\partial_2 h + \Sigma_{12} + \Sigma_{21} \Big] \bm{e}_3. \label{eq78b}
	\end{align}
\end{subequations}
Here, the explicit forms of $(W_{1},W_{2},W_{3})$ are given in~\eqref{Schur_frame_angular2}. The sum of~\eqref{eq78a} and~\eqref{eq78b} recovers the vorticity equation~\eqref{eq780}.
To the best of the author's knowledge, \eqref{eq78ab} has not been reported in the existing literature, which can be employed to understand spatiotemporal evolution and mutual interaction mechanisms of the vorticity-mode pair $(\bm{R}_{N},\bm{S}_{N})$. Although the expansion expressions in the Schur frame are already transparent, we aim to further cast them into forms analogous to the vorticity equation in vector form~\eqref{vv}, which is far from trivial. We notice that~\eqref{eq78a} and~\eqref{eq78b} can be rewritten as
\begin{subequations}\label{eq713}
	\begin{align}
		\frac{D\bm{R}_N}{Dt}
		&=-\vartheta \bm{R}_{N} \notag\\
		&\quad +2\psi W_{2}\bm{e}_{1}-2\psi W_{1}\bm{e}_{2}+2\alpha W_{2}\bm{e}_{3} \notag\\
		&\quad +2\psi\lambda_{r}\bm{e}_{3} \notag\\
		&\quad +2\partial_{2}\partial_{1}h\bm{e}_{3} \notag\\
		&\quad -2\Sigma_{21}\bm{e}_{3}, \label{eq:RNx} \\ 
		\frac{D\bm{S}_N}{Dt}
		&=-\vartheta\bm{S}_{N} \notag\\
		&\quad -2\psi W_{2}\bm{e}_{1}+2\psi W_{1}\bm{e}_{2}-2\alpha W_{2}\bm{e}_{3} \notag\\
		&\quad -\beta\psi\bm{e}_{1}+\alpha\psi\bm{e}_{2} \notag\\
		&\quad +\alpha\chi\bm{e}_{1}+\beta\chi\bm{e}_{2}+\gamma\lambda_{r}\bm{e}_{3} \notag\\
		&\quad +\alpha\gamma\bm{e}_{2} \notag\\
		&\quad-2\partial_{1}\partial_{2}h\bm{e}_{3} \notag\\
		&\quad +(\Sigma_{23} - \Sigma_{32})\bm{e}_{1} + (\Sigma_{31} - \Sigma_{13})\bm{e}_{2} + (\Sigma_{12} + \Sigma_{21})\bm{e}_{3}. \label{eq:SNx}
	\end{align}
\end{subequations}

From~\eqref{eq:D_matrix}, the strain-rate tensor $\bm{D}$ is further split as
\begin{subequations}\label{Dsplit}
	\begin{equation}
		\bm{D}=\bm{D}_{EL}+\bm{D}_{SH},
	\end{equation}
	\begin{equation}
		\bm{D}_{EL}\mapsto
		\begin{bmatrix}
			\chi & 0 & 0 \\
			0   & \chi & 0 \\
			0 & 0 & \lambda_r
		\end{bmatrix}, \qquad
		\bm{D}_{SH}\mapsto 
	\begin{bmatrix}
		0 & \dfrac12\gamma & -\dfrac12\beta \\[4pt]
		\dfrac12\gamma & 0 & \dfrac12\alpha \\[4pt]
		-\dfrac12\beta & \dfrac12\alpha & 0
	\end{bmatrix}.
	\end{equation}
\end{subequations}
\begin{figure}[t]
	\centering 
	\includegraphics[width=1.0\columnwidth,trim={3cm 3cm 1cm 3cm},clip]{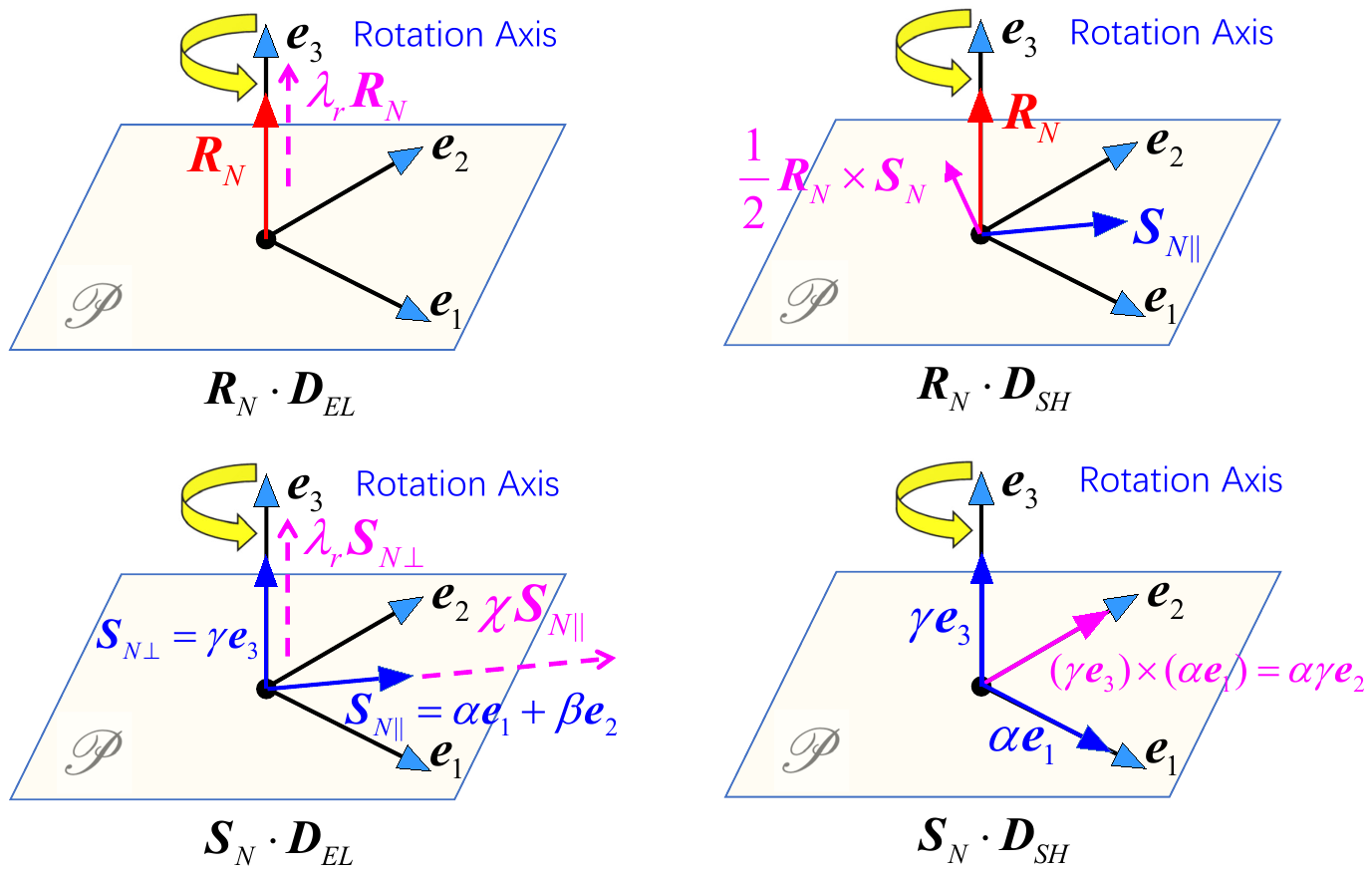}
	\caption{Illustration of the stretching and tilting mechanisms (magenta arrows) of $\bm{R}_{N}$ (red arrows) and $\bm{S}_{N}$ (blue arrows) under the action of the straining and shear strain-rate tensors $(\bm{D}_{EL},\bm{D}_{SH})$. Here, $\bm{e}_{3}$ denotes the local rotation axis, and $\mathscr{P}\equiv{\rm Span}\left\{\bm{e}_{1},\bm{e}_{2}\right\}$ defines the associated orthogonal plane. These three basis vectors form the orthonormal Schur frame $\left\{\bm{e}_{1},\bm{e}_{2},\bm{e}_{3}\right\}$. Yellow arrows mark the sense of rotation of $\bm{e}_{3}$.}
	\label{stretching_tilting_mechanisms}
\end{figure}
Here, $\bm{D}_{EL}$ is the straining strain-rate tensor and $\bm{D}_{SH}$ is the shear strain-rate tensor.
Then, it follows from~\eqref{IVD_general} and~\eqref{Dsplit} that
\begin{subequations}\label{eq715}
	\begin{align}
		\bm{R}_{N}\bm{\cdot}\bm{D}_{EL} &= 2\lambda_{r}\psi\bm{e}_{3}=\lambda_{r}\bm{R}_{N}, \label{eq715a}\\
		\bm{R}_{N}\bm{\cdot}\bm{D}_{SH} &= -\psi\beta\bm{e}_{1} + \psi\alpha\bm{e}_{2}=\frac{1}{2}\bm{R}_{N}\times\bm{S}_{N\parallel}, \label{eq715b}\\
		\bm{S}_{N}\bm{\cdot}\bm{D}_{EL}
		&=\chi\bm{S}_{N\parallel}+\lambda_{r}\bm{S}_{N\perp}, \label{eq715c}\\
		\bm{S}_{N}\bm{\cdot}\bm{D}_{SH} &= \alpha\gamma\bm{e}_{2}=(\gamma\bm{e}_3)\times(\alpha\bm{e}_{1}), \label{eq715d}
	\end{align}
\end{subequations}
where $\bm{S}_{N\parallel}=\bm{e}_{3}\times(\bm{S}_{N}\times\bm{e}_{3})=\alpha\bm{e}_{1}+\beta\bm{e}_{2}$ and $\bm{S}_{N\perp}=(\bm{S}_{N}\bm{\cdot}\bm{e}_{3})\bm{e}_{3}=\gamma\bm{e}_{3}$ are the characteristic spin components parallel to and perpendicular to the plane \(\mathscr{P}\), as shown in figure~\ref{stretching_tilting_mechanisms}.

By using~\eqref{Schur} and~\eqref{source_term_Sigma}, the terms associated with $\bm{\Sigma}$ in~\eqref{eq:RNx} and~\eqref{eq:SNx} evaluate to
\begin{subequations}\label{eq716}
	\begin{gather}
		-2\Sigma_{21}\bm{e}_{3} = \nu\nabla^2\bm{R}_{N} + \bm{J}[\bm{R}_{N}],\label{or1} \\
		(\Sigma_{23} - \Sigma_{32})\bm{e}_{1} + (\Sigma_{31} - \Sigma_{13})\bm{e}_{2} + (\Sigma_{12} + \Sigma_{21})\bm{e}_{3} = \nu\nabla^2\bm{S}_{N} + \bm{J}[\bm{S}_{N}].\label{or2}
	\end{gather}
\end{subequations}
By virtue of~\eqref{vv} and~\eqref{The fourth term}, one obtains
\begin{subequations}
\begin{equation}\label{Jcons}
	\bm{J}[\bm{S}_{N}]=\bm{\nabla}\times\bm{f}-\bm{J}[\bm{R}_{N}].
\end{equation}
If $\bm{f}=\bm{0}$, then $\bm{J}[\bm{R}_{N}]=-\bm{J}[\bm{S}_{N}]$ must hold exactly. From~\eqref{source_term_Sigma} and~\eqref{or1}, $\bm{J}[\bm{R}_{N}]$ is evaluated as
\begin{eqnarray}\label{JRexplicit}
	\bm{J}[\bm{R}_{N}]&=&-2\left(\bm{e}_{2}\bm{\cdot}\bm{\Sigma}\bm{\cdot}\bm{e}_{1}\right)\bm{e}_{3}-\nu\nabla^{2}\bm{R}_{N}\nonumber\\
	&=&-2\left(\nu_{b}+\frac{1}{3}\nu\right)\left(\bm{e}_{2}\bm{\cdot}\bm{\nabla\nabla}\vartheta\bm{\cdot}\bm{e}_{1}\right)\bm{e}_{3}-2\left(\bm{e}_{2}\bm{\cdot}\bm{\nabla}\bm{f}\bm{\cdot}\bm{e}_{1}\right)\bm{e}_{3}\nonumber\\
	& &-2\nu(\bm{e}_{2}\bm{\cdot}\nabla^{2}\bm{A}\bm{\cdot}\bm{e}_{1})\bm{e}_{3}-\nu\nabla^2\bm{R}_{N},
\end{eqnarray}
where the last two terms are further decomposed as
\begin{eqnarray}\label{JRexpand}
	& &-2\nu(\bm{e}_{2}\bm{\cdot}\nabla^{2}\bm{A}\bm{\cdot}\bm{e}_{1})\bm{e}_{3}-\nu\nabla^2\bm{R}_{N}\nonumber\\
	&=&\bm{J}_{1}[\bm{R}_{N}]+\bm{J}_{2}[\bm{R}_{N}]+\bm{J}_{3}[\bm{R}_{N}]+\bm{J}_{4}[\bm{R}_{N}]+\bm{J}_{5}[\bm{R}_{N}]+\bm{J}_{6}[\bm{R}_{N}]+\bm{J}_{7}[\bm{R}_{N}],
\end{eqnarray}
\begin{eqnarray}\label{JRexpand_terms}
	\bm{J}_{1}[\bm{R}_{N}]&\equiv&\left[2\nu(\nabla^2\bm{e}_{2})\bm{\cdot}\bm{A}\bm{\cdot}\bm{e}_{1}\right]\bm{e}_{3},\nonumber\\
	\bm{J}_{2}[\bm{R}_{N}]&\equiv&\left[2\nu\bm{e}_{2}\bm{\cdot}\bm{A}\bm{\cdot}(\nabla^2\bm{e}_{1})\right]\bm{e}_{3},\nonumber\\
	\bm{J}_{3}[\bm{R}_{N}]&\equiv&\left[4\nu(\bm{\nabla}\bm{e}_{2})\bm{:}\bm{\nabla}\bm{A}\bm{\cdot}\bm{e}_{1}\right]\bm{e}_{3},\nonumber\\
	\bm{J}_{4}[\bm{R}_{N}]&\equiv&\left[4\nu\bm{A}\bm{:}(\bm{e}_{2}\bm{\nabla}\bm{\cdot}\bm{\nabla}\bm{e}_{1})\right]\bm{e}_{3},\nonumber\\
	\bm{J}_{5}[\bm{R}_{N}]&\equiv&\left[4\nu \bm{e}_{2}\bm{\cdot}\bm{A\nabla}\bm{:}(\bm{e}_{1}\bm{\nabla})\right]\bm{e}_{3},\nonumber\\
	\bm{J}_{6}[\bm{R}_{N}]&\equiv&-4\nu\bm{\nabla}\psi\bm{\cdot}\bm{\nabla}\bm{e}_{3}\nonumber\\
	\bm{J}_{7}[\bm{R}_{N}]&\equiv&-2\nu\psi\nabla^{2}\bm{e}_{3}.
\end{eqnarray}
\end{subequations}
Here, $\bm{J}[\bm{R}_{N}]$ and $\bm{J}[\bm{S}_{N}]$ arise from the spatial derivatives of the Schur basis vectors (which are coupled to the VGT and its associated gradients), the dilatation, and the external forcing. Specifically, $\bm{J}_{1}-\bm{J}_{5}$ are associated with the gradients and Laplacians of $\bm{e}_{1}$ and $\bm{e}_{2}$, which are parallel to the rotation axis $\bm{e}_{3}$, whereas $\bm{J}_{6}$ and $\bm{J}_{7}$ are influenced by the spatial variation of $\bm{e}_{3}$. For a second-order tensor $\bm{T}$ and a third-order tensor $\bm{B}$, the double contraction reads $\bm{T}\bm{:}\bm{B}=T_{ij}B_{ijk}\,\bm{e}_{k}$. In practice, $\bm{J}[\bm{R}_{N}]$ and $\bm{J}[\bm{S}_{N}]$ can be evaluated directly using~\eqref{Jcons} and~\eqref{JRexplicit}, along with the identity $\nabla^2\bm{A}=\bm{\nabla}(\nabla^2\bm{u})$. Nevertheless,~\eqref{JRexpand} and~\eqref{JRexpand_terms} elucidate all physical mechanisms stemming from the spatial variation of the Schur frame, which will be further examined through the Lamb--Oseen vortex in~\S\ref{Decomposition of the viscous source pair}.

\begin{figure}[t]
	\centering 
	\includegraphics[width=1.2\columnwidth,trim={0cm 6cm 0cm 5cm},clip]{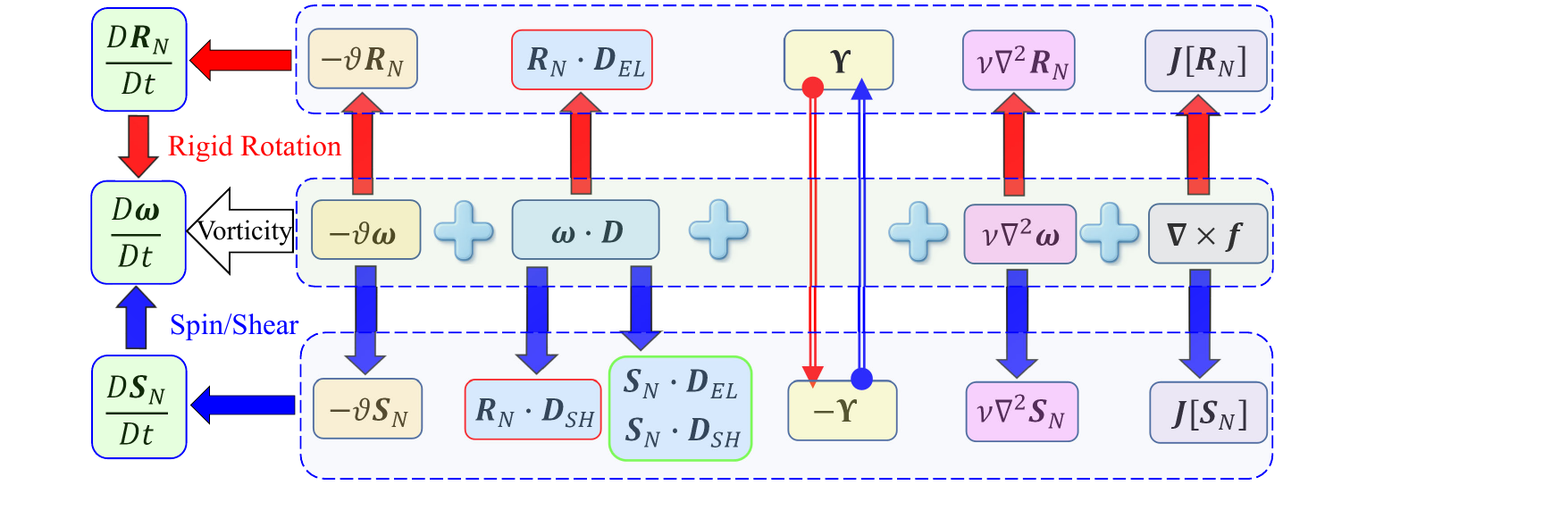}
	\caption{Relationship among the Lagrangian dynamics of the characteristic rigid-rotation mode $\bm{R}_{N}$~\eqref{eq:RNxx}, the characteristic spin/shear mode $\bm{S}_{N}$~\eqref{eq:SNxx}, and the vorticity $\bm{\omega}$~\eqref{vv}.} 
	\label{RSOmega}
\end{figure}

Substituting~\eqref{eq715} and~\eqref{eq716} into ~\eqref{eq:RNx} and~\eqref{eq:SNx} gives (see figure~\ref{RSOmega})
\begin{subequations}\label{eq717}
	\begin{align}
		\frac{D\bm{R}_N}{Dt}
		&=-\vartheta \bm{R}_{N}+\bm{R}_{N}\bm{\cdot}\bm{D}_{EL}+\bm{\Upsilon} +\nu\nabla^{2}\bm{R}_{N}+	\bm{J}[\bm{R}_{N}], \label{eq:RNxx} \\ 
		\frac{D\bm{S}_N}{Dt}
		&=-\vartheta\bm{S}_{N}+\bm{R}_{N}\bm{\cdot}\bm{D}_{SH} +\bm{S}_{N}\bm{\cdot}\bm{D} -\bm{\Upsilon} +\nu\nabla^2\bm{S}_{N} + 	\bm{J}[\bm{S}_{N}].\label{eq:SNxx}
	\end{align}
\end{subequations}
The physical interpretation of~\eqref{eq:RNxx} and~\eqref{eq:SNxx} are as follows.
\begin{itemize}
	\item On the right-hand side of~\eqref{eq:RNxx}, the first term accounts for the compressibility effect arising from the coupling between the dilatation $\vartheta$ and the characteristic rigid-rotation mode $\bm{R}_{N}$. It is widely recognized that the axial/tubular vortex stretching mechanism plays a crucial role in intensifying vortical energy and in facilitating the forward energy cascade from large-scale energetic coherent motions to small-scale shear-dominated motions in 3D turbulent flows~\citep{Pope2000}. In contrast to 2D flows, this mechanism, which is peculiar to 3D flows, is usually interpreted in terms of the vortex stretching term $\bm{\omega}\bm{\cdot}\bm{D}$ in the Helmholtz vorticity transport equation. However, vorticity alone cannot distinguish rigid-rotation motion from shear motion~\citep{Kolar2007VortexID,LiuCQ2018,Chen2026General,Chen2026Kinematic}. From a phenomenological morphology perspective, the basic
	forms of vortex structures at high Reynolds numbers typically manifest as axial vortices and shear layers, which often coexist within a single
	swirling system. Axial vortices are dominated by $\bm{R}_{N}$, whereas shear layers are dominated by $\bm{S}_{N}$. The second term $\bm{R}_{N}\bm{\cdot}\bm{D}_{EL}$, given by~\eqref{eq715a}, elucidates this mechanism in a more rational manner, since it is $\bm{R}_{N}$, rather than vorticity, that properly characterizes the rigid-rotation motion of an axial vortex: $\lambda_{r}>0$ corresponds to stretching and $\lambda_{r}<0$ to contraction along the local axis of rotation $\bm{e}_{3}$. The fourth term denotes the viscous diffusion of $\bm{R}_{N}$.
	\item On the right-hand side of~\eqref{eq:SNxx}, the first term is due to the coupling between the dilatation $\vartheta$ and the characteristic spin/shear mode $\bm{S}_{N}$. According to~\eqref{eq715b}, the second term represents the interaction between $\bm{R}_{N}$ and $\bm{D}_{SH}$, which tilts $\bm{S}_{N}$-lines (rather than $\bm{R}_{N}$-lines) towards the direction parallel to $\bm{R}_{N}\times\bm{S}_{N}/2$. The third term is decomposed as $\bm{S}_{N}\bm{\cdot}\bm{D}=\bm{S}_{N}\bm{\cdot}\bm{D}_{EL}+\bm{S}_{N}\bm{\cdot}\bm{D}_{SH}$. The term $\bm{S}_{N}\bm{\cdot}\bm{D}_{EL}$ in~\eqref{eq715c} reflects the stretching/contraction of in-plane spin component $\bm{S}_{N\parallel}=\alpha\bm{e}_{1}+\beta\bm{e}_{2}$ by $\chi$ and the axial component $\bm{S}_{N\perp}=\gamma\bm{e}_{3}$ by $\lambda_{r}$, whereas the term $\bm{S}_{N}\bm{\cdot}\bm{D}_{SH}$ in~\eqref{eq715d} denotes a quadratic self-coupling mechanism of the spin components along $\bm{e}_{1}$ and $\bm{e}_{3}$, which tilts the $\bm{S}_{N}$ line towards $\bm{e}_{2}$. The fifth term again represents the viscous diffusion of $\bm{S}_{N}$.  Figure~\ref{stretching_tilting_mechanisms} illustrates the stretching and tilting mechanisms for the characteristic vorticity modes $(\bm{R}_N,\bm{S}_{N})$ due to the straining and shear strain-rate tensors $(\bm{D}_{EL},\bm{D}_{SH})$. The superposition of the four contributions yields the well-known vortex stretching term for vorticity:
	\begin{equation}\label{omega_D}
		\bm{\omega}\bm{\cdot}\bm{D}=\bm{R}_{N}\bm{\cdot}\bm{D}_{EL}+\bm{R}_{N}\bm{\cdot}\bm{D}_{SH}+\bm{S}_{N}\bm{\cdot}\bm{D}_{EL}+\bm{S}_{N}\bm{\cdot}\bm{D}_{SH}.
	\end{equation}
	\item  As shown in figure~\ref{RSOmega}, the mutual transformation between $\bm{R}_{N}$ and $\bm{S}_{N}$ can be realized via the dynamical transfer term  $\bm{\Upsilon}$, which is expressed in the Schur frame as
	\begin{subequations}
			\begin{equation}\label{eq718}
			\bm{\Upsilon}\equiv2\psi W_{2}\bm{e}_{1}-2\psi W_{1}\bm{e}_{2}+2\alpha W_{2}\bm{e}_{3}+2\partial_{2}\partial_{1}h\bm{e}_{3}.
		\end{equation}
		The explicit form of~\eqref{eq718} follows directly from the substitution of~\eqref{eq:W1_new} and~\eqref{eq:W2_new}:
		\begin{equation}\label{explicit_Upsilon}
			\begin{aligned}
				\bm{\Upsilon} =\,&
				\left[ \frac{2\psi(\chi-\lambda_r)}{\Xi}\partial_3\partial_1h + \frac{2\psi^2}{\Xi}\partial_3\partial_2h \right]\bm{e}_1 \\
				&+ \left[ -\frac{2\psi(\psi+\gamma)}{\Xi}\partial_3\partial_1h + \frac{2\psi(\chi-\lambda_r)}{\Xi}\partial_3\partial_2h \right]\bm{e}_2 \\
				&+ \left[ 2\partial_2\partial_1h + \frac{2\alpha(\chi-\lambda_r)}{\Xi}\partial_3\partial_1h + \frac{2\alpha\psi}{\Xi}\partial_3\partial_2h \right]\bm{e}_3 \\
				&- \left[ \frac{2\psi(\chi-\lambda_r)}{\Xi}\Sigma_{31} + \frac{2\psi^2}{\Xi}\Sigma_{32} \right]\bm{e}_1 \\
				&+ \left[ \frac{2\psi(\psi+\gamma)}{\Xi}\Sigma_{31} - \frac{2\psi(\chi-\lambda_r)}{\Xi}\Sigma_{32} \right]\bm{e}_2 \\
				&- \left[ \frac{2\alpha(\chi-\lambda_r)}{\Xi}\Sigma_{31} + \frac{2\alpha\psi}{\Xi}\Sigma_{32} \right]\bm{e}_3.
			\end{aligned}
		\end{equation}
	\end{subequations}
Note that $\bm{\Upsilon}$ receives the contributions from the off-diagonal components of the enthalpy Hessian tensor $\bm{\nabla\nabla}h$ (replaced by the pressure Hessian tensor $\bm{\nabla\nabla}P$ in the incompressible limit) and the source term $\bm{\Sigma}$, where the weighting coefficients are formed from some dimensionless combinations of the VGT rotational invariants. Physically, these dynamical transfer mechanisms could be associated with the winding and unwinding behaviors of
sheet-like shear layers surrounding an axial vortex in a swirling system, which arise from strong
coupling of longitudinal and transverse fields~\citep{Chen2026General}. Compared with axial vortex stretching, this mechanism occurs in both 2D and 3D flows. Even with $\bm{\Upsilon}=\bm{0}$ and $\bm{f}=\bm{0}$, the spatial variation of the Schur frame keeps the Lagrangian dynamics of $\bm{R}_{N}$ and $\bm{S}_{N}$ coupled through the viscous source pair $(\bm{J}[\bm{R}_{N}],\bm{J}[\bm{S}_{N}])$.
\item For incompressible inviscid flow, by utilizing the vorticity and strain-rate evolution equations,~\citet{Ohkitani1995} showed that the pressure Hessian tensor $\bm{\nabla\nabla}P$ determines the second-order material evolution rate of vorticity through $D^2\bm{\omega}/Dt^2=-\bm{\nabla\nabla}P\bm{\cdot}\bm{\omega}$. In the present work, when considering incompressible viscous flow as a reduced case, although the pressure gradient $\bm{\nabla}P$ (which contains the first-order derivatives of pressure) does not appear in the vorticity transport equation, the components of the pressure Hessian tensor  (which contain the second-order derivatives of pressure, thereby raising the order of differentiation by one) reappear in the first-order evolution equations of  $(\bm{R}_{N},\bm{S}_{N})$. To the authors' knowledge, this distinct role of the pressure Hessian tensor has not been previously reported in the literature.
\item For 2D incompressible flow without external forcing,~\eqref{eq:RNxx} and~\eqref{eq:SNxx} are simplified to
\begin{subequations}\label{eq78abxx}
	\begin{align}
		\frac{D\bm{R}_N}{Dt}
		&=\bm{\Upsilon}+\nu\nabla^2\bm{R}_{N}+\bm{J}[\bm{R}_{N}], \label{eq78axx} \\[6pt]
		\frac{D\bm{S}_{N}}{Dt} 
		&=-\bm{\Upsilon}+\nu\nabla^2\bm{S}_{N}+\bm{J}[\bm{S}_{N}], \label{eq78bxx}
	\end{align}
	where the dynamical transfer term in~\eqref{eq718} reduces to
	\begin{equation}
		\bm{\Upsilon}=2\partial_{2}\partial_{1}P\bm{e}_{3}=2(\bm{e}_{1}\bm{\cdot}\bm{\nabla\nabla}P\bm{\cdot}\bm{e}_{2})\bm{e}_{3}.
	\end{equation}
    The viscous source terms in~\eqref{Jcons}~--~\eqref{JRexpand_terms} are calculated through
	\begin{eqnarray}\label{VVX}
	\bm{J}[\bm{R}_{N}]&=&-\bm{J}[\bm{S}_{N}]=2\nu(-\bm{e}_{2}\bm{\cdot}\nabla^{2}\bm{A}\bm{\cdot}\bm{e}_{1})\bm{e}_{3}-\nu\nabla^2\bm{R}_{N}\nonumber\\
	&=&\bm{J}_{1}[\bm{R}_{N}]+\bm{J}_{2}[\bm{R}_{N}]+\bm{J}_{3}[\bm{R}_{N}]+\bm{J}_{4}[\bm{R}_{N}]+\bm{J}_{5}[\bm{R}_{N}].
\end{eqnarray}
\end{subequations}
\end{itemize}
\subsection{Example: The Lamb-Oseen vortex}\label{Example: The Lamb-Oseen vortex}
\subsubsection{Analytical solutions and fundamental properties}
The Lamb--Oseen vortex is an unsteady planar axisymmetric analytical solution of the incompressible NS equations, describing viscous diffusion process of a point vortex initially located at the origin $O$~\citep{Oseen1912wirbel,Lamb1932hydrodynamics,Batchelor1967IntroFD}. In the cylindrical coordinate system $(r,\theta,z)$, with the orthonormal basis $\left\{\bm{e}_{r},\bm{e}_{\theta},\bm{e}_{z}\right\}$, the azimuthal velocity component $u_{\theta}(r,t)$ should satisfy
\begin{subequations}
\begin{equation}\label{LO1}
\frac{\partial u_\theta}{\partial t}=\nu\left(\frac{\partial^2 u_\theta}{\partial r^2}+\frac{1}{r}\frac{\partial u_\theta}{\partial r}-\frac{u_\theta}{r^2}\right),
\end{equation}
subjected to the initial and boundary conditions
\begin{equation}\label{LO2}
u_{\theta}(0,t)=0,~u_{\theta}(\infty,t)=0,~u_{\theta}(r,0)=\frac{\Gamma_0}{2\pi r},~u_{\theta}(r,\infty)=0.
\end{equation}
The assumptions include $u_r=u_z=0$, $\partial_{\theta}=0$, and $u_{\theta}=u_{\theta}(r,t)$. The constant $\Gamma_{0} (>0)$ characterizes the vortex strength, and $\nu$ is the kinematic viscosity.
\end{subequations}
\begin{figure}[t]
	\centering
	\subfigure[]{
		\begin{minipage}[t]{0.5\linewidth}
			\centering
			\includegraphics[width=1.0\columnwidth,trim={1cm 0cm 1cm 0cm},clip]{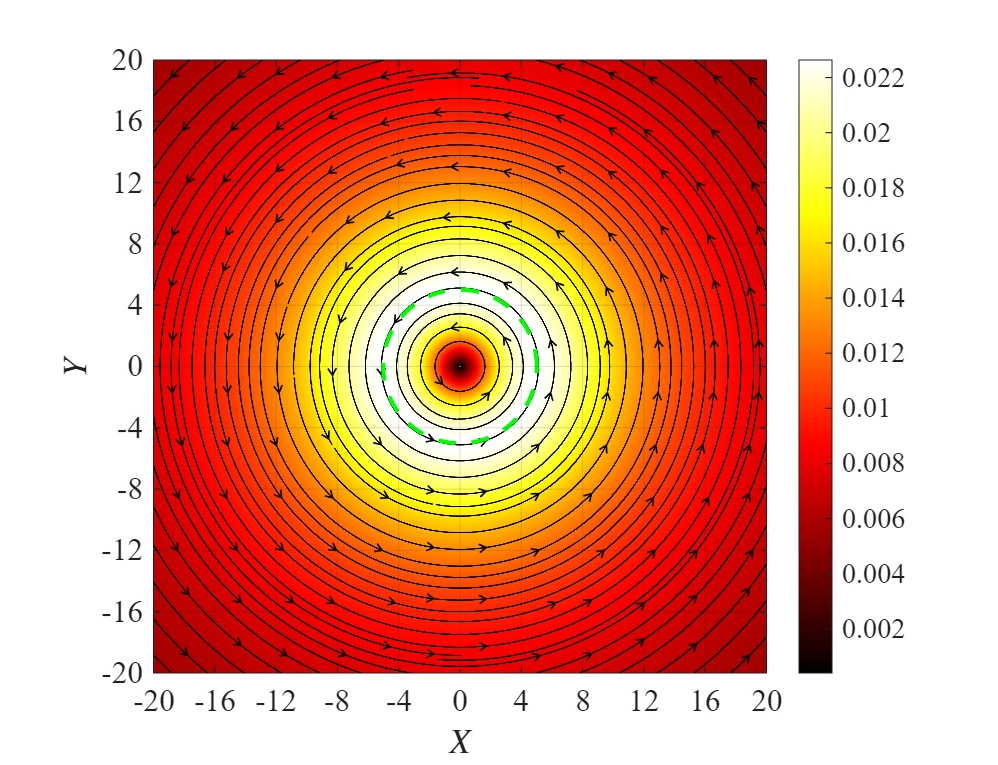}
			\label{LO_utheta}
		\end{minipage}%
	}%
	\subfigure[]{
		\begin{minipage}[t]{0.5\linewidth}
			\centering
			\includegraphics[width=1.0\columnwidth,trim={7cm 4.2cm 7cm 4.5cm},clip]{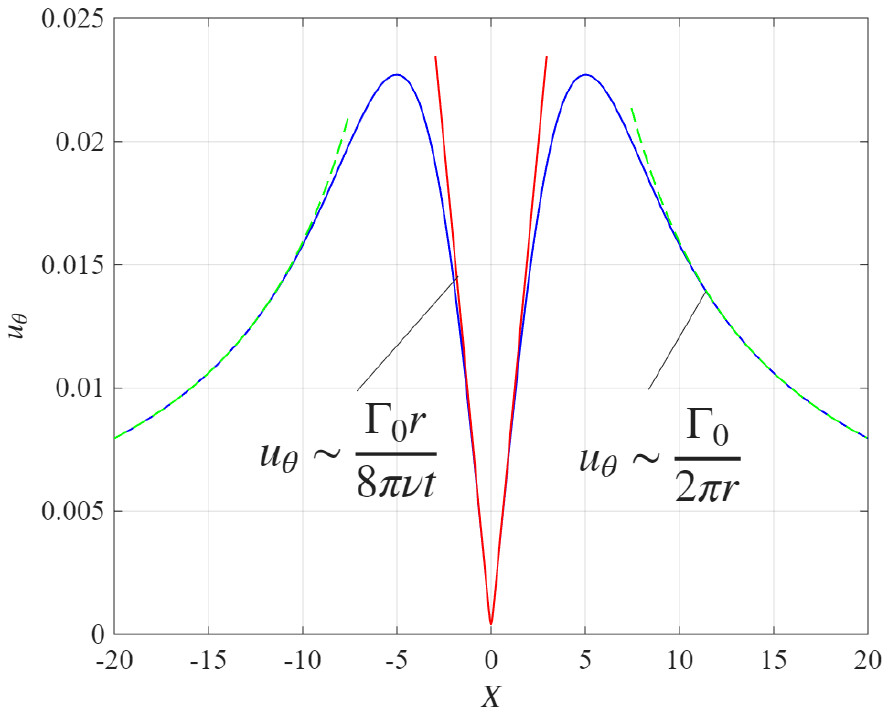}
			\label{LO_utheta_curve}
		\end{minipage}%
	}%
	
	\subfigure[]{
		\begin{minipage}[t]{0.5\linewidth}
			\centering
			\includegraphics[width=1.0\columnwidth,trim={0cm 0cm 0cm 0cm},clip]{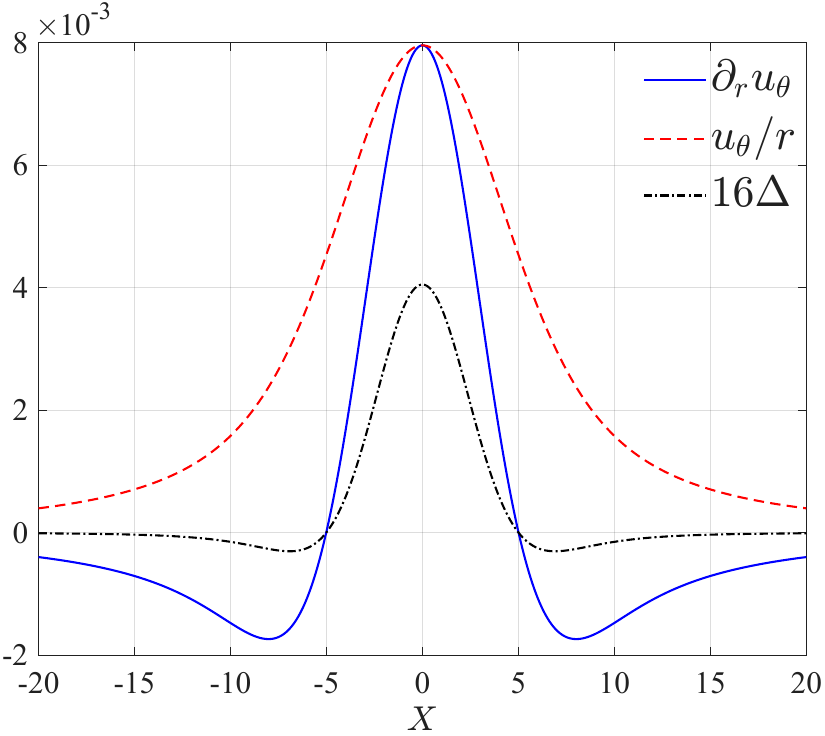}
			\label{duthetadr_uthetar_Delta}
		\end{minipage}%
	}%
	\caption{(a) Contour map of the azimuthal velocity component $u_{\theta}$ with superposed streamlines. The green dashed line marks the vortex core radius \(r_{c}=5.0128\), at which \(u_{\theta}\) reaches its maximum and \(\Delta=0\) holds. (b) Radial profile of $u_{\theta}$. (b) Radial profiles of $\partial_{r}u_{\theta}$, $u_{\theta}$ and the scaled discriminant $16\Delta$.}
	\label{Vorticity_limiting_regimes}
\end{figure}
\begin{subequations}
The solution of~\eqref{LO1} and~\eqref{LO2} yields the velocity field as
	\begin{equation}\label{LO3}
		u_{\theta}=\frac{\Gamma_0}{2\pi r}\left(1-\exp\left(-\frac{r^2}{4\nu t}\right)\right).
	\end{equation}
In all subsequent visualizations, we set $\Gamma=1$, $\nu=0.01$, and $t=500$, consistent with the parameters used by~\citet{Chu2025}.
As shown in figures~\ref{LO_utheta} and~\ref{LO_utheta_curve}, when $r\rightarrow 0^+$, $u_{\theta}\sim\Gamma_{0}r/(8\pi\nu t)$ (similar to the inviscid Rankine vortex-core solution), which approaches a rigid rotation with angular velocity $\Gamma_{0}/(8\pi\nu t)$; when $r\rightarrow+\infty$, $u_{\theta}\sim\Gamma_{0}/(2\pi r)$ (a potential vortex with strength $\Gamma_{0}$).
From~\eqref{LO3}, we obtain
\begin{equation}\label{LO4}
\frac{\partial u_\theta}{\partial r}
=
\frac{\Gamma_0}{2\pi r^2}
\left[
\left(
1+\frac{r^2}{2\nu t}
\right)
\exp\left(-\frac{r^2}{4\nu t}\right)
-1
\right],
\end{equation}
\begin{equation}\label{LO5}
\frac{u_{\theta}}{r}=\frac{\Gamma_0}{2\pi r^2}\left(1-\exp\left(-\frac{r^2}{4\nu t}\right)\right).
\end{equation}
\end{subequations}
Figure~\ref{duthetadr_uthetar_Delta} displays the radial profiles of $\partial_{r}u_{\theta}$ and $u_{\theta}/r$.
For $t>0$, both $\partial_{r}u_{\theta}$ and $u_{\theta}/r$ approaches $\Gamma_0/(8\pi\nu t)$ when $r\rightarrow 0^+$. Solving $\partial_{r}u_{\theta}=0$ yields the maximum velocity point $r_{c}=2.2418\sqrt{\nu t}$. By virtue of \eqref{LO4} and \eqref{LO5}, it is directly verified that
\begin{equation}\label{LO12}
	\frac{\partial u_\theta}{\partial r}<\frac{u_\theta}{r}\quad (r>0,~t>0,~\Gamma_{0}>0).
\end{equation}

Summing~\eqref{LO4} and~\eqref{LO5} yields the vorticity $\bm{\omega}=\omega_{z}\bm{e}_{z}$, where
\begin{eqnarray}\label{LO6}
\omega_{z}&=&\frac{1}{r}\frac{\partial}{\partial r}\left(ru_\theta\right)=\frac{\partial u_\theta}{\partial r}+\frac{u_\theta}{r}\nonumber\\
&=&\frac{\Gamma_0}{4\pi\nu t}\exp\left(-\frac{r^2}{4\nu t}\right).
\end{eqnarray}
For $r\rightarrow 0^+$, $\omega_{z}(0^+,t)=\Gamma_0/(4\pi\nu t)\rightarrow+\infty$ as $t\rightarrow 0^+$; for $r>0$, $\omega_{z}\rightarrow0$ as $t\rightarrow0^+$. Therefore, $\omega_{z}\rightarrow\Gamma_{0}\delta(r)/2\pi r$ as $t\rightarrow 0^+$, which represents a point vortex with the strength $\Gamma_{0}$ placed at $r=0$. Since $\omega_z>0$, the rotation axis is $\bm{e}_{3}=\bm{e}_{z}$. The total vorticity flux remains invariant: $\iint_{\mathbb R^2}\omega_{z}\,dxdy=\Gamma_{0}$.
Moreover, the radial pressure gradient $P=P(r,t)$ must balance the centripetal force, i.e., $\partial_{r}P=u_{\theta}^{2}/r$.
\begin{figure}[htbp]
	\centering
	\subfigure[$2\psi^+$]{
		\begin{minipage}[t]{0.5\linewidth}
			\centering
			\includegraphics[width=1.0\columnwidth,trim={1cm 0cm 1cm 0cm},clip]{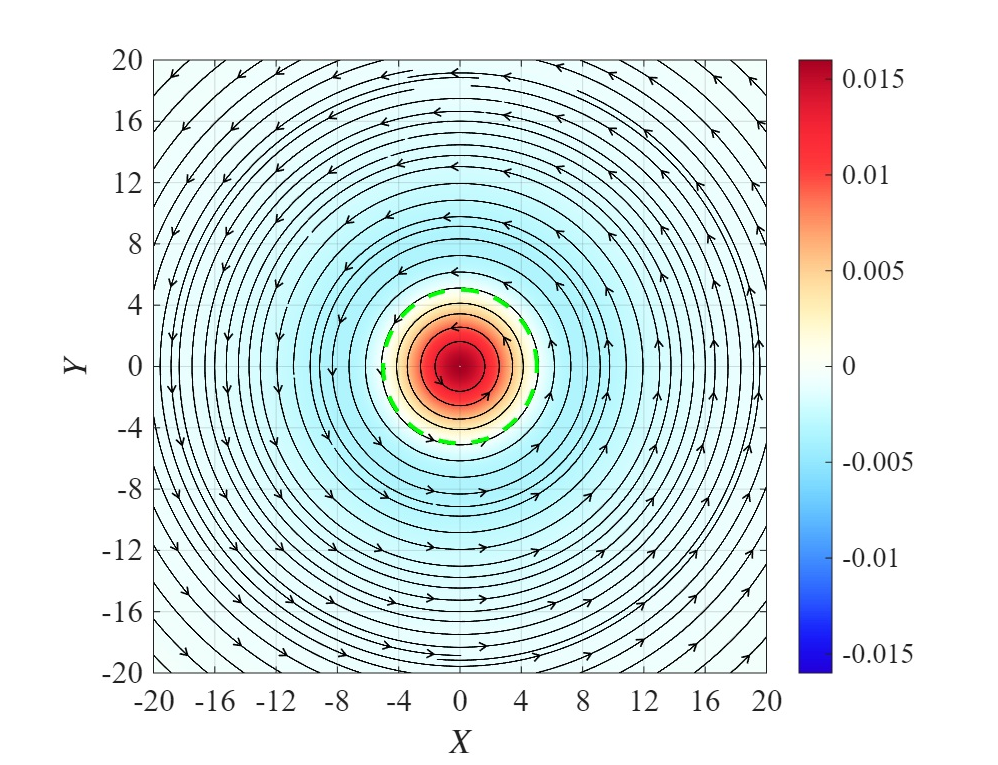}
			\label{LO_RNplus_map}
		\end{minipage}%
	}%
	\subfigure[$\gamma^+$]{
		\begin{minipage}[t]{0.5\linewidth}
			\centering
			\includegraphics[width=1.0\columnwidth,trim={1cm 0cm 1cm 0cm},clip]{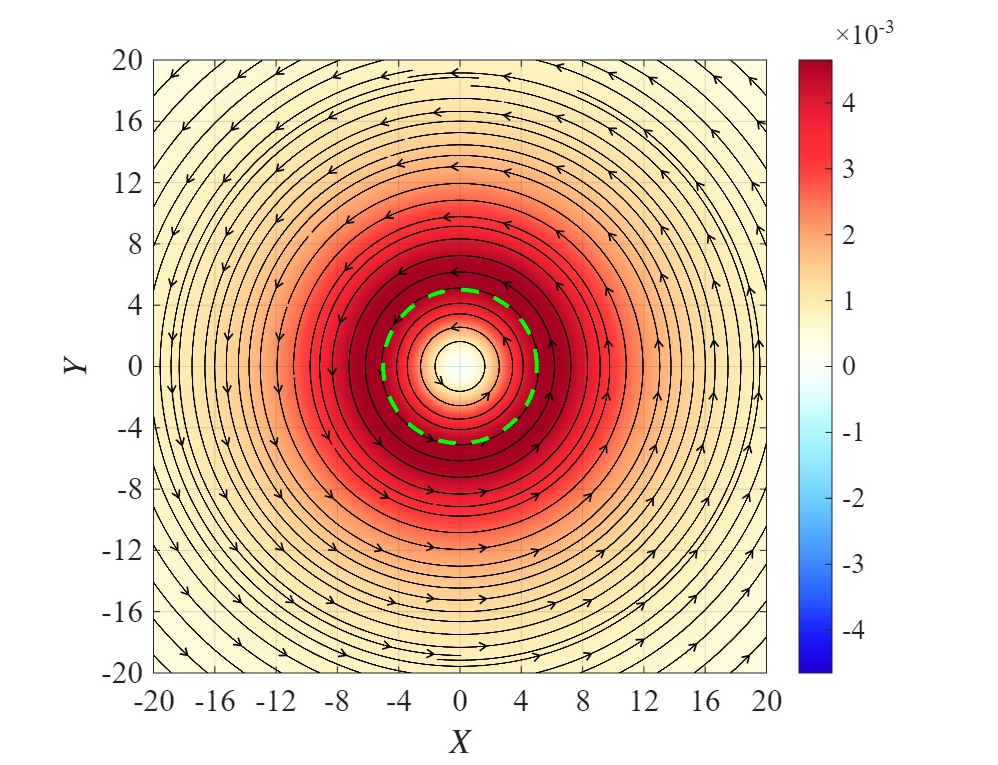}
			\label{LO_SNplus_map}
		\end{minipage}%
	}%
	
	\subfigure[$2\psi^-$]{
		\begin{minipage}[t]{0.5\linewidth}
			\centering
			\includegraphics[width=1.0\columnwidth,trim={1cm 0cm 1cm 0cm},clip]{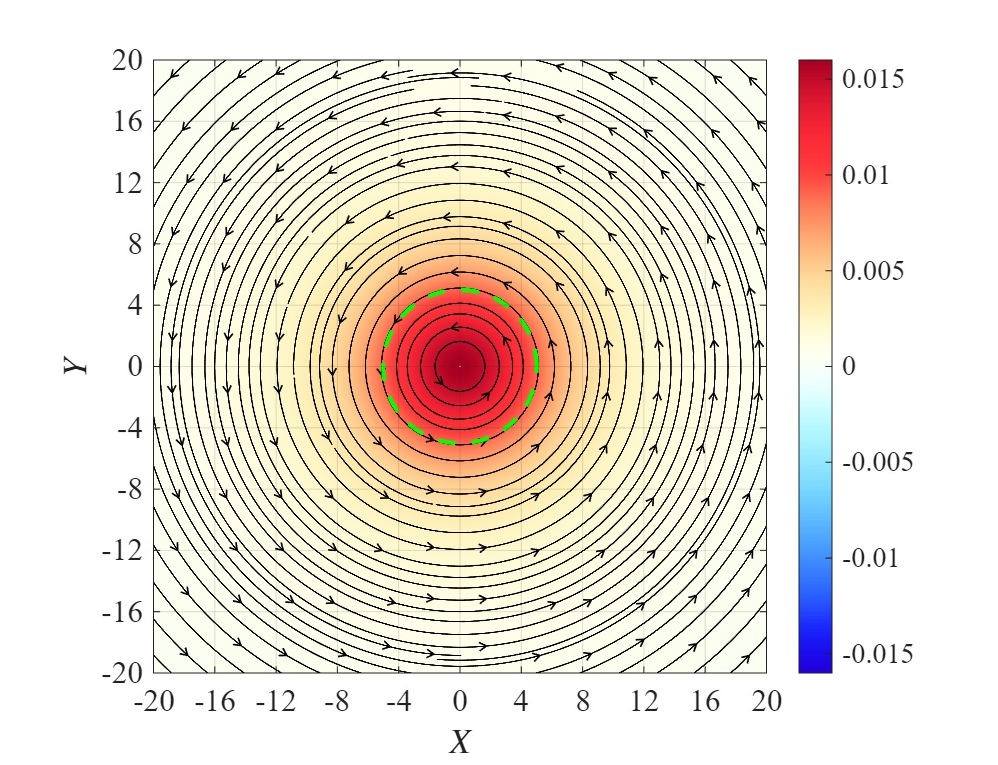}
			\label{LO_RNminus_map}
		\end{minipage}%
	}%
	\subfigure[$\gamma^-$]{
		\begin{minipage}[t]{0.5\linewidth}
			\centering
			\includegraphics[width=1.0\columnwidth,trim={1cm 0cm 1cm 0cm},clip]{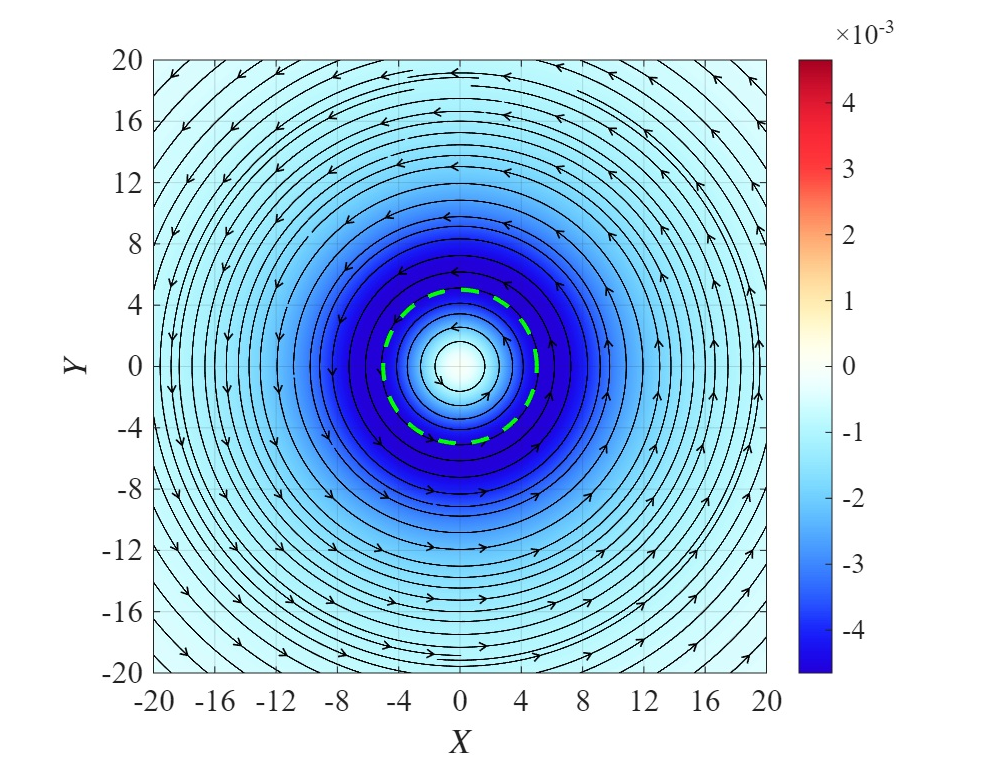}
			\label{LO_SNminus_map}
		\end{minipage}%
	}%
	
	\subfigure[$\omega_{3}$]{
		\begin{minipage}[t]{0.5\linewidth}
			\centering
			\includegraphics[width=1.0\columnwidth,trim={1cm 0cm 1cm 0cm},clip]{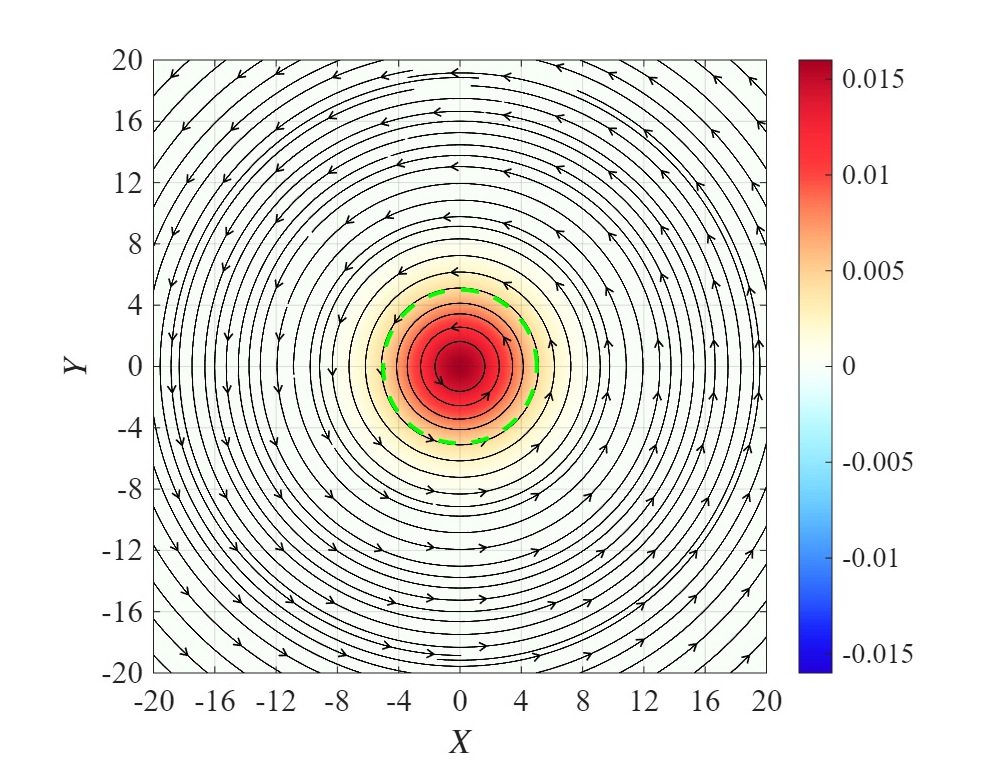}
			\label{LO_omegaz_map}
		\end{minipage}%
	}%
	\subfigure[$\Delta$]{
		\begin{minipage}[t]{0.5\linewidth}
			\centering
			\includegraphics[width=1.0\columnwidth,trim={1cm 0cm 1cm 0cm},clip]{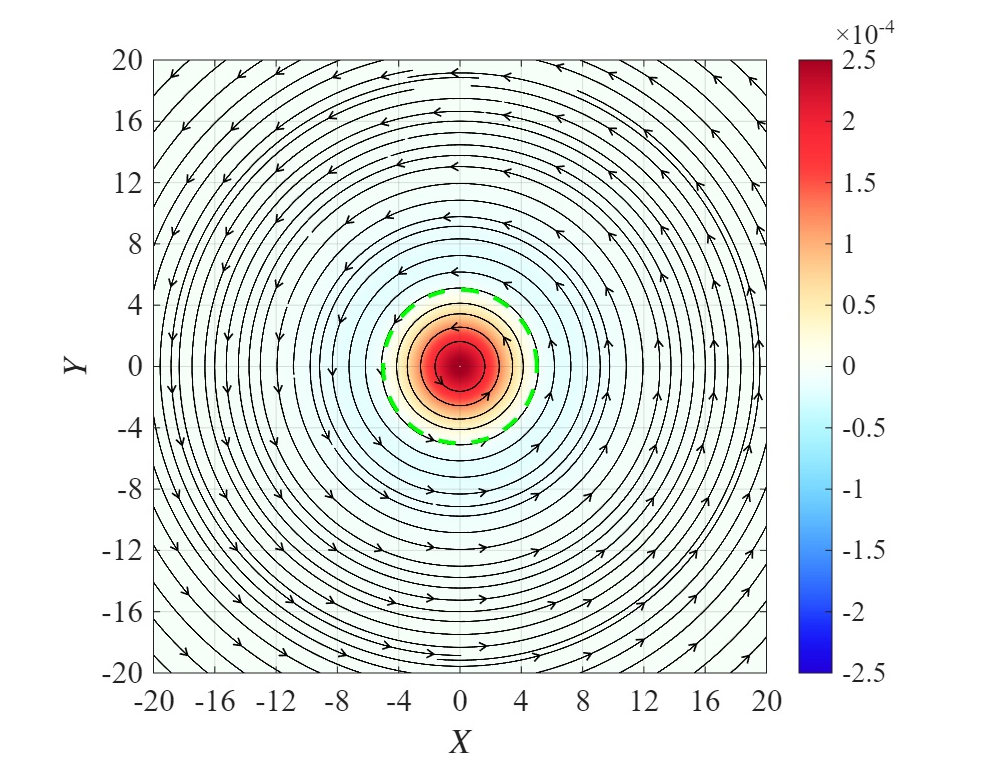}
			\label{LO_Delta_map}
		\end{minipage}%
	}%
	\caption{Instantaneous contour maps of (a) $2\psi^+$, (b) $\gamma^+$, (c) $2\psi^-$, (d) $\gamma^-$, (e) $\omega_{3}(=\omega_{z})$, and (f) $\Delta$. In each subfigure, the green dashed line marks the vortex core radius \(r_{c}=5.0128\), at which \(u_{\theta}\) reaches its maximum and \(\Delta=0\).}
	\label{plots_vorticity_modes}
\end{figure}
\subsubsection{Kinematic vorticity decompositions in the Lamb--Oseen vortex}
Under the orthonormal basis $\left\{\bm{e}_{r},\bm{e}_{\theta}\right\}$, the matrix of the VGT $\bm{A}$ reads
\begin{equation}\label{LO9}
\bm{A}=\dfrac{\partial u_\theta}{\partial r}\bm{e}_{r}\bm{e}_{\theta}-\dfrac{u_\theta}{r}\bm{e}_{\theta}\bm{e}_{r}\mapsto
	\begin{bmatrix}
		0& \dfrac{\partial u_\theta}{\partial r} \\
		-\dfrac{u_\theta}{r} & 0
	\end{bmatrix}.
\end{equation}
The corresponding discriminant is evaluated as
\begin{equation}\label{LO10}
\Delta=4\lambda_{\rm ci}^2=4\dfrac{\partial u_\theta}{\partial r}\dfrac{u_\theta}{r}.
\end{equation}
For \(u_{\theta}/r>0\), the sign of \(\Delta\) is the same as \(\partial_{r}u_{\theta}\). Therefore, $r_{c}=2.2418\sqrt{\nu t}$ is rationally defined as the vortex-core radius of the Lamb--Oseen vortex (see figure~\ref{duthetadr_uthetar_Delta}).
\begin{figure}[t]
	\centering
	\subfigure[]{
		\begin{minipage}[t]{0.5\linewidth}
			\centering
			\includegraphics[width=1.0\columnwidth,trim={0cm 0cm 0cm 0cm},clip]{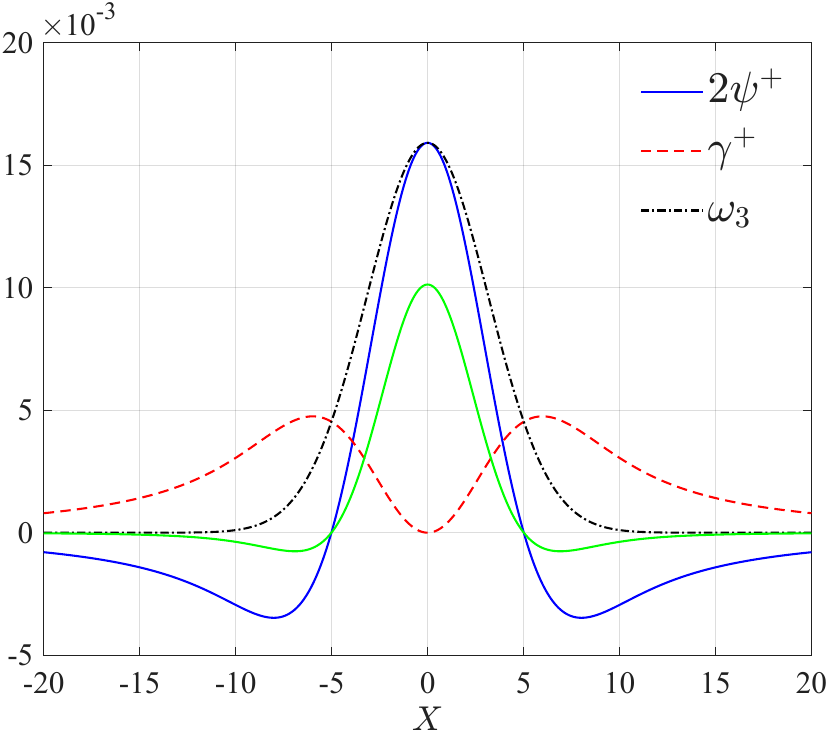}
			\label{LO_RSomega_plus_curve}
		\end{minipage}%
	}%
	\subfigure[]{
		\begin{minipage}[t]{0.5\linewidth}
			\centering
			\includegraphics[width=1.0\columnwidth,trim={0cm 0cm 0cm 0cm},clip]{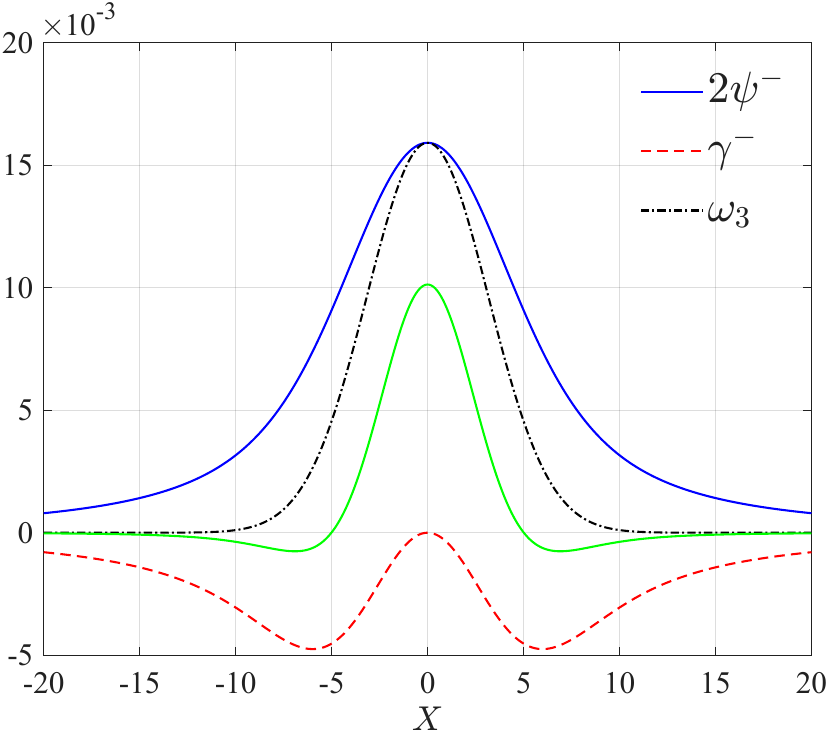}
			\label{LO_RSomega_minus_curve}
		\end{minipage}%
	}%
	
	\subfigure[]{
		\begin{minipage}[t]{0.5\linewidth}
			\centering
			\includegraphics[width=1.0\columnwidth,trim={0cm 0cm 0cm 0cm},clip]{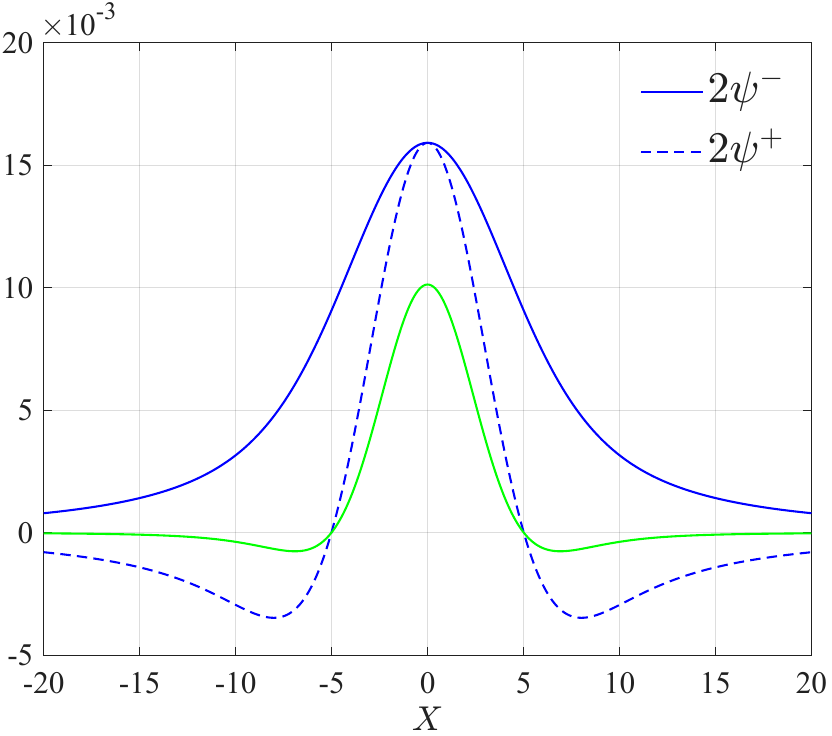}
			\label{LO_RNplus_RNminus_curve}
		\end{minipage}%
	}%
	\subfigure[]{
		\begin{minipage}[t]{0.5\linewidth}
			\centering
			\includegraphics[width=1.0\columnwidth,trim={0cm 0cm 0cm 0cm},clip]{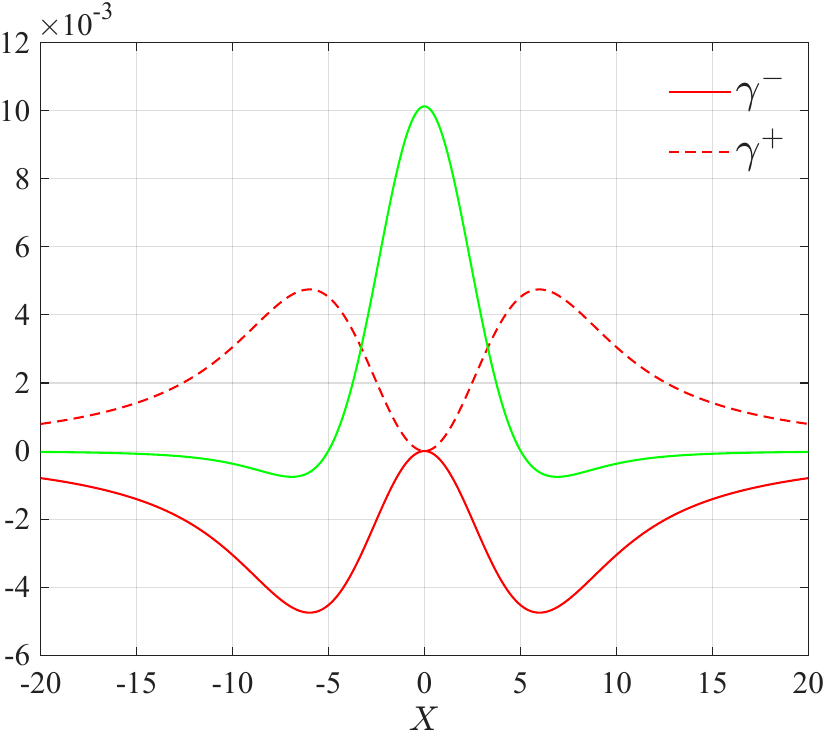}
			\label{LO_SNplus_SNminus_curve}
		\end{minipage}%
	}%
	\caption{The radial distributions of the characteristic vorticity modes. (a) $(2\psi^+,\gamma^+,\omega_3)$, (b) $(2\psi^-,\gamma^-,\omega_{3})$, (c) $(2\psi^-,2\psi^+)$, and (d) $(\gamma^-,\gamma^+)$. The green solid line indicates the scaled discriminant $40\Delta$.}
	\label{The radial distributions of the characteristic vorticity modes}
\end{figure}

By using~\eqref{LO9}, the commutative vorticity operators $(\bm{\Psi}_R,\bm{\Gamma}_{S})$ are expressed under $\left\{\bm{e}_{r},\bm{e}_{\theta}\right\}$ as
\begin{equation}\label{LO11}
	\bm{\Psi}_R \mapsto
	\begin{bmatrix}
		2\dfrac{\partial u_\theta}{\partial r} & 0 \\
		0 & 2\dfrac{u_\theta}{r} 
	\end{bmatrix}, \qquad
	\bm{\Gamma}_S \mapsto
	\begin{bmatrix}
		-\left(\dfrac{\partial u_\theta}{\partial r}-\dfrac{u_\theta}{r}\right) & 0 \\
		0 & \left(\dfrac{\partial u_\theta}{\partial r}-\dfrac{u_\theta}{r}\right)
	\end{bmatrix}.
\end{equation}
Since $\psi^->\psi^+$ holds for $\gamma\neq{0}$, the eigenvalues and eigenvectors of $\bm{\Psi}_R$ are identified as
\begin{eqnarray}\label{LO13}
	&\hat{\lambda}_{1}=2\dfrac{\partial u_\theta}{\partial r}=2\psi^+,~&\bm{\xi}_{1}=\bm{e}_{r}=\bm{e}_{1}^{-},\nonumber\\
	&\hat{\lambda}_{2}=2\dfrac{u_\theta}{r}=2\psi^{-},~&\bm{\xi}_{2}=\bm{e}_{\theta}=\bm{e}_{2}^{-}.
\end{eqnarray}
Accordingly, the eigenvalues and eigenvectors of $\bm{\Gamma}_S$ are given by
\begin{eqnarray}\label{LO14}
	&\check{\lambda}_{1}=\dfrac{u_\theta}{r}-\dfrac{\partial u_\theta}{\partial r} =\gamma^+,~&\bm{\xi}_{1}=\bm{e}_{r}=\bm{e}_{1}^{-},\nonumber\\
	&\check{\lambda}_{2}=\dfrac{\partial u_\theta}{\partial r}-\dfrac{u_\theta}{r}=\gamma^{-},~&\bm{\xi}_{2}=\bm{e}_{\theta}=\bm{e}_{2}^{-}.
\end{eqnarray}
By~\eqref{LO13} and~\eqref{LO14}, one can determine one Schur frame as $(\bm{e}_{1}^{-},\bm{e}_{2}^{-})=(\bm{e}_{r},\bm{e}_{\theta})$ constituted by the common eigenvectors $(\bm{\xi}_{1},\bm{\xi}_{2})$ of $(\bm{\Psi}_R,\bm{\Gamma}_{S})$. The remaining three Schur frames, one $(\bm{e}_{1}^{-},\bm{e}_{2}^{-})$ and two $(\bm{e}_{1}^{+},\bm{e}_{2}^{+})$, can be obtained by rotating the present Schur frame counterclockwise by $\pi/2$, $\pi$, and $3\pi/2$, respectively (see figure~\ref{Schur_frame_NND_OVD}). The eigenvalues, $(\hat{\lambda}_{1,2},\check{\lambda}_{1,2})$, along with $\bm{e}_{3}=\bm{e}_{z}$, just form the characteristic vorticity modes in the IVDs as shown below.
\begin{itemize}
	\item \textbf{Case A.} IVD-I with the Schur frame $(\bm{e}_{1}^{+},\bm{e}_{2}^{+})=(\bm{e}_{\theta},-\bm{e}_{r})$ or $(-\bm{e}_{\theta},\bm{e}_{r})$.
	\begin{subequations}
\begin{equation}
	\bm{\omega}=\bm{R}_{N}^{+}+\bm{S}_{N}^{+},~\bm{R}_{N}^{+}=2\psi^{+}\bm{e}_{3},~\bm{S}_{N}^{+}=\gamma^{+}\bm{e}_{3},
\end{equation}
    \item \textbf{Case B.} IVD-II with the Schur frame $(\bm{e}_{1}^{-},\bm{e}_{2}^{-})=(\bm{e}_{r},\bm{e}_{\theta})$ or $(-\bm{e}_{r},-\bm{e}_{\theta})$.
    \begin{equation}
    	\bm{\omega}=\bm{R}_{N}^{-}+\bm{S}_{N}^{-},~\bm{R}_{N}^{-}=2\psi^{-}\bm{e}_{3},~\bm{S}_{N}^{-}=\gamma^{-}\bm{e}_{3}.
    \end{equation}   		
    \end{subequations}
\end{itemize}
Figure~\ref{plots_vorticity_modes} shows the contour maps of $(2\psi^{\pm},\gamma^{\pm},\omega_{3})$ and $\Delta$, where all quantities display concentric radial distributions in the form of concentric annular rings. As demonstrated in figure~\ref{The radial distributions of the characteristic vorticity modes}, both $2\psi^{\pm}$ and $\omega_{3}$ achieve their maximum values at $r=0$. It is seen that $2\psi^{-}>0$ holds for all $r$. In the conventional vortex region $\Delta>0$, $2\psi^+$ is positive and becomes negative for $\Delta<0$. All vorticity modes decay to zero as $r\rightarrow+\infty$.

\subsubsection{Dynamics of the vorticity modes in the Lamb-Oseen vortex}
\begin{figure}[t]
	\centering
	\subfigure[]{
		\begin{minipage}[t]{0.5\linewidth}
			\centering
			\includegraphics[width=1.0\columnwidth,trim={7cm 4.9cm 7cm 4.5cm},clip]{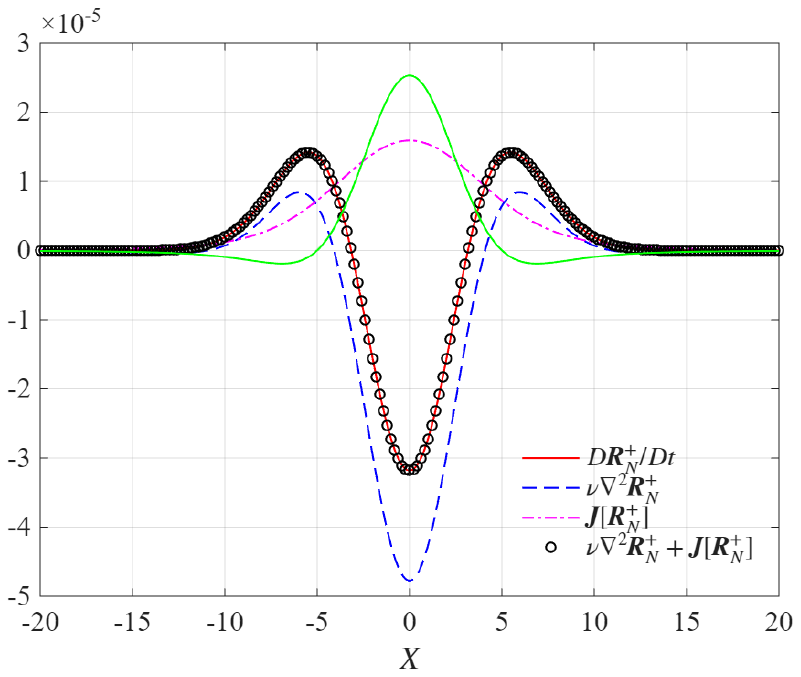}
			\label{LO_dyn_RNplus_curve}
		\end{minipage}%
	}%
		\subfigure[]{
			\begin{minipage}[t]{0.5\linewidth}
		\centering
		\includegraphics[width=1.0\columnwidth,trim={7cm 4.9cm 7cm 4.5cm},clip]{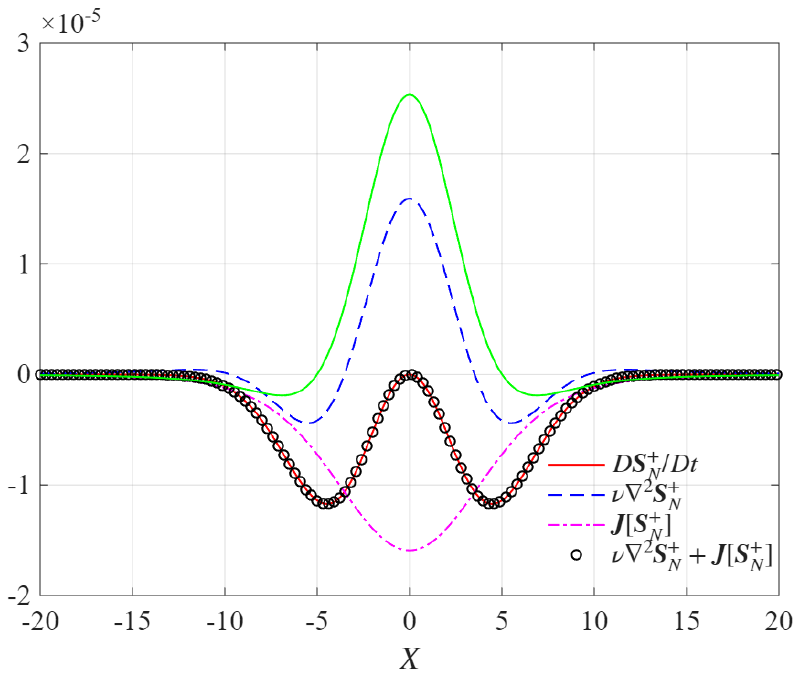}
		\label{LO_dyn_SNplus_curve}
	\end{minipage}%
	}%
	
		\subfigure[]{
			\begin{minipage}[t]{0.5\linewidth}
		\centering
		\includegraphics[width=1.0\columnwidth,trim={7cm 4.9cm 7cm 4.5cm},clip]{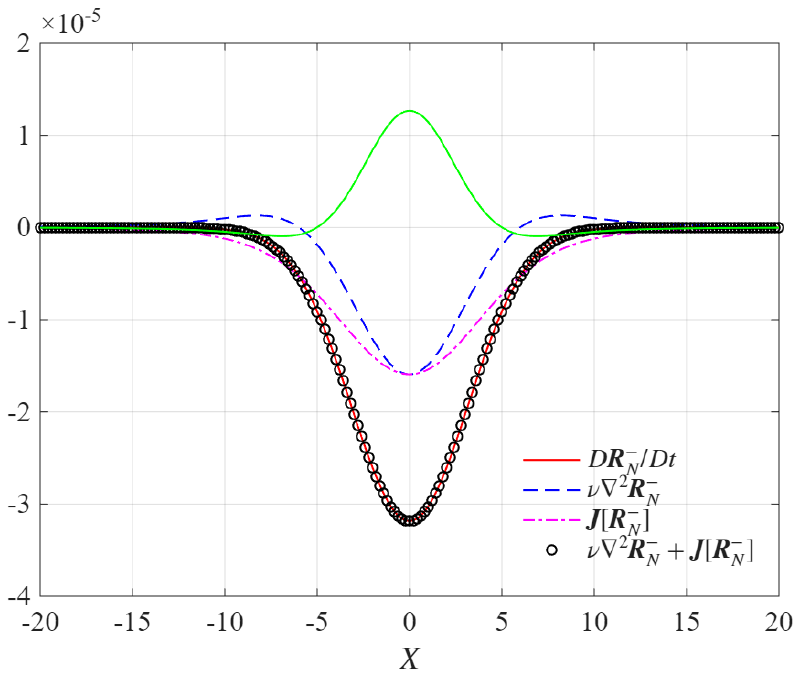}
		\label{LO_dyn_RNminus_curve}
	\end{minipage}%
	}%
			\subfigure[]{
		\begin{minipage}[t]{0.5\linewidth}
			\centering
			\includegraphics[width=1.0\columnwidth,trim={7cm 4.9cm 7cm 4.5cm},clip]{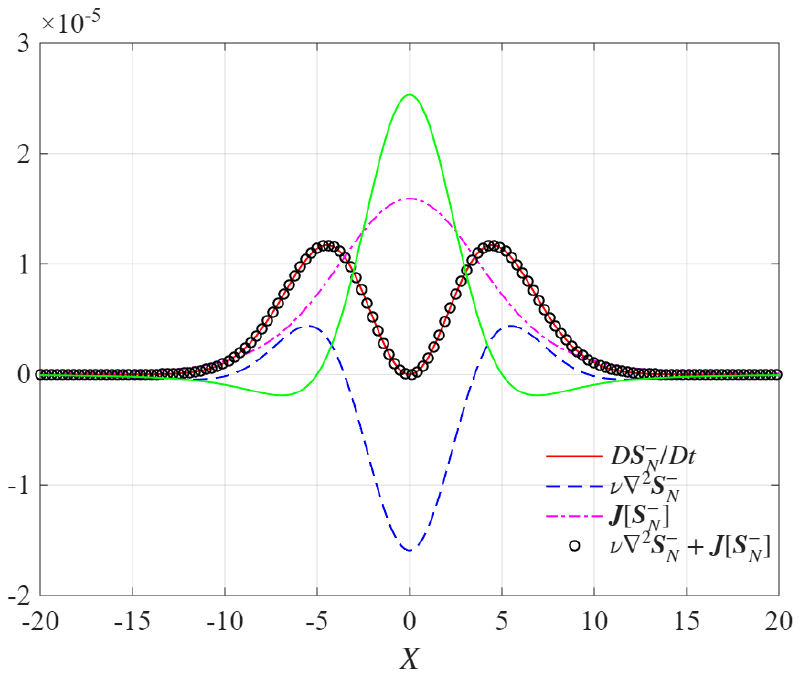}
			\label{LO_dyn_SNminus_curve}
		\end{minipage}%
	}%
	\caption{The radial distributions of the terms in the evolution equations for (a) $\bm{R}_{N}^{+}$~\eqref{eq78axx+}, (b) $\bm{S}_{N}^{+}$~\eqref{eq78bxx+}, (c) $\bm{R}_{N}^{-}$~\eqref{eq78axx-}, and (d) $\bm{S}_{N}^{-}$~\eqref{eq78bxx-}. In each subfigure, the green solid line denotes the discriminant \(\Delta\) scaled by an appropriate constant for improved visualization.}
	\label{Lagrangian_dynamics_comparison}
\end{figure}

Under the orthonormal basis $\left\{\bm{e}_{r},\bm{e}_{\theta}\right\}$, the pressure Hessian tensor is evaluated as
\begin{equation}
\bm{\nabla\nabla}P\mapsto
\begin{bmatrix}
	\dfrac{\partial^2 p}{\partial r^2}
	&
	\dfrac{1}{r}\dfrac{\partial^2 p}{\partial r\partial \theta}
	-\dfrac{1}{r^2}\dfrac{\partial p}{\partial \theta}
	\\[12pt]
	\dfrac{1}{r}\dfrac{\partial^2 p}{\partial r\partial \theta}
	-\dfrac{1}{r^2}\dfrac{\partial p}{\partial \theta}
	&
	\dfrac{1}{r}\dfrac{\partial p}{\partial r}
	+\dfrac{1}{r^2}\dfrac{\partial^2 p}{\partial \theta^2}
\end{bmatrix}
=\begin{bmatrix}
	\dfrac{\partial^2 p}{\partial r^2}
	&
0
	\\[12pt]
	0
	&
	\dfrac{1}{r}\dfrac{\partial p}{\partial r}
\end{bmatrix}.
\end{equation}
Therefore, the dynamical transfer term vanishes, $\bm{\Upsilon}=\bm{0}$, because the off-diagonal elements of $\bm{\nabla\nabla}P$ are zero.
As a result of the Laplacian of the velocity
\begin{equation}
	\nabla^2 \bm{u}
	=
	\left(
	\frac{\partial^2 u_\theta}{\partial r^2}
	+
	\frac{1}{r}\frac{\partial u_\theta}{\partial r}
	-
	\frac{u_\theta}{r^2}
	\right)
	\bm{e}_\theta,
\end{equation}
the Laplacian of the VGT evaluates to
\begin{eqnarray}
\nabla^2\bm{A}&=&\bm{\nabla}(\nabla^2\bm{u})\nonumber\\
&=&
\frac{\partial}{\partial r}
\left(
\frac{\partial^2 u_\theta}{\partial r^2}
+
\frac{1}{r}\frac{\partial u_\theta}{\partial r}
-
\frac{u_\theta}{r^2}
\right)
\bm{e}_r \bm{e}_\theta
-
\frac{1}{r}
\left(
\frac{\partial^2 u_\theta}{\partial r^2}
+
\frac{1}{r}\frac{\partial u_\theta}{\partial r}
-
\frac{u_\theta}{r^2}
\right)
\bm{e}_\theta \bm{e}_r\nonumber\\
&=&\left(
\frac{\partial^3 u_\theta}{\partial r^3}
+
\frac{1}{r}\frac{\partial^2 u_\theta}{\partial r^2}
-
\frac{2}{r^2}\frac{\partial u_\theta}{\partial r}
+
\frac{2u_\theta}{r^3}
\right)\bm{e}_r \bm{e}_\theta-
\frac{1}{r}
\left(
\frac{\partial^2 u_\theta}{\partial r^2}
+
\frac{1}{r}\frac{\partial u_\theta}{\partial r}
-
\frac{u_\theta}{r^2}
\right)
\bm{e}_\theta \bm{e}_r.
\end{eqnarray}
In addition, the following results are derived:
\begin{subequations}
	\begin{equation}
		\nabla^2\left(\frac{\partial u_\theta}{\partial r}\right)
		=
		\frac{\partial^3 u_\theta}{\partial r^3}
		+
		\frac{1}{r}\frac{\partial^2 u_\theta}{\partial r^2},
	\end{equation}
	\begin{equation}
		\nabla^2\left(\frac{u_\theta}{r}\right)
		=
		\frac{1}{r}
		\left(
		\frac{\partial^2 u_\theta}{\partial r^2}
		-
		\frac{1}{r}\frac{\partial u_\theta}{\partial r}
		+
		\frac{u_\theta}{r^2}
		\right),
	\end{equation}
	\begin{equation}
 \nabla^2 \omega_z	=
		\frac{\partial^3 u_\theta}{\partial r^3}
		+
		\frac{2}{r}\frac{\partial^2 u_\theta}{\partial r^2}
		-
		\frac{1}{r^2}\frac{\partial u_\theta}{\partial r}
		+
		\frac{u_\theta}{r^3}.
	\end{equation}
\end{subequations}
By employing~\eqref{LO1}, it follows that
\begin{subequations}
	\begin{gather}
		\frac{\partial}{\partial t}
		\left(
		\frac{u_{\theta}}{r}
		\right)
		=
		\frac{\nu}{r}
		\left(
		\frac{\partial^2 u_\theta}{\partial r^2}
		+
		\frac{1}{r}\frac{\partial u_\theta}{\partial r}
		-
		\frac{u_\theta}{r^2}
		\right),\\[6pt]
		\frac{\partial^2 u_\theta}{\partial r\partial t}
		=
		\nu
		\left(
		\frac{\partial^3 u_\theta}{\partial r^3}
		+
		\frac{1}{r}\frac{\partial^2 u_\theta}{\partial r^2}
		-
		\frac{2}{r^2}\frac{\partial u_\theta}{\partial r}
		+
		\frac{2u_\theta}{r^3}
		\right).
	\end{gather}
\end{subequations}
For 2D flows with $\bm{f}=\bm{0}$, $\bm{\omega}\bm{\cdot}\bm{A}=\bm{\omega}\bm{\cdot}\bm{D}=\bm{0}$ holds, so that~\eqref{vv} simplifies to
\begin{equation}
	\frac{D\bm{\omega}}{Dt}=\nu\nabla^{2}\bm{\omega}.
\end{equation}
\begin{itemize}
	\item \textbf{Case A.} The Lagrangian dynamics of $(\bm{R}_{N}^{+},\bm{S}_{N}^{+})$.\\
	Equations~\eqref{eq78axx} and~\eqref{eq78bxx} become
	\begin{subequations}\label{eq78abxx+}
		\begin{align}
		\frac{D\bm{R}_N^+}{Dt}
		&=\nu\nabla^2\bm{R}_{N}^{+}+\bm{J}[\bm{R}_{N}^{+}], \label{eq78axx+} \\[6pt]
		\frac{D\bm{S}_{N}^{+}}{Dt} 
		&=\nu\nabla^2\bm{S}_{N}^{+}+\bm{J}[\bm{S}_{N}^{+}].\label{eq78bxx+}
	\end{align}
	\end{subequations}
The terms in~\eqref{eq78axx+} are evaluated as
	\begin{subequations}\label{LOR123x}
	\begin{align}
		\frac{D\bm{R}_{N}^{+}}{Dt}
		&=
		\frac{\partial}{\partial t}
		\left(
		2\frac{\partial u_{\theta}}{\partial r}
		\right)
		\bm{e}_{z}=2\nu
		\left(
		\frac{\partial^3 u_\theta}{\partial r^3}
		+
		\frac{1}{r}\frac{\partial^2 u_\theta}{\partial r^2}
		-
		\frac{2}{r^2}\frac{\partial u_\theta}{\partial r}
		+
		\frac{2u_\theta}{r^3}
		\right)\bm{e}_z,\label{LOR1x}\\[6pt]
		\nu\nabla^2\bm{R}_{N}^{+}
		&=2\nu\left(
		\frac{\partial^3 u_\theta}{\partial r^3}
		+
		\frac{1}{r}\frac{\partial^2 u_\theta}{\partial r^2}\right)
		\bm{e}_{z},\label{LOR2x}\\[6pt]
		\bm{J}[\bm{R}_{N}^{+}]
		&=
		2\nu
		\left(
		\bm{e}_{r}\cdot\nabla^{2}\bm{A}\cdot\bm{e}_{\theta}
		\right)
		\bm{e}_{z}
		-
		\nu\nabla^2\bm{R}_{N}^{+}\nonumber\\[6pt]
		&=\frac{4\nu}{r^2}	\left(
		\frac{u_\theta}{r}-
		\frac{\partial u_\theta}{\partial r}
		\right)
		\bm{e}_{z}=4\nu\kappa^2\bm{S}_{N}^{+}.\label{LOR3x}
	\end{align}
\end{subequations}
	The terms in~\eqref{eq78bxx+} are evaluated as
		\begin{subequations}\label{LOS123x}
		\begin{align}
			\frac{D\bm{S}_{N}^{+}}{Dt}
			&=
			\frac{\partial}{\partial t}
			\left(
			\frac{u_\theta}{r}-	\frac{\partial u_{\theta}}{\partial r}
			\right)
			\bm{e}_{z}=-\nu
			\left(
			\frac{\partial^3 u_\theta}{\partial r^3}
			-
			\frac{3}{r^2}\frac{\partial u_\theta}{\partial r}
			+
			\frac{3u_\theta}{r^3}
			\right)\bm{e}_z,\label{LOS1x}\\[6pt]
			\nu\nabla^2\bm{S}_{N}^{+}
			&=-\nu
			\left(
			\frac{\partial^3 u_\theta}{\partial r^3}
			+
			\frac{1}{r^2}\frac{\partial u_\theta}{\partial r}
			-
			\frac{u_\theta}{r^3}
			\right)
			\bm{e}_z,\label{LOS2x}\\[6pt]
			\bm{J}[\bm{S}_{N}^{+}]
			&=
			-\frac{4\nu}{r^2}
			\left(
			\frac{u_\theta}{r}-\frac{\partial u_\theta}{\partial r}
			\right)
			\bm{e}_{z}
			=
			-4\nu\kappa^2\bm{S}_{N}^{+}.\label{LOS3x}
		\end{align}
	\end{subequations}
\end{itemize}
\begin{itemize}
	\item \textbf{Case B.} The Lagrangian dynamics of $(\bm{R}_{N}^{-},\bm{S}_{N}^{-})$.\\
		Equations~\eqref{eq78axx} and~\eqref{eq78bxx} become
	\begin{subequations}\label{eq78abxx-}
		\begin{align}
			\frac{D\bm{R}_N^-}{Dt}
			&=\nu\nabla^2\bm{R}_{N}^{-}+\bm{J}[\bm{R}_{N}^{-}], \label{eq78axx-} \\[6pt]
			\frac{D\bm{S}_{N}^{-}}{Dt} 
			&=\nu\nabla^2\bm{S}_{N}^{-}+\bm{J}[\bm{S}_{N}^{-}].\label{eq78bxx-}
		\end{align}
	\end{subequations}
	The terms in~\eqref{eq78axx} are evaluated as
	\begin{subequations}\label{LOR123}
		\begin{align}
			\frac{D\bm{R}_{N}^{-}}{Dt}
			&=
			\frac{\partial}{\partial t}
			\left(
			2\frac{u_{\theta}}{r}
			\right)
			\bm{e}_{z}=\frac{2\nu}{r}
			\left(
			\frac{\partial^2 u_\theta}{\partial r^2}
			+
			\frac{1}{r}\frac{\partial u_\theta}{\partial r}
			-
			\frac{u_\theta}{r^2}
			\right)
			\bm{e}_z,\label{LOR1}\\[6pt]
			\nu\nabla^2\bm{R}_{N}^{-}
			&=
			2\nu\frac{1}{r}
			\left(
			\frac{\partial^2 u_\theta}{\partial r^2}
			-
			\frac{1}{r}\frac{\partial u_\theta}{\partial r}
			+
			\frac{u_\theta}{r^2}
			\right)
			\bm{e}_{z},\label{LOR2}\\[6pt]
			\bm{J}[\bm{R}_{N}^{-}]
			&=
			2\nu
			\left(
			-\bm{e}_{\theta}\cdot\nabla^{2}\bm{A}\cdot\bm{e}_{r}
			\right)
			\bm{e}_{z}
			-
			\nu\nabla^2\bm{R}_{N}^{-}\nonumber\\[6pt]
			&=
			\frac{4\nu}{r^2}
			\left(
			\frac{\partial u_\theta}{\partial r}
			-
			\frac{u_\theta}{r}
			\right)
			\bm{e}_{z}
			=
			4\nu\kappa^2\bm{S}_{N}^{-}.\label{LOR3}
		\end{align}
	\end{subequations}
	The terms in~\eqref{eq78bxx} are evaluated as
	\begin{subequations}\label{LOS123}
		\begin{align}
			\frac{D\bm{S}_{N}^{-}}{Dt}
			&=
			\frac{\partial}{\partial t}
			\left(
			\frac{\partial u_{\theta}}{\partial r}
			-
			\frac{u_\theta}{r}
			\right)
			\bm{e}_{z}=\nu
			\left(
			\frac{\partial^3 u_\theta}{\partial r^3}
			-
			\frac{3}{r^2}\frac{\partial u_\theta}{\partial r}
			+
			\frac{3u_\theta}{r^3}
			\right)\bm{e}_z,\label{LOS1}\\[6pt]
			\nu\nabla^2\bm{S}_{N}^{-}
			&=
			\nu
			\left(
			\frac{\partial^3 u_\theta}{\partial r^3}
			+
			\frac{1}{r^2}\frac{\partial u_\theta}{\partial r}
			-
			\frac{u_\theta}{r^3}
			\right)
			\bm{e}_z,\label{LOS2}\\[6pt]
			\bm{J}[\bm{S}_{N}^{-}]
			&=
			-\frac{4\nu}{r^2}
			\left(
			\frac{\partial u_\theta}{\partial r}
			-
			\frac{u_\theta}{r}
			\right)
			\bm{e}_{z}
			=
			-4\nu\kappa^{2}\bm{S}_{N}^{-}.\label{LOS3}
		\end{align}
	\end{subequations}
\end{itemize}
Here, $\kappa\equiv1/r$ denotes the curvature of the Schur frame along a pathline. $\bm{J}[\bm{R}_{N}^{\pm}]$ and $\bm{J}[\bm{S}_{N}^{\pm}]$ describe the viscous coupling between the geometry of the Schur frame and the characteristic spin modes $\bm{S}_{N}^{\pm}$.  The terms in~\eqref{eq78abxx+} and~\eqref{eq78abxx-} are visualized in figures~\ref{visual1} and~\ref{visual2}, and their analytical expressions can be found in Appendix~\ref{app3}.
The physical balance among the terms in~\eqref{eq78abxx+} and~\eqref{eq78abxx-} is directly validated using the analytical expressions in~\eqref{LOR123x},~\eqref{LOS123x},~\eqref{LOR123}, and~\eqref{LOS123}, which are quantitatively compared in figure~\ref{Lagrangian_dynamics_comparison}.
In the inner core region (especially $\Delta>0$ near $r=0$), $D\bm{R}_{N}^{+}/Dt$ exhibits a highly negative peak dominated by negative $\nu\nabla^{2}\bm{R}_{N}^{+}$ over positive $\bm{J}\lbrack\bm{R}_{N}^{+}\rbrack$ (figure~\ref{LO_dyn_RNplus_curve}). In the vicinity of $r=0$, $D\bm{S}_{N}^{+}/Dt$ shows relatively low magnitude owing to the competition between positive $\nu\nabla^{2}\bm{S}_{N}^{+}$ and negative $\bm{J}\lbrack\bm{S}_{N}^{+}\rbrack$ with approximately equal magnitudes (figure~\ref{LO_dyn_SNplus_curve}). In the transition region (especially around $r=r_{c}$), the positive peaks of $D\bm{R}_{N}^{+}/Dt$ are contributed by both positive $\nu\nabla^{2}\bm{R}_{N}^{+}$ and positive $\bm{J}\lbrack\bm{R}_{N}^{+}\rbrack$ (figure~\ref{LO_dyn_RNplus_curve}). Near $r=r_{c}$, the double negative peaks of $D\bm{S}_{N}^{+}/Dt$ arise from negative $\nu\nabla^{2}\bm{S}_{N}^{+}$ and negative $\bm{J}\lbrack\bm{S}_{N}^{+}\rbrack$ (figure~\ref{LO_dyn_SNplus_curve}).
Being different from $\bm{R}_{N}^{+}$,  $D\bm{R}_{N}^{-}/Dt$ shows only one highly negative peak enhanced by both $\nu\nabla^{2}\bm{R}_{N}^{-}$ and $\bm{J}\lbrack\bm{R}_{N}^{-}\rbrack$ (figure~\ref{LO_dyn_RNminus_curve}). Since $\bm{S}_{N}^{+}=-\bm{S}_{N}^{-}$, each term in~\eqref{eq78bxx-} exhibits a variation trend precisely opposite to the counterpart in~\eqref{eq78bxx+} (figure~\ref{LO_dyn_SNminus_curve}).
\begin{figure}[htbp]
	\centering
	\subfigure[$D\bm{R}_{N}^{+}/Dt$]{
		\begin{minipage}[t]{0.5\linewidth}
			\centering
			\includegraphics[width=1.0\columnwidth,trim={1cm 0cm 1cm 0cm},clip]{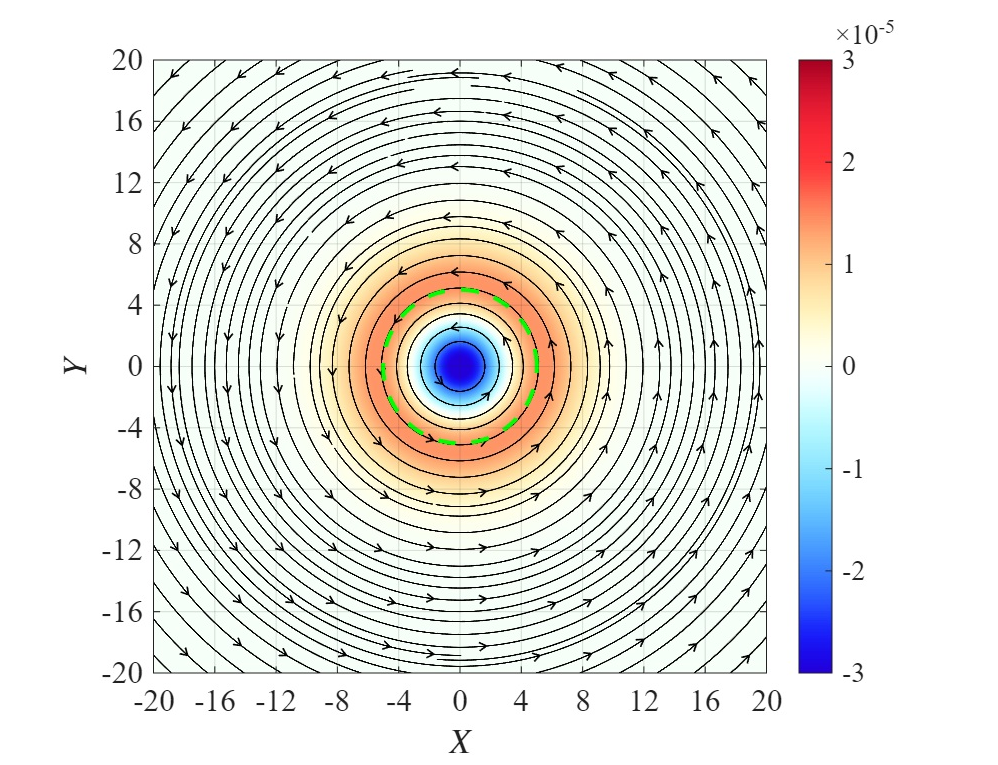}
			\label{LO_DRNpDt_map}
		\end{minipage}%
	}%
		\subfigure[$\nu\nabla^{2}\bm{R}_{N}^{+}$]{
		\begin{minipage}[t]{0.5\linewidth}
			\centering
			\includegraphics[width=1.0\columnwidth,trim={1cm 0cm 1cm 0cm},clip]{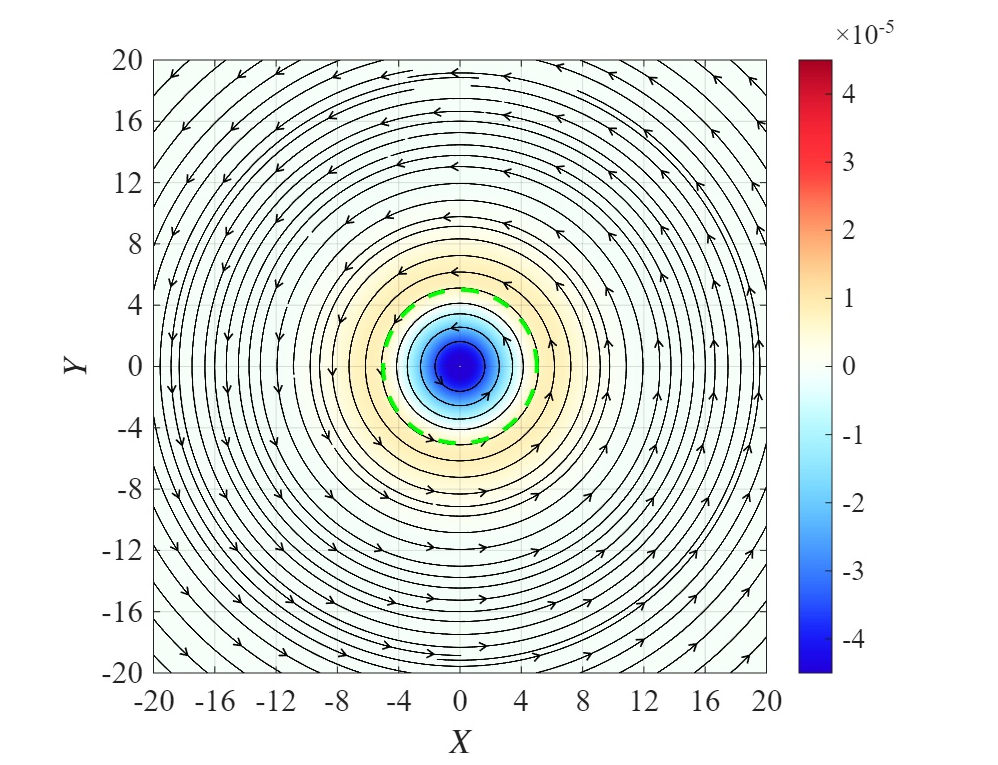}
			\label{LO_nuLapRNp_map}
		\end{minipage}%
	}%
	
        \subfigure[$\bm{J}\lbrack\bm{R}_{N}^{+}\rbrack$]{
		\begin{minipage}[t]{0.5\linewidth}
			\centering
			\includegraphics[width=1.0\columnwidth,trim={1cm 0cm 1cm 0cm},clip]{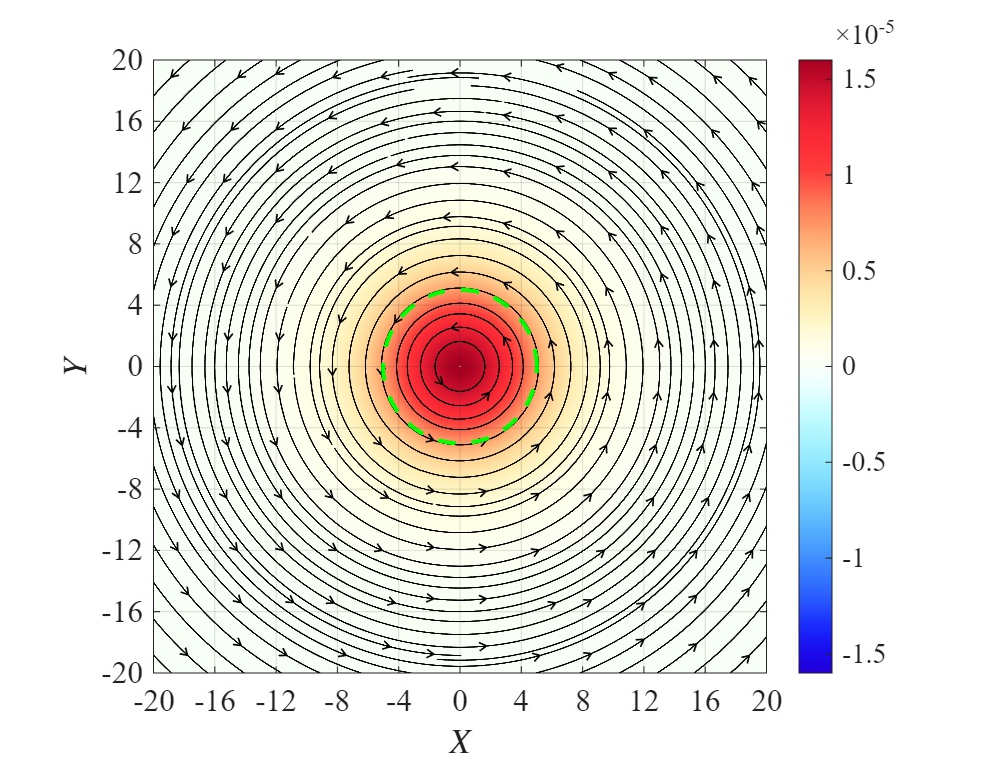}
			\label{LO_JRNp_map}
		\end{minipage}%
	}%
		\subfigure[$D\bm{S}_{N}^{+}/Dt$]{
		\begin{minipage}[t]{0.5\linewidth}
			\centering
			\includegraphics[width=1.0\columnwidth,trim={1cm 0cm 1cm 0cm},clip]{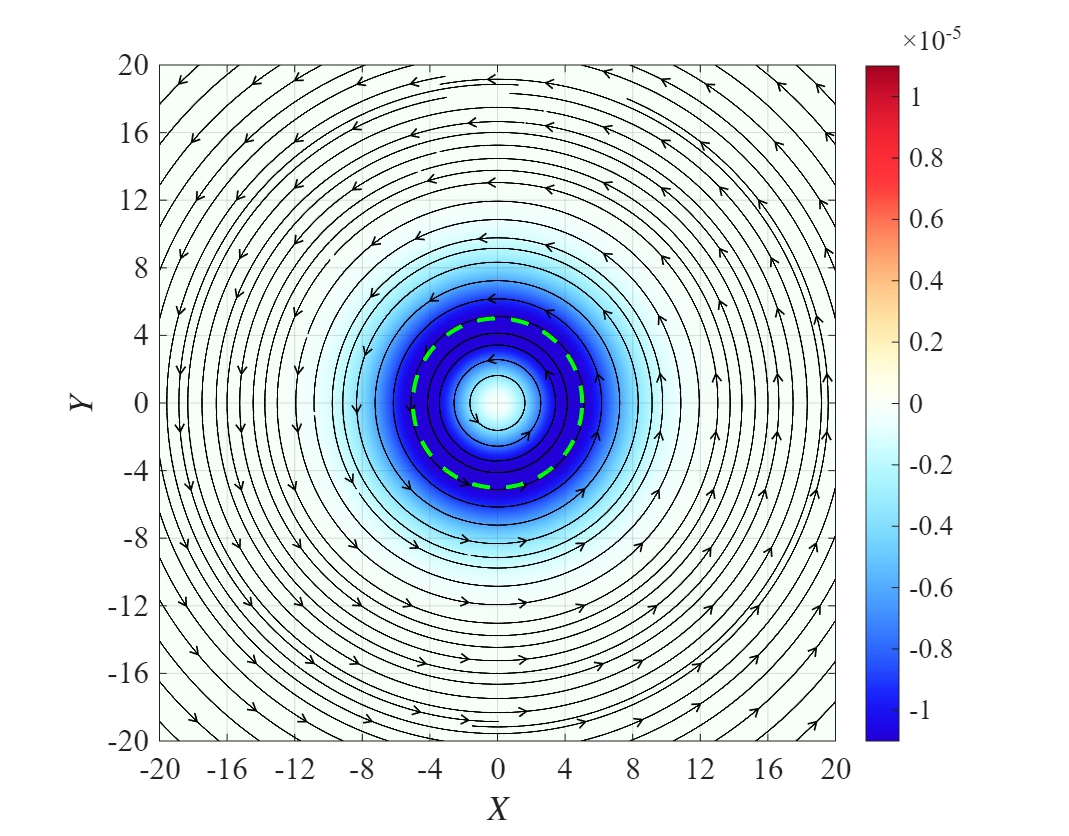}
			\label{LO_DSNpDt_map}
		\end{minipage}%
	}%
	
			\subfigure[$\nu\nabla^2\bm{S}_{N}^{+}$]{
		\begin{minipage}[t]{0.5\linewidth}
			\centering
			\includegraphics[width=1.0\columnwidth,trim={1cm 0cm 1cm 0cm},clip]{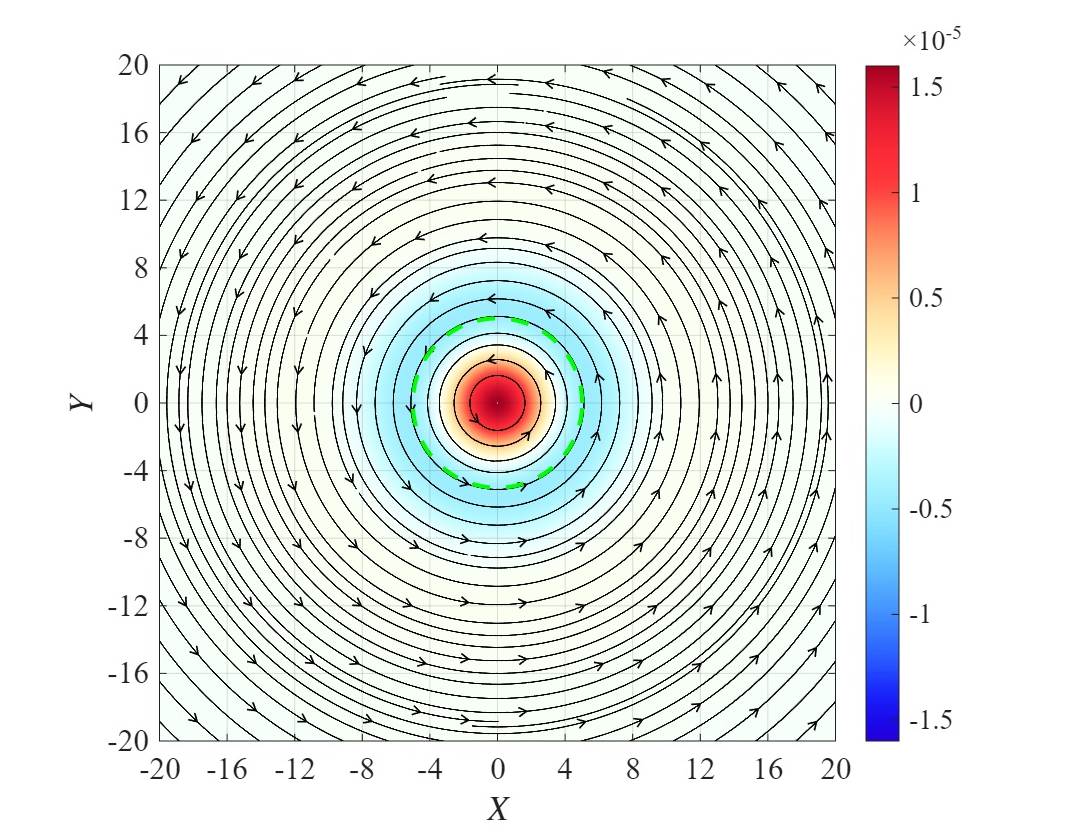}
			\label{LO_nuLapSNp_map}
		\end{minipage}%
	}%
				\subfigure[$\bm{J}\lbrack\bm{S}_{N}^{+}\rbrack$]{
		\begin{minipage}[t]{0.5\linewidth}
			\centering
			\includegraphics[width=1.0\columnwidth,trim={1cm 0cm 1cm 0cm},clip]{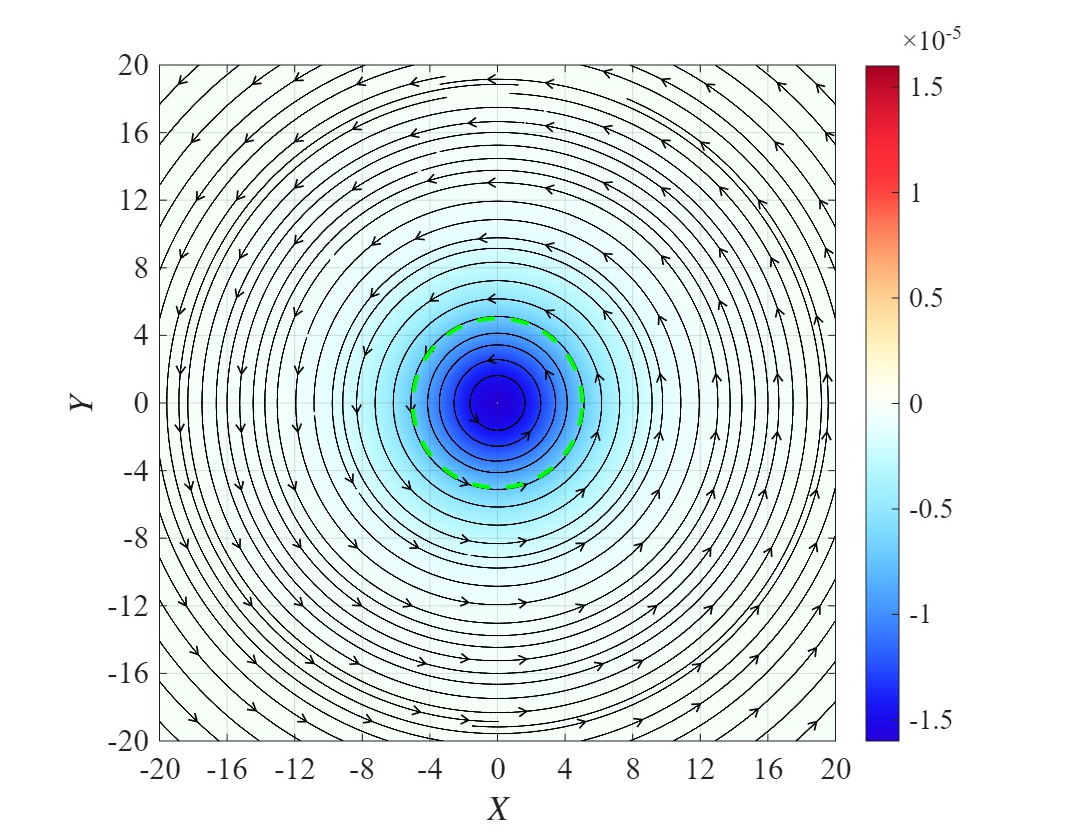}
			\label{LO_JSNp_map}
		\end{minipage}%
	}%
	\caption{Instantaneous contour maps of (a) $D\bm{R}_{N}^{+}/Dt$, (b) $\nu\nabla^{2}\bm{R}_{N}^{+}$, (c) $\bm{J}[\bm{R}_{N}^{+}]$, (d) $D\bm{S}_{N}^{+}/Dt$, (e) $\nu\nabla^{2}\bm{S}_{N}^{+}$, and (f) $\bm{J}[\bm{S}_{N}^{+}]$. In each subfigure, the green dashed line denotes the vortex--core radius $r_{c}=5.0128$, where $u_{\theta}$ attains its maximum and $\Delta=0$.}
	\label{visual1}
\end{figure}
\begin{figure}[htbp]
	\centering
	\subfigure[$D\bm{R}_{N}^{-}/Dt$]{
		\begin{minipage}[t]{0.5\linewidth}
			\centering
			\includegraphics[width=1.0\columnwidth,trim={1cm 0cm 1cm 0cm},clip]{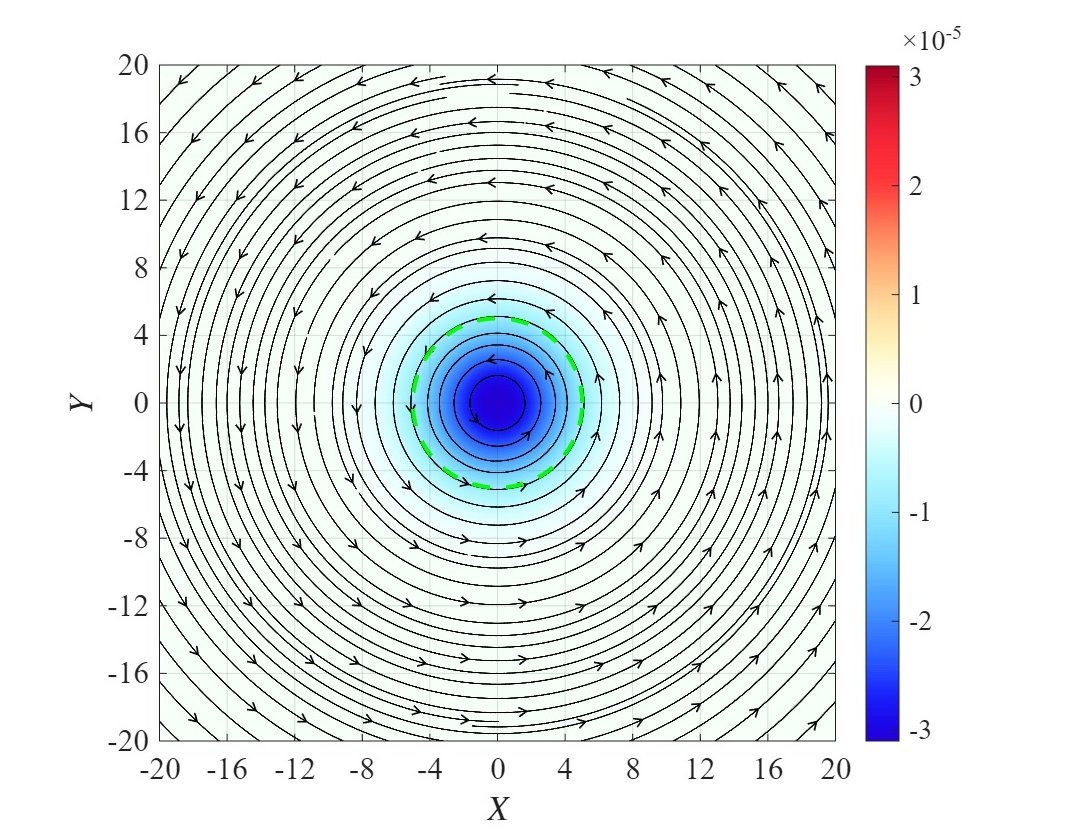}
			\label{LO_DRNmDt_map}
		\end{minipage}%
	}%
	\subfigure[$\nu\nabla^{2}\bm{R}_{N}^{-}$]{
		\begin{minipage}[t]{0.5\linewidth}
			\centering
			\includegraphics[width=1.0\columnwidth,trim={1cm 0cm 1cm 0cm},clip]{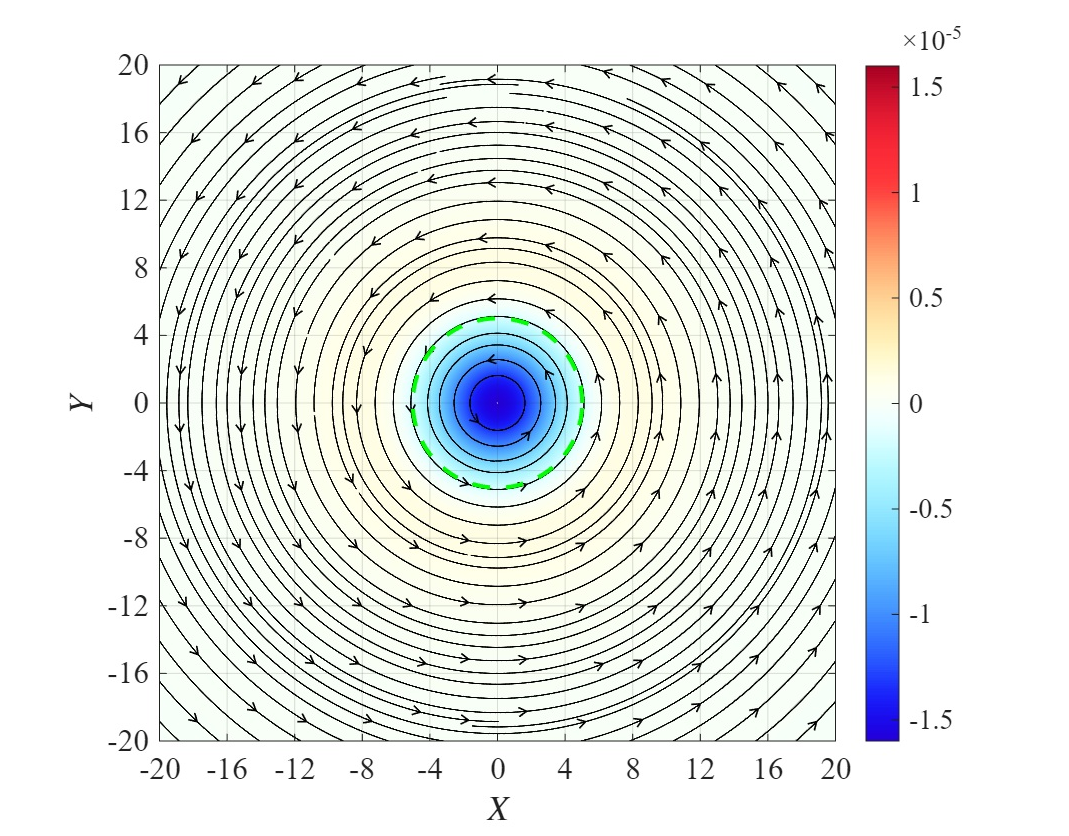}
			\label{LO_nuLapRNm_map}
		\end{minipage}%
	}%
	
	\subfigure[$\bm{J}\lbrack\bm{R}_{N}^{-}\rbrack$]{
		\begin{minipage}[t]{0.5\linewidth}
			\centering
			\includegraphics[width=1.0\columnwidth,trim={1cm 0cm 1cm 0cm},clip]{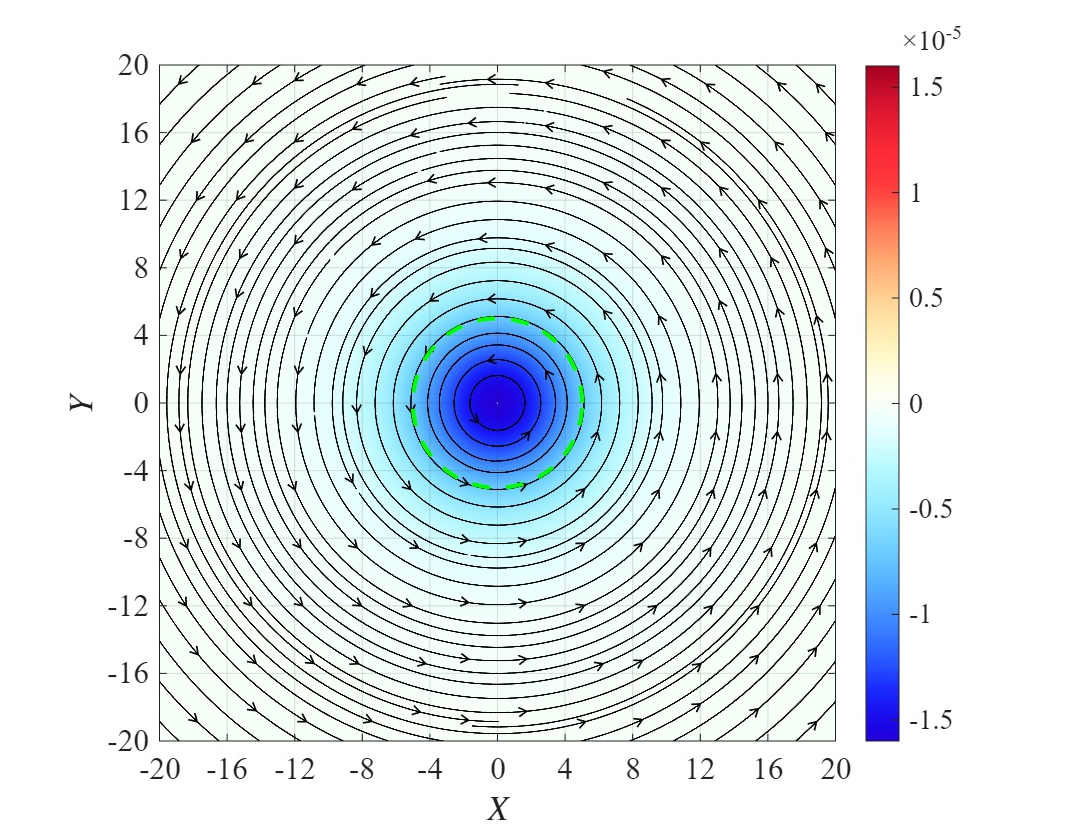}
			\label{LO_JRNm_map}
		\end{minipage}%
	}%
	\subfigure[$D\bm{S}_{N}^{-}/Dt$]{
		\begin{minipage}[t]{0.5\linewidth}
			\centering
			\includegraphics[width=1.0\columnwidth,trim={1cm 0cm 1cm 0cm},clip]{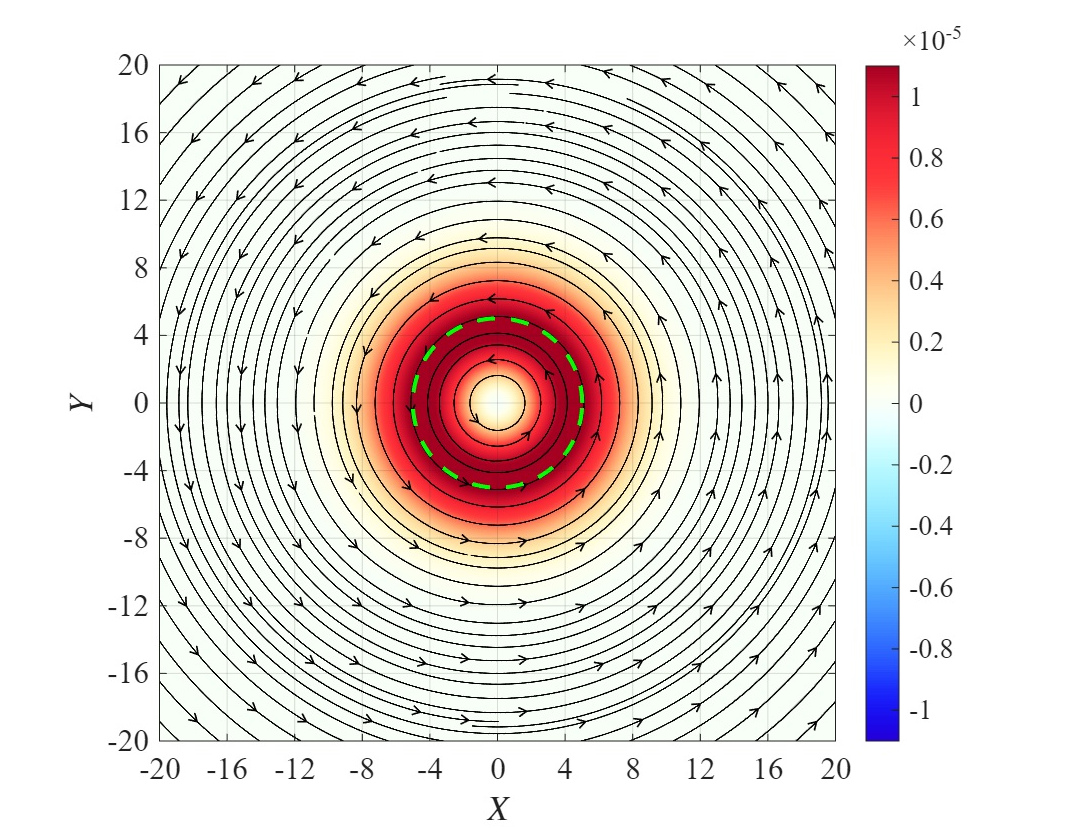}
			\label{LO_DSNmDt_map}
		\end{minipage}%
	}%
	
	\subfigure[$\nu\nabla^2\bm{S}_{N}^{-}$]{
		\begin{minipage}[t]{0.5\linewidth}
			\centering
			\includegraphics[width=1.0\columnwidth,trim={1cm 0cm 1cm 0cm},clip]{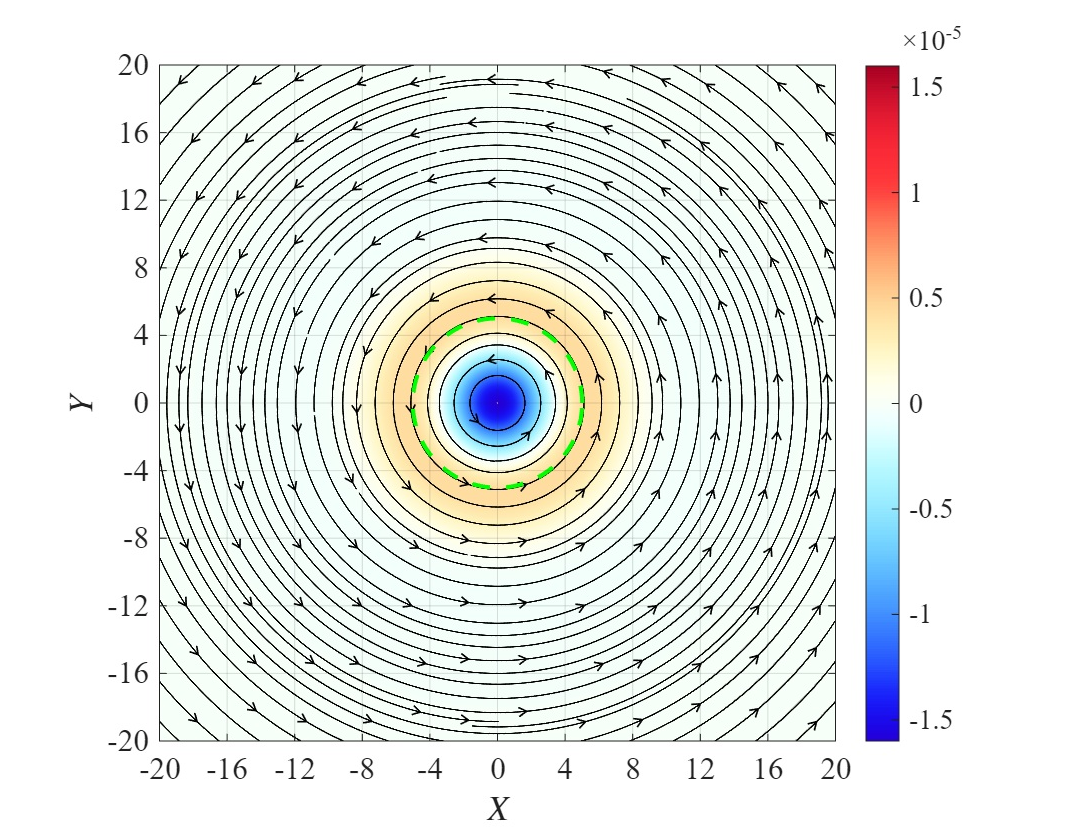}
			\label{LO_nuLapSNm_map}
		\end{minipage}%
	}%
	\subfigure[$\bm{J}\lbrack\bm{S}_{N}^{-}\rbrack$]{
		\begin{minipage}[t]{0.5\linewidth}
			\centering
			\includegraphics[width=1.0\columnwidth,trim={1cm 0cm 1cm 0cm},clip]{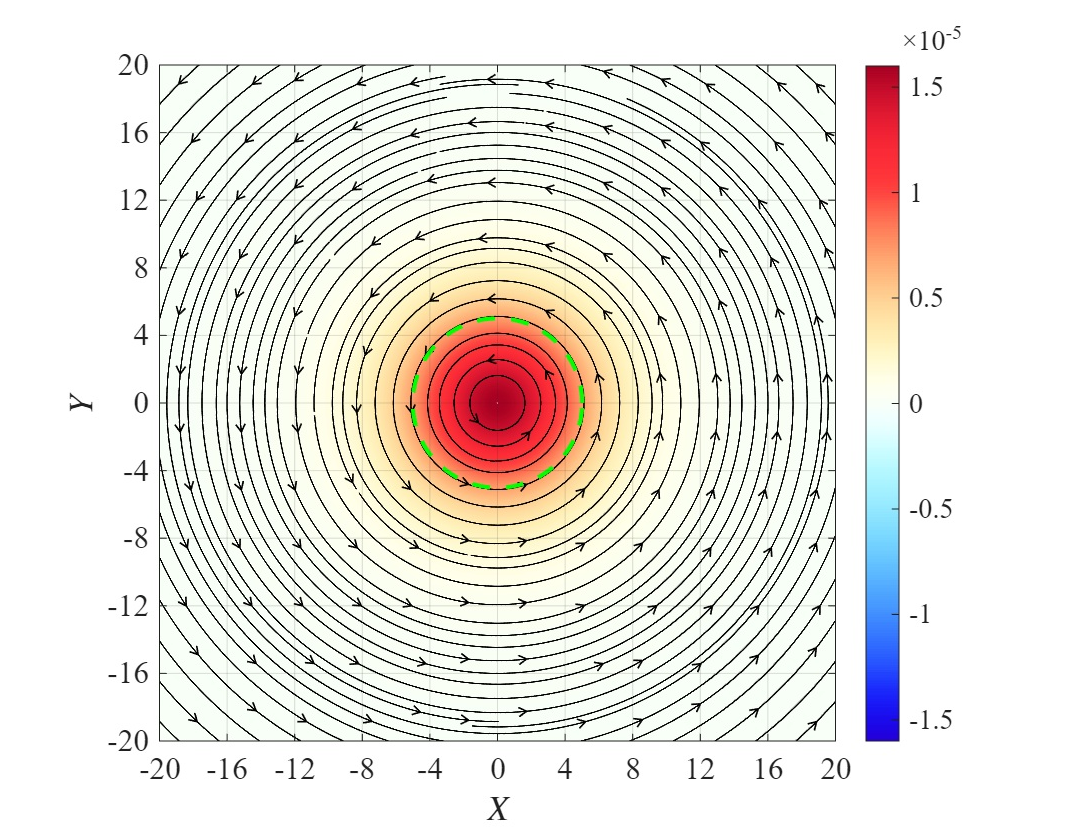}
			\label{LO_JSNm_map}
		\end{minipage}%
	}%
\caption{Instantaneous contour maps of (a) $D\bm{R}_{N}^{-}/Dt$, (b) $\nu\nabla^{2}\bm{R}_{N}^{-}$, (c) $\bm{J}[\bm{R}_{N}^{-}]$, (d) $D\bm{S}_{N}^{-}/Dt$, (e) $\nu\nabla^{2}\bm{S}_{N}^{-}$, and (f) $\bm{J}[\bm{S}_{N}^{-}]$. In each subfigure, the green dashed line denotes the vortex--core radius $r_{c}=5.0128$, where $u_{\theta}$ attains its maximum and $\Delta=0$.}
	\label{visual2}
\end{figure}

\subsubsection{Decomposition of the viscous source pair}\label{Decomposition of the viscous source pair}
The following fundamental relations can be directly derived:
\begin{subequations}
\begin{equation}
	\bm{\nabla}\bm{e}_{r}=\frac{1}{r}\bm{e}_{\theta}\bm{e}_{\theta},~\bm{\nabla}\bm{e}_{\theta}=-\frac{1}{r}\bm{e}_{\theta}\bm{e}_{r},
\end{equation} 
\begin{equation}
	\bm{e}_{r}\bm{\nabla}=\bm{\nabla}\bm{e}_{r}^{\rm T}=\frac{1}{r}\bm{e}_{\theta}\bm{e}_{\theta},~\bm{e}_{\theta}\bm{\nabla}=\bm{\nabla}\bm{e}_{\theta}^{\rm T}=-\frac{1}{r}\bm{e}_{r}\bm{e}_{\theta},
\end{equation} 
\begin{equation}
	\nabla^2\bm{e}_{\theta}=-\frac{1}{r^2}\bm{e}_{\theta},~\nabla^2\bm{e}_{r}=-\frac{1}{r^2}\bm{e}_{r},
\end{equation}
\begin{equation}
	\frac{\partial\bm{A}}{\partial\theta}
	=\left(\frac{\partial u_\theta}{\partial r}
	-\frac{u_\theta}{r}\right)\left(\bm{e}_{\theta}\bm{e}_{\theta}-\bm{e}_{r}\bm{e}_{r}\right),
\end{equation}
\begin{equation}
	\frac{\partial\bm{A}}{\partial r}=\frac{\partial^2 u_\theta}{\partial r^2}\bm{e}_{r}\bm{e}_{\theta}+\frac{1}{r}\left(\frac{u_\theta}{r}-\frac{\partial u_\theta}{\partial r}\right)\bm{e}_{\theta}\bm{e}_{r},
\end{equation}
\begin{equation}
	\bm{\nabla}\bm{A}=\frac{\partial^2 u_\theta}{\partial r^2} \bm{e}_r \bm{e}_r \bm{e}_\theta + \frac{1}{r}\left( \frac{u_\theta}{r} -  \frac{\partial u_\theta}{\partial r} \right) \bm{e}_r \bm{e}_\theta \bm{e}_r + \frac{1}{r} \left( \frac{\partial u_\theta}{\partial r} - \frac{u_\theta}{r} \right) (\bm{e}_\theta \bm{e}_\theta \bm{e}_\theta - \bm{e}_\theta \bm{e}_r \bm{e}_r),
\end{equation}
\begin{equation}
	\bm{A}\bm{\nabla}=\frac{\partial^2 u_\theta}{\partial r^2} \bm{e}_r \bm{e}_\theta \bm{e}_r + \left( \frac{u_\theta}{r^2} - \frac{1}{r} \frac{\partial u_\theta}{\partial r} \right) \bm{e}_\theta \bm{e}_r \bm{e}_r + \frac{1}{r} \left( \frac{\partial u_\theta}{\partial r} - \frac{u_\theta}{r} \right) (\bm{e}_\theta \bm{e}_\theta \bm{e}_\theta - \bm{e}_r \bm{e}_r \bm{e}_\theta).
\end{equation}	
\end{subequations}
\begin{itemize}
	\item \textbf{Case A.} For the vorticity modes $(\bm{R}_{N}^{+},\bm{S}_{N}^{+})$ and the Schur frame $(\bm{e}_{1}^{+},\bm{e}_{2}^{+})=(\bm{e}_{\theta},-\bm{e}_{r})$ or $(-\bm{e}_{\theta},\bm{e}_{r})$, it follows from~\eqref{VVX} that
	\begin{equation}\label{S1}
		\bm{J}[\bm{R}_{N}^+]=\bm{J}_{1}[\bm{R}_{N}^+]+\bm{J}_{2}[\bm{R}_{N}^+]+\bm{J}_{3}[\bm{R}_{N}^+]+\bm{J}_{4}[\bm{R}_{N}^+]+\bm{J}_{5}[\bm{R}_{N}^+].
	\end{equation}
	The terms on the right-hand side of~\eqref{S1} are evaluated as
	\begin{eqnarray}\label{S2}
		\bm{J}_{1}[\bm{R}_{N}^+]&\equiv&-\left[2\nu(\nabla^2\bm{e}_{r})\bm{\cdot}\bm{A}\bm{\cdot}\bm{e}_{\theta}\right]\bm{e}_{z}=\frac{2\nu}{r^2}\frac{\partial u_\theta}{\partial r}\bm{e}_{z},\nonumber\\
		\bm{J}_{2}[\bm{R}_{N}^+]&\equiv&-\left[2\nu\bm{e}_{r}\bm{\cdot}\bm{A}\bm{\cdot}(\nabla^2\bm{e}_{\theta})\right]\bm{e}_{z}=\frac{2\nu}{r^2}\frac{\partial u_\theta}{\partial r}\bm{e}_{z},\nonumber\\
		\bm{J}_{3}[\bm{R}_{N}^+]&\equiv&-\left[4\nu(\bm{\nabla}\bm{e}_{r})\bm{:}\bm{\nabla}\bm{A}\bm{\cdot}\bm{e}_{\theta}\right]\bm{e}_{z}=\frac{4\nu}{r^2}\left(\frac{u_\theta}{r}-\frac{\partial u_\theta}{\partial r}\right)\bm{e}_{z},\nonumber\\
		\bm{J}_{4}[\bm{R}_{N}^+]&\equiv&-\left[4\nu\bm{A}\bm{:}(\bm{e}_{r}\bm{\nabla}\bm{\cdot}\bm{\nabla}\bm{e}_{\theta})\right]\bm{e}_{z}=-\frac{4\nu}{r^2}\frac{u_\theta}{r}\bm{e}_z,\nonumber\\
		\bm{J}_{5}[\bm{R}_{N}^+]&\equiv&-\left[4\nu \bm{e}_{r}\bm{\cdot}\bm{A\nabla}\bm{:}(\bm{e}_{\theta}\bm{\nabla})\right]\bm{e}_{z}=\frac{4\nu}{r^2}\left(\frac{u_\theta}{r}-\frac{\partial u_\theta}{\partial r}\right)\bm{e}_{z}.
	\end{eqnarray}
    As $r\rightarrow0^+$, we obtain
    \begin{equation}
    	\begin{aligned}
    		&\bm{J}_1[\bm{R}_{N}^{+}] = \bm{J}_2[\bm{R}_{N}^{+}] \sim \frac{\Gamma_0}{4\pi t\, r^2}\bm{e}_z \to +\infty, \\
    		&\bm{J}_3[\bm{R}_{N}^{+}] = \bm{J}_5[\bm{R}_{N}^{+}] \to \frac{\Gamma_0}{8\pi \nu t^2}\bm{e}_z, \\
    		&\bm{J}_4[\bm{R}_{N}^{+}] \sim -\frac{\Gamma_0}{2\pi t\, r^2}\,\bm{e}_z \to -\infty.
    	\end{aligned}
    \end{equation}
Note that $\bm{J}_4[\bm{R}_{N}^{+}]$ diverges twice as fast as $\bm{J}_{1}[\bm{R}_{N}^{+}]$ and $\bm{J}_{2}[\bm{R}_{N}^{+}]$, with the opposite sign, while $\bm{J}_3[\bm{R}_{N}^{+}]$ and $\bm{J}_5[\bm{R}_{N}^{+}]$ remain finite and positive at the origin. Hence, $\bm{J}[\bm{R}_{N}^{+}]$ is always finite.
Substituting~\eqref{S2} into~\eqref{S1} recovers~\eqref{LOR3x}.
	\item \textbf{Case B.} For the vorticity modes $(\bm{R}_{N}^{-},\bm{S}_{N}^{-})$ and the Schur frame $(\bm{e}_{1}^{-},\bm{e}_{2}^{-})=(\bm{e}_{r},\bm{e}_{\theta})$ or $(-\bm{e}_{r},-\bm{e}_{\theta})$, it follows from~\eqref{VVX} that
	\begin{equation}\label{S3}
		\bm{J}[\bm{R}_{N}^-]=\bm{J}_{1}[\bm{R}_{N}^-]+\bm{J}_{2}[\bm{R}_{N}^-]+\bm{J}_{3}[\bm{R}_{N}^-]+\bm{J}_{4}[\bm{R}_{N}^-]+\bm{J}_{5}[\bm{R}_{N}^-].
	\end{equation}
	The terms on the right-hand side of~\eqref{S3} are evaluated as
	\begin{eqnarray}\label{S4}
		\bm{J}_{1}[\bm{R}_{N}^-]&\equiv&\left[2\nu(\nabla^2\bm{e}_{\theta})\bm{\cdot}\bm{A}\bm{\cdot}\bm{e}_{r}\right]\bm{e}_{z}=\frac{2\nu}{r^2}\frac{u_\theta}{r}\bm{e}_{z},\nonumber\\
		\bm{J}_{2}[\bm{R}_{N}^-]&\equiv&\left[2\nu\bm{e}_{\theta}\bm{\cdot}\bm{A}\bm{\cdot}(\nabla^2\bm{e}_{r})\right]\bm{e}_{z}=\frac{2\nu}{r^2}\frac{u_\theta}{r}\bm{e}_{z},\nonumber\\
		\bm{J}_{3}[\bm{R}_{N}^-]&\equiv&\left[4\nu(\bm{\nabla}\bm{e}_{\theta})\bm{:}\bm{\nabla}\bm{A}\bm{\cdot}\bm{e}_{r}\right]\bm{e}_{z}=\frac{4\nu}{r^2}\left( \frac{\partial u_\theta}{\partial r} - \frac{u_\theta}{r} \right)\bm{e}_z,\nonumber\\
		\bm{J}_{4}[\bm{R}_{N}^-]&\equiv&\left[4\nu\bm{A}\bm{:}(\bm{e}_{\theta}\bm{\nabla}\bm{\cdot}\bm{\nabla}\bm{e}_{r})\right]\bm{e}_{z}=-\frac{4\nu}{r^2}\frac{\partial u_\theta}{\partial r}\bm{e}_z,\nonumber\\
		\bm{J}_{5}[\bm{R}_{N}^-]&\equiv&\left[4\nu \bm{e}_{\theta}\bm{\cdot}\bm{A\nabla}\bm{:}(\bm{e}_{r}\bm{\nabla})\right]\bm{e}_{z}=\frac{4\nu}{r^2}\left( \frac{\partial u_\theta}{\partial r} - \frac{u_\theta}{r} \right)\bm{e}_z.
	\end{eqnarray}
	  As $r\rightarrow0^+$, we obtain
	\begin{equation}
		\begin{aligned}
			&\bm{J}_1[\bm{R}_{N}^{-}] = \bm{J}_2[\bm{R}_{N}^{-}] \sim \frac{\Gamma_0}{4\pi t\, r^2}\bm{e}_z \to +\infty, \\
			&\bm{J}_3[\bm{R}_{N}^{-}] = \bm{J}_5[\bm{R}_{N}^{-}] \to -\frac{\Gamma_0}{8\pi \nu t^2}\bm{e}_z, \\
			&\bm{J}_4[\bm{R}_{N}^{-}] \sim -\frac{\Gamma_0}{2\pi t\, r^2}\,\bm{e}_z \to -\infty.
		\end{aligned}
	\end{equation}
	Note that $\bm{J}_4[\bm{R}_{N}^{-}]$ diverges twice as fast as $\bm{J}_{1}[\bm{R}_{N}^{-}]$ and $\bm{J}_{2}[\bm{R}_{N}^{-}]$, with the opposite sign, while $\bm{J}_3[\bm{R}_{N}^{-}]$ and $\bm{J}_5[\bm{R}_{N}^{-}]$ remain finite and positive at the origin. Hence, $\bm{J}[\bm{R}_{N}^{-}]$ is always finite.
	Substituting~\eqref{S4} into~\eqref{S3} recovers~\eqref{LOR3}.
\end{itemize}

\section{Lagrangian dynamics of the commutative vorticity operators}\label{UU4}
\subsection{General theory for three-dimensional flows}
In a given Schur frame $(\bm{e}_{1},\bm{e}_{2},\bm{e}_{3})$, the two cases from \S\ref{Operator-form vorticity decomposition} can be cast into a unified form:
	\begin{equation}
	\bm{\Psi}_R \mapsto 
	\begin{bmatrix}
		2(\psi+\gamma) & 0 \\
		0 & 2\psi 
	\end{bmatrix}, \qquad
	\bm{\Gamma}_S \mapsto 
	\begin{bmatrix}
		-\gamma & 0 \\
		0 & \gamma 
	\end{bmatrix}.
\end{equation}
Consequently, we have the entity expressions as
\begin{subequations}\label{three_Matrices}
	\begin{align}
		\bm{\Lambda} &\equiv \omega_{3}\bm{\sigma}_{0}
		= \omega_{3}(\bm{e}_{1}\bm{e}_{1}+\bm{e}_{2}\bm{e}_{2}), \label{Lambda} \\
		\bm{\Psi}_{R} &= 2(\psi+\gamma)\bm{e}_{1}\bm{e}_{1}
		+ 2\psi\bm{e}_{2}\bm{e}_{2}, \label{Psi_R} \\
		\bm{\Gamma}_{S} &= -\gamma\bm{e}_{1}\bm{e}_{1}
		+ \gamma\bm{e}_{2}\bm{e}_{2}. \label{Gamma_S}
	\end{align}
\end{subequations}
Then, by using~\eqref{ss11} and~\eqref{new1}, their material derivatives are evaluated as
\begin{subequations}\label{eq:Gamma_evolution}
	\begin{align}
		\frac{D\bm{\Lambda}}{Dt} 
		&= 
		\bigl(-W_1\beta + W_2\alpha -2\chi\omega_{3} + \Sigma_{12} - \Sigma_{21}\bigr)
		(\bm{e}_1\bm{e}_1+\bm{e}_2\bm{e}_2) \notag \\
		&\quad -W_2\omega_{3}(\bm{e}_3\bm{e}_1+\bm{e}_1\bm{e}_3) + W_1\omega_{3}(\bm{e}_3\bm{e}_2+\bm{e}_2\bm{e}_3), \label{eq:Lambda_final} \\
		\frac{D\boldsymbol{\Psi}_{R}}{Dt}
		&= 2\left[-W_1\beta - 2\chi(\psi+\gamma)- \partial_1\partial_2 h+ \Sigma_{12}\right]\,\boldsymbol e_1\boldsymbol e_1 \notag \\
		&\quad + 2\left[W_2\alpha - 2\chi\psi + \partial_2\partial_1 h- \Sigma_{21}\right]\,\boldsymbol e_2\boldsymbol e_2 \notag \\
		&\quad + 2W_3\gamma\,(\boldsymbol e_1\boldsymbol e_2+\boldsymbol e_2\boldsymbol e_1) \notag \\
		&\quad - 2W_2(\psi+\gamma)\,(\boldsymbol e_3\boldsymbol e_1+\boldsymbol e_1\boldsymbol e_3) \notag \\
		&\quad + 2W_1\psi\,(\boldsymbol e_3\boldsymbol e_2+\boldsymbol e_2\boldsymbol e_3), \label{eq:Psi_R_evolution} \\
		\frac{D\bm{\Gamma}_S}{Dt}
		&= \left[-W_1\beta - W_2\alpha - 2\chi\gamma - 2\partial_1\partial_2 h + \Sigma_{12} + \Sigma_{21}\right]
		\left(-\bm e_1\bm e_1 + \bm e_2\bm e_2\right) \notag \\
		&\quad - 2\gamma W_3 (\bm e_1\bm e_2 + \bm e_2\bm e_1)
		+ \gamma W_2 (\bm e_3\bm e_1 + \bm e_1\bm e_3)
		+ \gamma W_1 (\bm e_3\bm e_2 + \bm e_2\bm e_3). \label{eq:GammaS_evolution}
	\end{align}
\end{subequations}
The matrices of $D(\bm{\Lambda},\bm{\Psi}_{R},\bm{\Gamma}_{S})/Dt$ are given as follows.
\begin{itemize}
	\item\underline{The material derivative of $\bm{\Lambda}$}
\begin{eqnarray}
	\dfrac{D\bm{\Lambda}}{Dt}\mapsto\begin{bmatrix}
		\dfrac{D\omega_{3}}{Dt} & 0 & -\omega_3 W_2 \\
		0 & \dfrac{D\omega_{3}}{Dt} & \omega_3 W_1 \\
		-\omega_3 W_2 & \omega_3 W_1 & 0
	\end{bmatrix},
\end{eqnarray}
where the expression of $D\omega_{3}/Dt$ is given in~\eqref{g3}:
\begin{align*}
	\frac{D\omega_3}{Dt} &= -W_{1}\beta+W_{2}\alpha+F_{12}-F_{21}\notag\\
	&= -W_1\beta + W_2\alpha - 2\chi\omega_3 + \Sigma_{12} - \Sigma_{21}.
\end{align*}
Therefore, it follows that
\begin{subequations}\label{M1}
	\begin{align}
		\frac{D\bm{\Lambda}}{Dt}
		&= -2\chi\bm{\Lambda}+\mathbf{F}_{\mathrm{W}}+\mathbf{F}_{\mathrm{\Sigma}}, \label{M1a} \\[4pt]
		-2\chi\bm{\Lambda}
		&\mapsto
		\begin{bmatrix}
			-2\chi\omega_{3} & 0 & 0 \\
			0 & -2\chi\omega_{3} & 0 \\
			0 & 0 & 0
		\end{bmatrix}, \label{M1b} \\[4pt]
		\mathbf{F}_{\mathrm{W}}
		&\mapsto
		\begin{bmatrix}
			-W_{1}\beta+W_{2}\alpha & 0 & -\omega_3 W_2 \\
			0 & -W_{1}\beta+W_{2}\alpha & \omega_3 W_1 \\
			-\omega_3 W_2 & \omega_3 W_1 & 0
		\end{bmatrix}, \label{M1c} \\[4pt]
		\mathbf{F}_{\mathrm{\Sigma}}
		&\mapsto
		\begin{bmatrix}
			\Sigma_{12}-\Sigma_{21} & 0 & 0 \\
			0 & \Sigma_{12}-\Sigma_{21} & 0 \\
			0 & 0 & 0
		\end{bmatrix}
		= \nu\nabla^2\bm{\Lambda}+\bm{J}[\bm{\Lambda}]. \label{M1d}
	\end{align}
\end{subequations}
	\item\underline{The material derivative of $\bm{\Psi}_{R}$}
\begin{eqnarray}
	\dfrac{D\bm{\Psi}_R}{Dt}\mapsto
	\begin{bmatrix}
		2\dfrac{D\phi}{Dt} & 2W_3\gamma & -2W_2\phi \\
		2W_3\gamma &2\dfrac{D\psi}{Dt} & 2W_1\psi \\
		-2W_{2}\phi & 2W_1\psi & 0
	\end{bmatrix},
\end{eqnarray}
where the expressions of $D(\phi,\psi)/Dt$ are given in~\eqref{eq:new_psi} and~\eqref{DphiDt}, respectively:
\begin{align*}
	\frac{D\phi}{Dt}
	&= -W_{1}\beta + F_{12}= -W_1\beta - 2\chi\phi- \partial_1\partial_2 h+ \Sigma_{12}, \\
	\frac{D\psi}{Dt}
	&= W_{2}\alpha - F_{21}= W_2\alpha - 2\chi\psi + \partial_2\partial_1 h - \Sigma_{21}.
\end{align*}
Therefore, it follows that
\begin{subequations}\label{L1}
	\begin{align}
		\frac{D\bm{\Psi}_R}{Dt}
		&= -2\chi\bm{\Psi}_{R}+\hat{\mathbf{F}}_{\mathrm{W}}+\hat{\mathbf{F}}_{\mathrm{h}}+\hat{\mathbf{F}}_{\mathrm{\Sigma}}, \label{L1a} \\[4pt]
		-2\chi\bm{\Psi}_{R}
		&\mapsto
		\begin{bmatrix}
			-4\chi\phi & 0 & 0 \\
			0 & -4\chi\psi & 0 \\
			0 & 0 & 0
		\end{bmatrix}, \label{L1b} \\[4pt]
		\hat{\mathbf{F}}_{\mathrm{W}}
		&\mapsto
		\begin{bmatrix}
			-2W_{1}\beta & W_{1}\alpha-W_{2}\beta & -2W_2\phi \\
			W_{1}\alpha-W_{2}\beta & 2W_{2}\alpha & 2W_1\psi \\
			-2W_{2}\phi & 2W_1\psi & 0
		\end{bmatrix}, \label{L1c} \\[4pt]
	\hat{\mathbf{F}}_{\mathrm{h}}
		&\mapsto	
		\begin{bmatrix}
			-2\partial_{1}\partial_{2}h & \partial_{1}^{2}h-\partial_{2}^{2}h & 0 \\
			\partial_{1}^{2}h-\partial_{2}^{2}h & 2\partial_{1}\partial_{2}h & 0 \\
			0 & 0 & 0
		\end{bmatrix}, \label{L1d} \\[4pt]
	\hat{\mathbf{F}}_{\mathrm{\Sigma}}
		&\mapsto
		\begin{bmatrix}
			2\Sigma_{12} & \Sigma_{22}-\Sigma_{11} & 0 \\
			\Sigma_{22}-\Sigma_{11} & -2\Sigma_{21} & 0 \\
			0 & 0 & 0
		\end{bmatrix}
		= \nu\nabla^2\bm{\Psi}_{R}+\bm{J}[\bm{\Psi}_{R}]. \label{L1e}
	\end{align}
\end{subequations}
	\item\underline{The material derivative of $\bm{\Gamma}_{S}$}
\begin{eqnarray}
	\dfrac{D\bm{\Gamma}_S}{Dt}\mapsto
	\begin{bmatrix}
		-\dfrac{D\gamma}{Dt} & -2\gamma W_3 & \gamma W_2 \\
		-2\gamma W_3 & \dfrac{D\gamma}{Dt} & \gamma W_1 \\
		\gamma W_2 & \gamma W_1 & 0
	\end{bmatrix},
\end{eqnarray}
where the expression of $D\gamma/Dt$ is given in~\eqref{eq:new_gamma}:
\begin{align*}
	\dfrac{D\gamma}{Dt} &= -W_1\beta - W_2\alpha + F_{12} + F_{21}\\
	&= -W_1\beta - W_2\alpha - 2\chi\gamma - 2\partial_1\partial_2 h + \Sigma_{12} + \Sigma_{21}.
\end{align*}
Therefore, it follows that
\begin{subequations}\label{N1}
	\begin{align}
		\frac{D\bm{\Gamma}_S}{Dt}
		&= -2\chi\bm{\Gamma}_{S}+\check{\mathbf{F}}_{\mathrm{W}}+\check{\mathbf{F}}_{\mathrm{h}}+\check{\mathbf{F}}_{\mathrm{\Sigma}}, \label{N1a} \\[4pt]
		-2\chi\bm{\Gamma}_{S}
		&\mapsto
		\begin{bmatrix}
			2\chi\gamma & 0 & 0 \\
			0 & -2\chi\gamma & 0 \\
			0 & 0 & 0
		\end{bmatrix}, \label{N1b} \\[4pt]
		\check{\mathbf{F}}_{\mathrm{W}}
		&\mapsto
		\begin{bmatrix}
			W_{1}\beta+W_{2}\alpha & -W_{1}\alpha+W_{2}\beta & \gamma W_{2} \\
			-W_{1}\alpha+W_{2}\beta & -W_{1}\beta-W_{2}\alpha & \gamma W_{1} \\
			\gamma W_{2} & \gamma W_{1} & 0
		\end{bmatrix}, \label{N1c} \\[4pt]
		\check{\mathbf{F}}_{\mathrm{h}}
		&\mapsto
		\begin{bmatrix}
			2\partial_{1}\partial_{2}h & \partial_{2}^{2}h-\partial_{1}^{2}h & 0 \\
			\partial_{2}^{2}h-\partial_{1}^{2}h & -2\partial_{1}\partial_{2}h & 0 \\
			0 & 0 & 0
		\end{bmatrix}, \label{N1d} \\[4pt]
		\check{\mathbf{F}}_{\mathrm{\Sigma}}
		&\mapsto
		\begin{bmatrix}
			-\Sigma_{12}-\Sigma_{21} & \Sigma_{11}-\Sigma_{22} & 0 \\
			\Sigma_{11}-\Sigma_{22} & \Sigma_{12}+\Sigma_{21} & 0 \\
			0 & 0 & 0
		\end{bmatrix}
		= \nu\nabla^{2}\bm{\Gamma}_{S}+\bm{J}[\bm{\Gamma}_{S}]. \label{N1e}
	\end{align}
\end{subequations}
From~\eqref{M1},~\eqref{L1} and~\eqref{N1}, the following properties follow:	
\begin{subequations}
	\begin{equation}
	\mathbf{F}_{\rm W}=\hat{\mathbf{F}}_{\rm W}+\check{\mathbf{F}}_{\rm W},~\hat{\mathbf{F}}_{\rm h}=-\check{\mathbf{F}}_{\rm h},
	\end{equation}
	\begin{equation}
		\mathbf{F}_{\rm \Sigma}=\hat{\mathbf{F}}_{\rm \Sigma}+\check{\mathbf{F}}_{\rm \Sigma},~\bm{J}[\bm{\Lambda}]=\bm{J}[\bm{\Psi}_{R}]+\bm{J}[\bm{\Gamma}_{S}].
	\end{equation}
\end{subequations}
Therefore, $\hat{\mathbf{F}}_{\rm h}$ and $\check{\mathbf{F}}_{\rm h}$ mediate the mutual transformation between $\bm{\Psi}_{R}$ and $\bm{\Gamma}_{S}$ via the components of the enthalpy Hessian tensor $\bm{\nabla\nabla}h$ projected onto $\mathscr{P}$. The off-plane components of $\bm{\nabla\nabla}h$ are embodied in the angular velocity of the Schur frame along a pathline within $({\mathbf{F}}_{\rm W},\hat{\mathbf{F}}_{\rm W},\check{\mathbf{F}}_{\rm W})$. The triplet $(\bm{J}[\bm{\Lambda}],\bm{J}[\bm{\Psi}_{R}],\bm{J}[\bm{\Gamma}_{S}])$ has been introduced to represent the spatial variations of the Schur basis vectors, the dilatation, and the external forcing.
\end{itemize}

\subsection{General theory for two-dimensional flows}
For 2D flows, it holds that $\alpha=\beta=\lambda_{r}=0$ and  $\vartheta=2\chi$. The partial derivative along $\bm{e}_{3}$ is $\partial_{3}=\bm{e}_{3}\bm{\cdot}\bm{\nabla}=0$. In the local Cartesian coordinate system $(x_1,x_2,x_3)$ induced by the Schur frame $(\bm{e}_{1},\bm{e}_{2},\bm{e}_{3})$ at the point $\mathfrak{P}$, the enthalpy Hessian is expanded as $\bm{\nabla\nabla}h=\bm{e}_{i}\bm{e}_{j}\partial_{i}\partial_{j}h$ where the components are $\partial_{i}\partial_{j}h=\bm{e}_{i}\bm{\cdot}\bm{\nabla\nabla}h\bm{\cdot}\bm{e}_{j}~(i,j=1,2)$. Since $W_{1}=W_{2}=0$, the equation of motion of the Schur frame in~\eqref{ss11} is simplified to
\begin{equation*}
	\frac{D\bm{e}_{1}}{Dt}=W_{3}\bm{e}_{2},~~\frac{D\bm{e}_{2}}{Dt}=-W_{3}\bm{e}_{1},~~\frac{D\bm{e}_{3}}{Dt}=\bm{0},
\end{equation*}
where $W_{3}=(F_{22}-F_{11})/(2\gamma)$ is the $\bm{e}_{3}$-component of the angular velocity of the Schur frame. The term in~\eqref{The fourth term} becomes $\star\left(2\mathscr{A}(\bm{\Sigma})\right)=(\Sigma_{12}-\Sigma_{21})\bm{e}_{3}=(\nu\nabla^2\omega_{3}+\partial_{1}f_{2}-\partial_{2}f_{1})\bm{e}_{3}$.

\begin{itemize}
	\item\underline{The material derivative of $\bm{\Lambda}$}
	\begin{equation}
		\dfrac{D\bm{\Lambda}}{Dt}
		\mapsto
		\begin{bmatrix}
			\dfrac{D\omega_3}{Dt} & 0 \\
			0 &\dfrac{D\omega_{3}}{Dt}
		\end{bmatrix}
		=\begin{bmatrix}
		F_{12}-F_{21} & 0 \\
		0 &F_{12}-F_{21}
		\end{bmatrix},
	\end{equation}
where the expression of $D\omega_{3}/Dt$ is
\begin{align*}
	\frac{D\omega_3}{Dt} =F_{12}-F_{21}
	=  - 2\chi\omega_3 + \Sigma_{12} - \Sigma_{21}.
\end{align*}
Therefore, ${D\bm{\Lambda}}/{Dt}$ can be decomposed as
\begin{subequations}
	\begin{align}
		\dfrac{D\bm{\Lambda}}{Dt} &= -2\chi\bm{\Lambda}+\mathbf{F}_{\Sigma},\\
		-2\chi\bm{\Lambda} &\mapsto \begin{bmatrix}
			-\vartheta\omega_{3} & 0 \\
			0 &-\vartheta\omega_{3}
		\end{bmatrix},\\
		\mathbf{F}_{\Sigma} &\mapsto \begin{bmatrix}
			\Sigma_{12}-\Sigma_{21} & 0 \\
			0 &\Sigma_{12}-\Sigma_{21}
		\end{bmatrix}.
	\end{align}
\end{subequations}
The first term represents vorticity-dilatation coupling while the second term represents the effect due to the viscous diffusion of vorticity and the curl of external forcing.
\end{itemize}
\begin{itemize}
	\item\underline{The material derivative of $\bm{\Psi}_{R}$}
	\begin{eqnarray}
		\frac{D\bm{\Psi}_R}{Dt}
		\mapsto
		\begin{bmatrix}
			2\dfrac{D\phi}{Dt} & 2\gamma W_3 \\
			2\gamma W_3 &2\dfrac{D\psi}{Dt}
		\end{bmatrix}
		=\begin{bmatrix}
			2F_{12} & F_{22}-F_{11} \\
			F_{22}-F_{11} & -2F_{21}
		\end{bmatrix},
	\end{eqnarray}
	where $2\gamma W_{3}=F_{22}-F_{11}=\partial_1^2 h - \partial_2^2 h + \Sigma_{22} - \Sigma_{11}$. The expressions of $D(\phi,\psi)/Dt$ are 
	\begin{align*}
	\frac{D\phi}{Dt}&=F_{12}= -2\chi\phi - \partial_{1}\partial_{2} h + \Sigma_{12},\\
		\frac{D\psi}{Dt}&=-F_{21}=-2\chi\psi+\partial_{1}\partial_{2}h-\Sigma_{21}.
	\end{align*}
	Therefore, ${D\bm{\Psi}_{R}}/{Dt}$ can be decomposed as
\begin{subequations}
	\begin{align}
		\dfrac{D\bm{\Psi}_{R}}{Dt}
		&= -2\chi\bm{\Psi}_{R}+\hat{\mathbf{F}}_{\rm h}+\hat{\mathbf{F}}_{\rm\Sigma},\\
		-2\chi\bm{\Psi}_{R}
		&\mapsto
		\begin{bmatrix}
			-4\chi\phi & 0 \\
			0 &-4\chi\psi
		\end{bmatrix},\\
	    \hat{\mathbf{F}}_{\rm h}
		&\mapsto
		\begin{bmatrix}
			-2\partial_{1}\partial_{2}h & \partial_{1}^2h-\partial_{2}^2h \\
			\partial_{1}^2h-\partial_{2}^2h&2\partial_{1}\partial_{2}h
		\end{bmatrix},\\
		\hat{\mathbf{F}}_{\rm \Sigma}
		&\mapsto
		\begin{bmatrix}
			2\Sigma_{12} & \Sigma_{22}-\Sigma_{11} \\
			\Sigma_{22}-\Sigma_{11} &-2\Sigma_{21}
		\end{bmatrix}.
	\end{align}
\end{subequations}
\end{itemize}
\begin{itemize}
	\item\underline{The material derivative of $\bm{\Gamma}_{S}$}
	\begin{eqnarray}
		\frac{D\bm{\Gamma}_S}{Dt}\mapsto
		\begin{bmatrix}
		-\frac{D\gamma}{Dt} & -2\gamma W_3 \\
		-2\gamma W_3 & \frac{D\gamma}{Dt}
		\end{bmatrix}
		=	\begin{bmatrix}
			-(F_{12}+F_{21}) & F_{11}-F_{22} \\
			F_{11}-F_{22} & F_{12}+F_{21}
		\end{bmatrix},
	\end{eqnarray}
	where
	\begin{equation*}
		\frac{D\gamma}{Dt}=F_{12}+F_{21}=-2\chi\gamma - 2\partial_{1}\partial_{2}h+ \Sigma_{12} + \Sigma_{21}.
	\end{equation*}
	Therefore, ${D\bm{\Gamma}_{S}}/{Dt}$ can be decomposed as
\begin{subequations}
	\begin{align}
		\dfrac{D\bm{\Gamma}_{S}}{Dt}
		&= -2\chi\bm{\Gamma}_{S}+\check{\mathbf{F}}_{\rm h}+\check{\mathbf{F}}_{\rm\Sigma},\\
		-2\chi\bm{\Gamma}_{S}
		&\mapsto
		\begin{bmatrix}
			2\chi\gamma & 0 \\
			0 &-2\chi\gamma
		\end{bmatrix},\\
		\check{\mathbf{F}}_{\rm h}
		&\mapsto
		\begin{bmatrix}
			2\partial_{1}\partial_{2}h & \partial_{2}^2h-\partial_{1}^2h \\
			\partial_{2}^2h-\partial_{1}^2h&-2\partial_{1}\partial_{2}h
		\end{bmatrix},\\
		\check{\mathbf{F}}_{\rm\Sigma}
		&\mapsto
		\begin{bmatrix}
			-\Sigma_{12}-\Sigma_{21} & \Sigma_{11}-\Sigma_{22} \\
			\Sigma_{11}-\Sigma_{22} &\Sigma_{12}+\Sigma_{21}
		\end{bmatrix}.
	\end{align}
\end{subequations}
\end{itemize}

\section{Restricted Euler dynamics for incompressible flow}\label{restricted_Euler}
\subsection{Restricted Euler equation}
For incompressible flow without external forcing,~\eqref{VGT_eq2} is simplified as
\begin{equation}\label{E1}
	\frac{D\bm{A}}{Dt}=-\bm{A}^{2}-\bm{\nabla\nabla}{P}+\nu\nabla^2\bm{A}.
\end{equation}
The nonlocality of pressure complicates velocity gradient dynamics. The pressure Hessian tensor can be decomposed into local (i.e., the isotropic pressure effect) and nonlocal (i.e., the deviatoric pressure effect) contributions, yielding~\citep{Cantwell1992,Johnson2024Multiscale}
\begin{eqnarray}\label{E2}
\bm{\nabla\nabla}P=\frac{1}{3}\nabla^{2}P\bm{I}+\left(\bm{\nabla\nabla}P-\frac{1}{3}\nabla^{2}P\bm{I}\right).
\end{eqnarray}
Since the pressure satisfies the Poisson equation
\begin{eqnarray}\label{E3}
	\nabla^2 P=-{\rm tr}(\bm{A}^2)=2\mathcal{Q},
\end{eqnarray}
the isotropic contribution in~\eqref{E2} can be equivalently expressed as
\begin{eqnarray}\label{E4}
\frac{1}{3}\nabla^{2}P\bm{I}=-\frac{1}{3}{\rm tr}(\bm{A}^2)\bm{I}=\frac{2}{3}\mathcal{Q}\bm{I}.
\end{eqnarray}
By neglecting the nonlocal contribution of the pressure Hessian and viscous term along Lagrangian trajectories (i.e., the homogeneous assumption), one obtains the restricted Euler equation~\citep{Vieillefosse1982Local,Vieillefosse1984Internal,Cantwell1992}
\begin{equation}\label{E5}
	\frac{D\bm{A}}{Dt}=-\bm{A}^{2}-\frac{1}{3}\nabla^{2}P\bm{I}=-\bm{A}^2+\frac{1}{3}{\rm tr}(\bm{A}^2)\bm{I},
\end{equation}
subjected to the incompressibility condition $\vartheta=\bm{\nabla}\bm{\cdot}\bm{u}=0$. This equation was introduced as a conceptual model for elucidating gradient self-amplification effect.

\subsection{Evolution equations for the rotational invariants of the velocity gradient tensor}
Using a method similar to that employed in~\textsection\textsection \ref{UU1} and~\ref{UU2}, we obtain
\begin{subequations}\label{resa}
\begin{align}
	\dfrac{D\chi}{Dt} &= \frac{1}{2}(W_{1}\alpha+W_{2}\beta) - \chi^2 + \psi(\psi+\gamma) - \frac{1}{3}\nabla^2 P, \label{eq:chi} \\
	\dfrac{D\psi}{Dt} &= W_{2}\alpha - 2\chi\psi, \label{eq:psi} \\
	\dfrac{D\gamma}{Dt} &= - W_{1}\beta -  W_{2}\alpha - 2\chi\gamma, \label{eq:gamma} \\
	\dfrac{D\beta}{Dt} &= W_{1}(2\psi+\gamma)-W_3\alpha  - \beta(\chi+\lambda_r)+ \alpha(\psi+\gamma), \label{eq:beta} \\
	\dfrac{D\alpha}{Dt} &=  -W_{2} (2\psi+\gamma) +W_{3}\beta - \beta\psi - \alpha(\chi+\lambda_r), \label{eq:alpha} \\
	\dfrac{D\lambda_{r}}{Dt} &= -\beta W_2 - \alpha W_1 - \lambda_r^2 - \frac{1}{3}\nabla^2{P}. \label{eq:lambda_rxx}
\end{align}
\end{subequations}
where the angular velocity of the Schur frame must satisfy three constraints:
\begin{subequations}
	\label{eq:constraints}
	\begin{align}
		(\chi - \lambda_r) W_2 + \psi W_1 &= 0, \label{eq:c1} \\
		(\psi+\gamma) W_2 + (\lambda_r - \chi) W_1 &= 0, \label{eq:c2} \\
		2W_3\gamma + \beta W_2 - \alpha W_1 &= 0. \label{eq:c3}
	\end{align}
\end{subequations}
In the conventional vortex region where $\Delta>0$, $\Xi=(\chi-\lambda_{r})^2+\Delta/4$, as defined in~\eqref{Xifactor}, is strictly positive. Consequently,~\eqref{eq:c1} and~\eqref{eq:c2} admit only the trivial solution $W_{1}=W_{2}=0$. Further, if $\gamma\neq 0$, it follows from~\eqref{eq:c3} that $W_{3}=0$ as well. Under these conditions, the Schur frame $(\bm{e}_{1},\bm{e}_{2},\bm{e}_{3})$ remains invariant along any pathline of the restricted Euler system. As a result, the evolution equations for the rotation invariants in~\eqref{resa} reduce to
\begin{subequations}\label{resb}
	\begin{align}
		\dfrac{D\chi}{Dt} &= - \chi^2 + \psi(\psi+\gamma) - \frac{1}{3}\nabla^2 P, \label{eq:chix} \\
		\dfrac{D\psi}{Dt} &= - 2\chi\psi, \label{eq:psix} \\
		\dfrac{D\gamma}{Dt} &=  - 2\chi\gamma, \label{eq:gammax} \\
		\dfrac{D\beta}{Dt} &=   - \beta(\chi+\lambda_r)+ \alpha(\psi+\gamma), \label{eq:betax} \\
		\dfrac{D\alpha}{Dt} &= - \beta\psi - \alpha(\chi+\lambda_r), \label{eq:alphax} \\
		\dfrac{D\lambda_{r}}{Dt} &=- \lambda_r^2 - \frac{1}{3}\nabla^2{P}. \label{eq:lambda_rx}
	\end{align}
\end{subequations}
Due to $\lambda_{r}=-2\chi$, the pressure Laplacian is expressed as
\begin{eqnarray}\label{ppp}
	\nabla^2 P=2\mathcal{Q}=\frac{1}{2}\Delta-6\chi^2=\frac{1}{2}\Delta-\frac{3}{2}\lambda_{r}^2,
\end{eqnarray}
It follows that $(\chi,\lambda_{r})$ in~\eqref{eq:chix} and~\eqref{eq:lambda_rx} must evolve according to
\begin{subequations}\label{resc}
	\begin{align}
		\dfrac{D\chi}{Dt} &= \chi^2+\frac{1}{12}\Delta>0, \label{eq:chi1} \\
		\dfrac{D\lambda_{r}}{Dt} &=-\frac{1}{2} \lambda_r^2 - \frac{1}{6}\Delta<0. \label{eq:lambda_r1}
	\end{align}
\end{subequations}
From~\eqref{eq:psix} and~\eqref{eq:gammax}, the 2D discriminant $\Delta=4\psi(\psi+\gamma)$ satisfies the equation
\begin{eqnarray}\label{evolve_Delta}
	\dfrac{D\Delta}{Dt}=-4\chi\Delta.
\end{eqnarray}
Using~\eqref{eq:Q11},~\eqref{eq:R11},~\eqref{ppp},~\eqref{eq:chi1}, and~\eqref{evolve_Delta}, the evolution equations for the Vieillefosse-Cantwell invariant pair $(\mathcal{Q},\mathcal{R})$ are derived as
\begin{subequations}
	\begin{align}
		\frac{D\mathcal{Q}}{Dt}
		&= -2\chi\bigl(5\mathcal{Q}+12\chi^2-\nabla^{2}P\bigr) = -3\mathcal{R}, \label{eq:Q_evol} \\
		\frac{D\mathcal{R}}{Dt}
		&= \frac{1}{3}\mathcal{Q}\nabla^{2}P = \frac{2}{3}\mathcal{Q}^2. \label{eq:R_evol}
	\end{align}
\end{subequations}
Equations~\eqref{eq:Q_evol} and~\eqref{eq:R_evol} are consistent with those reported in the literature~\citep{Vieillefosse1984Internal,Cantwell1992,Meneveau2011Lagrangian}. Here, the Laplacian of pressure $\nabla^2P$ is kept in~\eqref{resb}, which allows its contribution to be explicitly discernible in~\eqref{eq:Q_evol} and~\eqref{eq:R_evol}.

\section{Conclusions and discussions}\label{Conclusions}
\begin{enumerate}
\item The Lagrangian dynamics of the real Schur form of the VGT $\bm{A}\equiv\bm{\nabla}\bm{u}$ is found to satisfy a Heisenberg-type-like equation in the Schur frame, supplemented by an additional source term $F$ [\eqref{ss4} and~\eqref{ss4a}]. Starting from the Navier-Stokes equations for compressible Newtonian fluids, the evolution equations for a set of VGT rotational invariants $(\chi,\lambda_{r},\gamma,\psi,\alpha,\beta)$ are derived within the conventional vortex region defined by the positive discriminant (i.e., $\Delta>0$) [\eqref{new1}]. Along a pathline, the angular velocity of the Schur basis vectors is explicitly determined by the components of the enthalpy Hessian tensor $\bm{\nabla\nabla}h$ and the source term $\bm{\Sigma}$, with the weighting coefficients obtained from the elementary algebraic manipulation of the rotational invariants [\eqref{Schur_frame_angular2}]. In the incompressible limiting case, $\bm{\nabla\nabla}h$ should be replaced by the pressure Hessian tensor $\bm{\nabla\nabla}P$. The theory is also applicable for $\Delta\leq 0$, and discussions on a few degenerate cases are documented in \S\ref{Remarks on degenerate cases}.
\item We derive the evolution equations for the derived invariants of the VGT, including the 2D discriminant $\Delta$~\eqref{g1}, the imaginary part of the complex eigenvalue pair $\lambda_{\rm ci}$~\eqref{g2}, the axial vorticity component $\omega_{3}$~\eqref{g3}, the characteristic rigid-rotation mode $\phi$~\eqref{DphiDt}, and the principal invariants $(\mathcal{P},\mathcal{Q},\mathcal{R})$ [\eqref{g4},~\eqref{eq:DQ} and~\eqref{eq:R_evolution}]. When the enthalpy Hessian tensor $\bm{\nabla\nabla}h$ and the source term $\bm{\Sigma}$ are neglected, it is shown that the relative material evolution rates of $(\Delta,\lambda_{\mathrm{ci}},\omega_{3},\phi)$ are proportional to the normal straining factor $-\chi$, and thus decrease for $\chi>0$. For incompressible flow, retaining only the isotropic pressure-Hessian contribution and neglecting the source term $\bm{\Sigma}$ consistently yields the Vieillefosse--Cantwell closed dynamical system for $(\mathcal{Q},\mathcal{R})$ [\eqref{eq7p16} and~\eqref{eq7p18}].
\item We derive the evolution equations for both the characteristic vorticity modes $(\bm{R}_{N},\bm{S}_{N})$ [\eqref{eq:RNxx} and~\eqref{eq:SNxx}], which provide a new theoretical tool for studying the interaction and transformation between rigid rotation $\bm{R}_{N}$ and shear $\bm{S}_{N}$. The equations are expressed as (figure~\ref{RSOmega})
\begin{subequations}
	\begin{align}
		\frac{D\bm{R}_N}{Dt}
		&=-\vartheta \bm{R}_{N}+\bm{R}_{N}\bm{\cdot}\bm{D}_{EL}+\bm{\Upsilon} +\nu\nabla^{2}\bm{R}_{N}+	\bm{J}[\bm{R}_{N}],\nonumber \\ 
		\frac{D\bm{S}_N}{Dt}
		&=-\vartheta\bm{S}_{N}+\bm{R}_{N}\bm{\cdot}\bm{D}_{SH} +\bm{S}_{N}\bm{\cdot}\bm{D} -\bm{\Upsilon} +\nu\nabla^2\bm{S}_{N} + 	\bm{J}[\bm{S}_{N}].\nonumber
	\end{align}
\end{subequations}
The theory reveals all stretching and tilting mechanisms of the distinct vorticity modes due to their interactions with the straining and shear strain-rate tensors $(\bm{D}_{EL},\bm{D}_{SH})$ (figure~\ref{stretching_tilting_mechanisms}). The well-known axial vortex stretching/contraction mechanism is properly embodied in the term $\bm{R}_{N}\bm{\cdot}\bm{D}_{EL}=\lambda_{r}\bm{R}_{N}$~\eqref{eq715a} in the evolution equation of $\bm{R}_{N}$. The interaction between $\bm{R}_{N}$ and $\bm{D}_{SH}$ tilts the $\bm{S}_{N}$-lines towards the direction parallel to $\bm{R}_{N}\bm{\cdot}\bm{D}_{SH}=\bm{R}_{N}\times\bm{S}_{N}/2$~\eqref{eq715b}. The term $\bm{S}_{N}\bm{\cdot}\bm{D}_{EL}$~\eqref{eq715c} accounts for the stretching/contraction mechanism of the in-plane spin component $\bm{S}_{N\parallel}=\alpha\bm{e}_{1}+\beta\bm{e}_{2}$ by $\chi$ and the axial spin component $\bm{S}_{N\perp}=\gamma\bm{e}_{3}$ by $\lambda_{r}$, whereas the term $\bm{S}_{N}\bm{\cdot}\bm{D}_{SH}$~\eqref{eq715d} denotes a quadratic self-coupling mechanism of the spin components along $\bm{e}_{1}$ and $\bm{e}_{3}$, which tilts the $\bm{S}_{N}$-lines towards $\bm{e}_{2}$. Summation over these four terms recovers the classical vortex stretching term~\eqref{omega_D}.
\item In the present formulation, the dynamical transfer term $\bm{\Upsilon}$ [\eqref{eq718} and~\eqref{explicit_Upsilon}] uncovers a previously unrecognized role of the enthalpy Hessian tensor $\bm{\nabla\nabla}h$ (or the pressure Hessian tensor $\bm{\nabla\nabla}P$ in incompressible limit) in the mutual transformation and redistribution of the two vorticity modes, through its off-diagonal elements in the Schur frame and the longitudinal-transverse coupling with the rotational invariants. Physically, this term elucidates the
mechanism of mutual transformation between axial vortices and surrounding shear layers coexisting
within a swirling flow system, a process occurring in both two- and three-dimensional flows. Even if $\bm{\Upsilon}=\bm{0}$, the Lagrangian dynamics of the vorticity modes are coupled through the viscous source pair $(\bm{J}[\bm{R}_N],\bm{J}[\bm{S}_{N}])$ [\eqref{Jcons}~--~\eqref{JRexpand_terms}], arising from the spatial derivatives of the Schur basis vectors (which are coupled
to the VGT and its associated gradients), the dilatation, and the external forcing. As an example, we investigate the kinematics and dynamics of the characteristic vorticity modes $(\bm{R}_{N}^{\pm},\bm{S}_{N}^{\pm})$ for the Lamb--Oseen vortex (\S\ref{Example: The Lamb-Oseen vortex}). The analysis not only validates the proposed theory but also unveils the elegant form of the viscous source pair $(\bm{J}[\bm{R}_N^{\pm}],\bm{J}[\bm{S}_{N}^{\pm}])$ due to the coupling between the Schur frame geometry and the characteristic spin modes, namely, $\bm{J}[\bm{R}_N^{\pm}]=4\nu\kappa^{2}\bm{S}_{N}^{\pm}=-\bm{J}[\bm{S}_N^{\pm}]$.
\item For incompressible flow, it is noteworthy that the pressure gradient $\bm{\nabla}P$ vanishes in the Helmholtz equation for $D\bm{\omega}/Dt$, whereas the pressure Hessian tensor $\bm{\nabla\nabla}P$ reappears in the evolution equation for $D^2\bm{\omega}/Dt^2$. The present study reveals that the components
of $\bm{\nabla\nabla}P$ enter both the equations for $D\bm{R}_{N}/Dt$ and $D\bm{S}_{N}/Dt$.
\item We derive the evolution equations for a pair of commutative vorticity operators $(\bm{\Psi}_{R},\bm{\Gamma}_{S})$ (\textsection\ref{UU4}). It is shown that $\hat{\mathbf{F}}_{\rm h}$ and $\check{\mathbf{F}}_{\rm h}(=-\hat{\mathbf{F}}_{\rm h})$ are responsible for the mutual transformation between $\bm{\Psi}_{R}$ and $\bm{\Gamma}_{S}$ through the components of the enthalpy Hessian tensor $\bm{\nabla\nabla}h$ (or the pressure Hessian tensor $\bm{\nabla\nabla}P$ in incompressible limit) projected onto $\mathscr{P}$.
\item As a reduced case, we discuss the restricted Euler dynamics for incompressible flow by retaining only the isotropic pressure Hessian contribution (\textsection\ref{restricted_Euler}). It is found that in the region where $\Delta>0$ and $\gamma\neq 0$, the Schur frame remains invariant along any pathline. The effects of pressure anisotropy and
viscous diffusion, which are neglected in the restricted Euler system, are responsible for regularizing the
finite-time singularity that would otherwise arise.
\item To the author's knowledge, kinematic decompositions of the VGT and vorticity have been developed from various perspectives and applied to complex flow diagnostics, including flow structures and turbulence statistics. However, a complete Lagrangian theory for the VGT rotational invariants, the Schur frame, the characteristic vorticity modes, and the commutative vorticity operators have not yet been reported in the existing literature. Although the present work derives the theory and demonstrates its initial application to the Lamb--Oseen vortex, its extension to 3D turbulent flows remains to be explored in future studies.
\end{enumerate}

\section*{Declarations}
\begin{itemize}
	\item \textbf{Funding} 	This work was funded by the National Natural Science
	Foundation of China (Grant No. 12402262).
	\item \textbf{Conflict of interest} The authors have no conflicts to disclose.
	\item \textbf{Data availability} The data that support the findings of this study are available from the corresponding author upon reasonable request.
	\item \textbf{Author contributions}	Tao Chen was involved in conceptualization, formal analysis, funding acquisition, investigation, visualization, methodology, the original draft preparation, review and editing of the manuscript.
	\item \textbf{Author ORCID} Tao Chen https://orcid.org/0000-0001-6838-204X.
\end{itemize}

\backmatter


\begin{appendices}
	\section{Notation herein and connections with published literature}\label{Notations herein and connections with published literature}
	\begin{figure}[t]
		\centering
		\subfigure[]{
			\begin{minipage}[t]{0.5\linewidth}
				\centering
				\includegraphics[width=1.0\columnwidth,trim={6cm 6cm 7cm 6cm},clip]{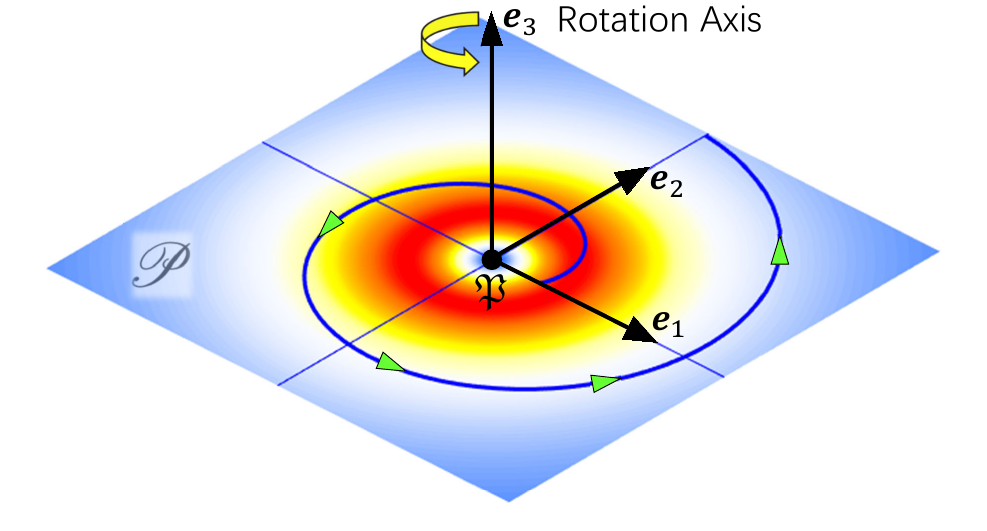}
				\label{Schur_frame_present}
			\end{minipage}%
		}%
		\subfigure[]{
			\begin{minipage}[t]{0.5\linewidth}
				\centering
				\includegraphics[width=1.0\columnwidth,trim={6cm 6cm 7cm 4.5cm},clip]{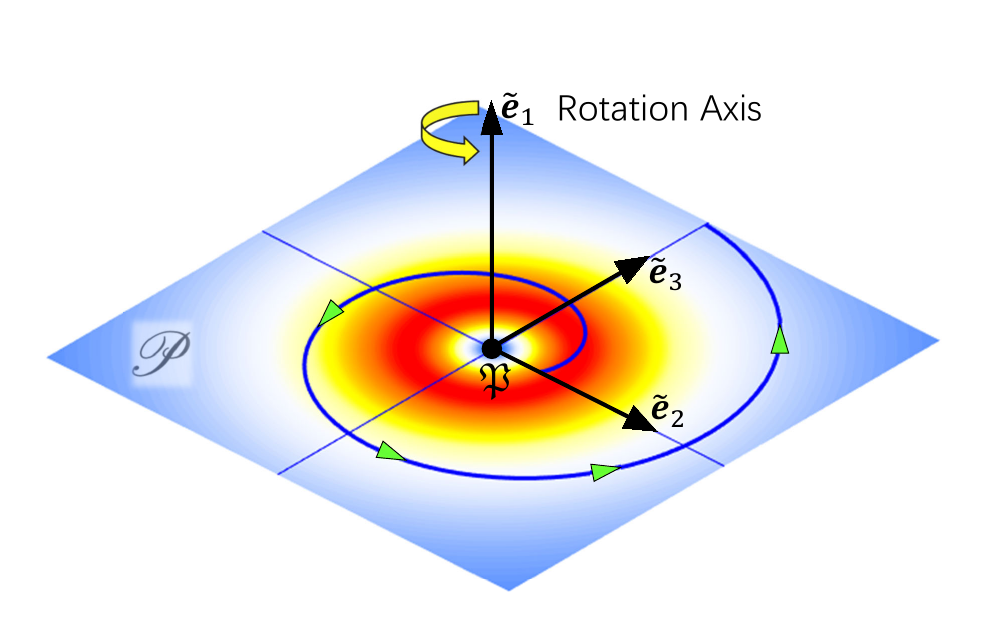}
				\label{Schur_frame_KH}
			\end{minipage}%
		}%
		\caption{The Schur frames at the point $\mathfrak{P}$ in (a) the present work and (b)~\citet{Kronborg2023}: $\bm{e}_{3}=\tilde{\bm{e}}_{1}$, $\bm{e}_{2}=\tilde{\bm{e}}_{3}$, and $\bm{e}_{1}=\tilde{\bm{e}}_{2}$. The rotation axis $\bm{e}_{3}$ is specified as the real eigenvector of the VGT that satisfies $\omega_{3}\equiv\bm{\omega}\bm{\cdot}\bm{e}_{3}>0$. The blue curve $\mathscr{C}$ denotes a typical swirling streamline of the velocity field relative to $\mathfrak{P}$, with the direction indicated by the green arrows.}
		\label{Schur_frame_comparison}
	\end{figure}
	In this appendix, we clarify the connections between the notation used in the present study and that adopted in the literature. The real Schur form~\eqref{Schur} follows the notation in~\citet{LiZhen2014} and~\citet{LiZhen2024}, which is intrinsically consistent with that used by~\citet{LiuCQ2025} and~\citet{Kronborg2023}. Any apparent differences in the matrix
	representation arise solely from differing conventions for the Schur frame basis vectors and for row and column matrix indices.
	
	Figure~\ref{Schur_frame_comparison} compares the Schur frames used in this study (figure~\ref{Schur_frame_present}) and~\citet{Kronborg2023} (figure~\ref{Schur_frame_KH}) at the point $\mathfrak{P}$. The rotation axis $\bm{e}_{3}$ is specified as the real eigenvector of the VGT satisfying $\omega_{3}>0$ and $\bm{e}_{3}\bm{\cdot}\bm{A}=\lambda_{r}\bm{e}_{3}$. In figure~\ref{Schur_frame_present}, the Schur frame $\left\{\bm{e}_{1},\bm{e}_{2},\bm{e}_{3}\right\}$ at $\mathfrak{P}$ induces a Cartesian coordinate system $(x_{1},x_{2},x_{3})$ in which the VGT at $\mathfrak{P}$ is expressed as 
	\begin{eqnarray}
		\bm{A}\equiv\bm{\nabla u}=\left(\bm{e}_{j}\frac{\partial}{\partial x_j}\right)\left(u_{i}\bm{e}_{i}\right)=\frac{\partial u_i}{\partial x_j}\bm{e}_{j}\bm{e}_{i}=\partial_{j}u_{i}\bm{e}_{j}\bm{e}_{i}.
	\end{eqnarray}
	With the convention that $j$ denotes the row index and $i$ the column index, we arrive at the real Schur form~\eqref{Schur}, i.e.,
	\begin{eqnarray}\label{Schur_X1}
		\bm{A}\mapsto A=\left[\partial_{j}u_{i}\right]=\begin{bmatrix}
			\chi & \psi + \gamma & -\beta \\
			-\psi & \chi & \alpha \\
			0 & 0 & \lambda_r
		\end{bmatrix}.
	\end{eqnarray}
	The matrix $A_{\rm Liu}$ used by~\citet{LiuCQ2025} is in fact ${A}^{\rm T}$, i.e.,
	\begin{equation}\label{Schur_X2}
		\bm{A}\mapsto A_{\rm Liu}=A^{\rm T}=\begin{bmatrix}
			\chi & -\psi & 0 \\
			\psi+\gamma & \chi & 0 \\
			-\beta & \alpha & \lambda_r
		\end{bmatrix}.
	\end{equation}
	In figure~\ref{Schur_frame_KH}, the Schur frame at $\mathfrak{P}$ adopted by~\citet{Kronborg2023} is $\left\{\tilde{\bm{e}}_{1},\tilde{\bm{e}}_{2},\tilde{\bm{e}}_{3}\right\}=\left\{\bm{e}_{3},\bm{e}_{1},\bm{e}_{2}\right\}$, which induces a Cartesian coordinate system $(\tilde{x}_{1},\tilde{x}_{2},\tilde{x}_{3})=(x_{3},x_{1},x_{2})$. Accordingly, the VGT at $\mathfrak{P}$ is expressed as
	\begin{eqnarray}
		\bm{A}\equiv\bm{\nabla u}=\left(\tilde{\bm{e}}_{j}\frac{\partial}{\partial \tilde{x}_j}\right)\left(\tilde{u}_{i}\tilde{\bm{e}}_{i}\right)=\frac{\partial \tilde{u}_i}{\partial \tilde{x}_j}\tilde{\bm{e}}_{j}\tilde{\bm{e}}_{i}=\tilde{\partial}_{j}\tilde{u}_{i}\tilde{\bm{e}}_{j}\tilde{\bm{e}}_{i}.
	\end{eqnarray}
	With the convention that $j$ denotes the row index and $i$ the column index, we obtain
	\begin{equation}\label{Schur_X4}
		\bm{A}\mapsto\tilde{A}=\left[\tilde{\partial}_{j}\tilde{u}_{i}\right]
		=\begin{bmatrix}
			\lambda_{r}& 0 & 0\\
			-\beta & \chi&\psi+\gamma\\
			\alpha&  -\psi& \chi
		\end{bmatrix}.
	\end{equation}
	The matrix $A_{\rm KH}$ used by~\citet{Kronborg2023} is in fact $\tilde{A}^{\rm T}$, i.e.,
	\begin{equation}\label{Schur_X5}
		\bm{A}\mapsto A_{\rm KH}=\tilde{A}^{\rm T}
		=\begin{bmatrix}
			\lambda_{r}& -\beta & \alpha\\
			0 & \chi&-\psi\\
			0&  \psi+\gamma& \chi
		\end{bmatrix}.
	\end{equation}
	The original matrix in~\citet[Eq. (16)]{Kronborg2023} is written as (The breve superscript here is used only for distinction)
	\begin{equation}\label{Schur_X6}
		\bm{A}\mapsto{A}_{\rm KH}	=\begin{bmatrix}
			\breve{\lambda}_{1}& \breve{\varepsilon} & \breve{\zeta}\\
			0 & \breve{a}&\breve{\beta}\\
			0&  \breve{\gamma}& \breve{a}
		\end{bmatrix},
	\end{equation}
	Comparing~\eqref{Schur_X5} and~\eqref{Schur_X6} yields the relations among the rotation invariants:
	\begin{equation}
		(\breve{\lambda}_{1},\breve{\varepsilon},\breve{\zeta},\breve{a},\breve{\beta},\breve{\gamma})
		=
		(\lambda_r,-\beta,\alpha,\chi,-\psi,\psi+\gamma).
	\end{equation}
	Therefore, the parameter $\breve{b}$ in~\citet[Eq. (15)]{Kronborg2023} is
	\begin{equation}
		\breve{b}=\sqrt{-\breve{\beta}\breve{\gamma}}=\sqrt{\psi(\psi+\gamma)}=\lambda_{\rm ci}.
	\end{equation}
\section{Proof of~\eqref{eq7p3b}}\label{app1}
In an orthonormal frame $(\bm{e}_{1},\bm{e}_{2},\bm{e}_{3})$, which need not be the Schur frame, we have
\begin{eqnarray}\label{A1}
\bm{D}\bm{\cdot}\bm{\varOmega}+\bm{\varOmega}\bm{\cdot}\bm{D}=(D_{im}\varOmega_{mj}+\varOmega_{im}D_{mj})\bm{e}_{i}\bm{e}_{j}.
\end{eqnarray}
Then, acting the Hodge star operator on~\eqref{A1} gives
\begin{eqnarray}\label{A2}
\star\left(\bm{D}\bm{\cdot}\bm{\varOmega}+\bm{\varOmega}\bm{\cdot}\bm{D}\right)&=&\frac{1}{2}e_{kij}(D_{im}\varOmega_{mj}+\varOmega_{im}D_{mj})\bm{e}_{k}\nonumber\\
&=&\frac{1}{2}e_{kij}D_{im}\varOmega_{mj}\bm{e}_{k}+\frac{1}{2}e_{kij}\varOmega_{im}D_{mj}\bm{e}_{k}\nonumber\\
&=&\frac{1}{2}e_{kij}D_{im}\varOmega_{mj}\bm{e}_{k}+\frac{1}{2}e_{kji}\varOmega_{jm}D_{mi}\bm{e}_{k}\nonumber\\
&=&e_{kij}D_{im}\varOmega_{mj}\bm{e}_{k}.
\end{eqnarray}
Because of
\begin{eqnarray}\label{A3}
\varOmega_{mj}=\frac{1}{2}e_{mjp}\omega_{p},
\end{eqnarray}
it follows from~\eqref{A2} that
\begin{eqnarray}
\star\left(\bm{D}\bm{\cdot}\bm{\varOmega}+\bm{\varOmega}\bm{\cdot}\bm{D}\right)
&=&\frac{1}{2}e_{jki}e_{jpm}D_{im}\omega_{p}\bm{e}_{k}\nonumber\\
&=&\frac{1}{2}\left(\delta_{kp}\delta_{im}-\delta_{ip}\delta_{km}\right)D_{im}\omega_{p}\bm{e}_{k}\nonumber\\
&=&\frac{1}{2}D_{ii}\omega_{k}\bm{e}_{k}-\frac{1}{2}D_{ik}\omega_{i}\bm{e}_{k}\nonumber\\
&=&\frac{1}{2}\vartheta\bm{\omega}-\frac{1}{2}\bm{\omega}\bm{\cdot}\bm{D}.
\end{eqnarray}
This completes the proof of~\eqref{eq7p3b}.
\section{Explicit expressions for the terms in~\eqref{eq78abxx+}~--~\eqref{LOS123}}\label{app3}
	\begin{subequations}\label{LOR123xap}
	\begin{align}
		\frac{D\bm{R}_{N}^{+}}{Dt}
		&= \frac{\Gamma_0}{4\pi\nu t^2}\left(\frac{r^2}{2\nu t}-1\right)
		\exp\!\left(-\frac{r^2}{4\nu t}\right)\,\bm{e}_z,\label{LOR1xap}\\[6pt]
	\nu\nabla^{2}\bm{R}_{N}^{+}
	&=
	\left[
	-\frac{4\nu\Gamma_{0}}{\pi r^{4}}
	+
	\left(
	\frac{\Gamma_{0}r^{2}}{8\pi\nu^{2}t^{3}}
	-\frac{\Gamma_{0}}{4\pi\nu t^{2}}
	+\frac{\Gamma_{0}}{\pi t\,r^{2}}
	+\frac{4\nu\Gamma_{0}}{\pi r^{4}}
	\right)	\exp\!\left(-\frac{r^2}{4\nu t}\right)
	\right]\bm{e}_{z}.,\label{LOR2xap}\\[6pt]
	\bm{J}[\bm{R}_{N}^{+}]
	&=\left[
	\frac{4\nu \Gamma_0}{\pi r^4}
	-
	\left(
	\frac{4\nu \Gamma_0}{\pi r^4}
	+
	\frac{\Gamma_0}{\pi t r^2}
	\right)
	\exp\left(-\frac{r^2}{4\nu t}\right)
	\right]\bm{e}_z.\label{LOR3xap}
	\end{align}
\end{subequations}
\begin{subequations}\label{LOS123x-final}
	\begin{align}
		\frac{D\bm S_N^+}{Dt}
		&=
		-\frac{\Gamma_0 r^2}{16\pi \nu^2 t^3}
		\exp\left(-\frac{r^2}{4\nu t}\right)\bm{e}_z
	,\\[6pt]
		\nu\nabla^2\bm S_N^+
		&=
	\left[
	\frac{4\nu \Gamma_0}{\pi r^4}
	-
	\left(
	\frac{\Gamma_0}{\pi t r^2}
	+
	\frac{4\nu \Gamma_0}{\pi r^4}
	+
	\frac{\Gamma_0 r^2}{16\pi \nu^2 t^3}
	\right)
	\exp\left(-\frac{r^2}{4\nu t}\right)
	\right]\bm{e}_z,\\[6pt]
		\bm J[\bm S_N^+]
		&=\left[
		-\frac{4\nu \Gamma_0}{\pi r^4}
		+
		\left(
		\frac{4\nu \Gamma_0}{\pi r^4}
		+
		\frac{\Gamma_0}{\pi t r^2}
		\right)
		\exp\left(-\frac{r^2}{4\nu t}\right)
		\right]\bm{e}_z.
	\end{align}
\end{subequations}
\begin{subequations}\label{LOR123_final}
	\begin{align}
		\frac{D\bm{R}_{N}^{-}}{Dt}
		&=
		-\frac{\Gamma_0}{4\pi\nu t^2}
		\exp\!\left(-\frac{r^2}{4\nu t}\right)\bm{e}_z,\label{LOR1_final}\\[6pt]
		\nu\nabla^2\bm{R}_{N}^{-}
		&=
		\left[
		\frac{4\nu\Gamma_0}{\pi r^4}
		-
		\left(
		\frac{4\nu\Gamma_0}{\pi r^4}
		+
		\frac{\Gamma_0}{\pi t r^2}
		+
		\frac{\Gamma_0}{4\pi\nu t^2}
		\right)
		\exp\!\left(-\frac{r^2}{4\nu t}\right)
		\right]\bm{e}_z,\label{LOR2_final}\\[6pt]
		\bm{J}[\bm{R}_{N}^{-}]
		&=
		\left[
		-\frac{4\nu\Gamma_0}{\pi r^4}
		+
		\left(
		\frac{4\nu\Gamma_0}{\pi r^4}
		+
		\frac{\Gamma_0}{\pi t r^2}
		\right)
		\exp\!\left(-\frac{r^2}{4\nu t}\right)
		\right]\bm{e}_z.\label{LOR3_final}
	\end{align}
\end{subequations}
\begin{subequations}\label{LOS123x-finalnew}
	\begin{align}
		\frac{D\bm{S}_{N}^{-}}{Dt}
		&=
		\frac{\Gamma_0 r^2}{16\pi \nu^2 t^3}
		\exp\left(-\frac{r^2}{4\nu t}\right)\bm{e}_z,\\[6pt]
		\nu\nabla^2\bm{S}_{N}^{-}
		&=
		\left[
		-\frac{4\nu \Gamma_0}{\pi r^4}
		+
		\left(
		\frac{\Gamma_0}{\pi t r^2}
		+
		\frac{4\nu \Gamma_0}{\pi r^4}
		+
		\frac{\Gamma_0 r^2}{16\pi \nu^2 t^3}
		\right)
		\exp\left(-\frac{r^2}{4\nu t}\right)
		\right]\bm{e}_z,\\[6pt]
		\bm{J}[\bm{S}_{N}^{-}]
		&=
		\left[
		\frac{4\nu \Gamma_0}{\pi r^4}
		-
		\left(
		\frac{4\nu \Gamma_0}{\pi r^4}
		+
		\frac{\Gamma_0}{\pi t r^2}
		\right)
		\exp\left(-\frac{r^2}{4\nu t}\right)
		\right]\bm{e}_z.
	\end{align}
\end{subequations}




\end{appendices}


\bibliography{sn-bibliography}

@book{Wu2006vorticity,
  title={Vorticity and Vortex Dynamics},
  author={Wu, J.-Z. and Ma, H.-Y. and Zhou, M.-D.},
  year={2006},
  publisher={Springer},
  address={Berlin}
}

@article{Chen2025Kinematic,
  title={Kinematic vorticity decomposition for generic two-dimensional flow},
  author={Chen, T. and Liu, T.},
  journal={Physics of Fluids},
  volume={37},
  year={2025}
}

@article{Chen2026General,
  title={A general kinematic theory of fluid-element rotation and intrinsic vorticity decompositions},
  author={Chen, T. and Wu, J.-Z. and Mao, F. and Liu, T.},
  journal={Journal of Fluid Mechanics},
  volume={1032},
  pages={A55},
  year={2026}
}

@article{Chen2026Kinematic,
  title={Kinematic vorticity decompositions and operator commutativity},
  author={Chen, T.},
  journal={Physics of Fluids},
  volume={38},
  pages={027140},
  year={2026}
}

@article{LiuCQ2018,
  author    = {Liu, Chao and Gao, Yang and Tian, Shuying and Dong, Xiang},
  title     = {Rortex-A new vortex vector definition and vorticity tensor and vector decompositions},
  journal   = {Physics of Fluids},
  volume    = {30},
  pages     = {035103},
  year      = {2018},
}

@inbook{LiuCQ2020,
  author    = {Liu, C. and Wang, Y.},
  title     = {Liutex and third generation of vortex definition and identification},
  booktitle = {An Invited Workshop from Chaos 2020},
  editor    = {Liu, C. and Wang, Y.},
  publisher = {Springer Nature},
  address   = {Cham, Switzerland},
  pages     = {1--479},
  year      = {2020},
}

@article{XuWQ2019,
  author    = {Xu, W. and Gao, Y. and Deng, Y. and Liu, J. and Liu, C.},
  title     = {{An explicit expression for the calculation of the Rortex vector}},
  journal   = {Physics of Fluids},
  volume    = {31},
  pages     = {095102},
  year      = {2019},
}

@article{Klein1910wirbeln,
  author    = {Klein, F.},
  title     = {{\"U}ber die Bildung von Wirbeln in reibungslosen Fl{\"u}ssigkeiten},
  journal   = {Zeitschrift f{\"u}r angewandte Mathematik und Physik},
  volume    = {59},
  pages     = {259--262},
  year      = {1910}
}

@article{Kaden1931aufwicklung,
  author    = {Kaden, H.},
  title     = {Aufwicklung einer unstabilen Unstetigkeitsfl{\"a}che},
  journal   = {Ingenieur-Archiv},
  volume    = {2},
  pages     = {140--168},
  year      = {1931}
}

@article{Betz1950wirbel,
  author    = {Betz, A.},
  title     = {Wie entsteht ein Wirbel in einer wenig z{\"a}hen Fl{\"u}ssigkeit},
  journal   = {Naturwissenschaften},
  volume    = {37},
  pages     = {193--196},
  year      = {1950}
}

@article{Cantwell1992,
  author    = {Cantwell, B. J.},
  title     = {Exact solution of a restricted {Euler} equation for the velocity gradient tensor},
  journal   = {Physics of Fluids A},
  volume    = {4},
  pages     = {782--793},
  year      = {1992},
}

@article{Cantwell1993,
  author = {Brian J. Cantwell},
  title = {On the behavior of velocity gradient tensor invariants in direct numerical simulations of turbulence},
  journal = {Physics of Fluids A},
  volume = {5},
  pages = {2008},
  year = {1993},
}

@article{Vieillefosse1982Local,
  author    = {Vieillefosse, P.},
  title     = {Local interaction between vorticity and shear in a perfect incompressible fluid},
  journal = {Journal de Physique (Paris)},
  volume    = {43},
  pages     = {837--842},
  year      = {1982},
}

@article{Vieillefosse1984Internal,
  author    = {Vieillefosse, P.},
  title     = {Internal motion of a small element of fluid in an inviscid flow},
  journal   = {Physica A},
  volume    = {125},
  pages     = {150--162},
  year      = {1984},
}

@article{Johnson2024Multiscale,
  author    = {Johnson, Perry L. and Wilczek, Michael},
  title     = {Multiscale velocity gradients in turbulence},
  journal   = {Annual Review of Fluid Mechanics},
  volume    = {56},
  pages     = {463--490},
  year     = {2024},
}

@article{Meneveau2011Lagrangian,
  author    = {Meneveau, C.},
  title     = {Lagrangian dynamics and models of the velocity gradient tensor in turbulent flows},
  journal   = {Annual Review of Fluid Mechanics},
  volume    = {43},
  pages     = {219--245},
  year      = {2011},
}

@article{Chong1990General,
  author    = {Chong, M. S. and Perry, A. E. and Cantwell, B. J.},
  title     = {A general classification of three‑dimensional flow fields},
  journal   = {Physics of Fluids A},
  volume    = {2},
  pages     = {765--777},
  year      = {1990},
}

@article{Soria1994Study,
  author    = {Soria, J. and Sondergaard, R. and Cantwell, B. J. and Chong, M. S. and Perry, A. E.},
  title     = {A study of the fine‑scale motions of incompressible time‑developing mixing layers},
  journal   = {Physics of Fluids},
  volume    = {6},
  pages     = {871--884},
  year      = {1994},
}

@article{Chacin2000Dynamics,
  author    = {Chacin, J. M. and Cantwell, B. J.},
  title     = {{Dynamics of a low Reynolds number turbulent boundary layer}},
  journal   = {Journal of Fluid Mechanics},
  volume    = {404},
  pages     = {87--115},
  year      = {2000},
}

@book{Tsinober2009Informal,
  author    = {Tsinober, A.},
  title     = {An Informal Conceptual Introduction to Turbulence},
  series    = {Fluid Mechanics and Its Applications},
  volume    = {92},
  publisher = {Springer},
  address   = {Dordrecht},
  year      = {2009},
}

@inbook{Cauchy1841Memoire,
  author    = {Cauchy, Augustin‑Louis},
  title     = {Mémoire sur les dilatations, les condensations et les rotations produites par un changement de forme dans un système de points matériels},
  booktitle = {Exercices d'analyse et de physique mathématique},
  volume    = {2},
  pages     = {302--330},
  publisher = {Bachelier},
  address   = {Paris},
  year      = {1841}
}

@article{Stokes1845InternalFriction,
  author    = {Stokes, G. G.},
  title     = {On the theories of the internal friction of fluids in motion, and of the equilibrium and motion of elastic solids},
  journal   = {Trans. Camb. Phil. Soc.},
  volume    = {8},
  pages     = {287--319},
  year      = {1845}
}

@book{Truesdell1954KinematicsVorticity,
  author    = {Truesdell, C.},
  title     = {The Kinematics of Vorticity},
  publisher = {Indiana University Press},
  address   = {Bloomington, Indiana},
  year      = {1954},
  series    = {Indiana University Publications Science Series},
  number    = {14}
}

@book{Batchelor1967IntroFD,
  author    = {Batchelor, G. K.},
  title     = {An introduction to fluid dynamics},
  publisher = {Cambridge University Press},
  address   = {Cambridge},
  year      = {1967},
}

@inproceedings{Kolar2004,
  title={{2D velocity-field analysis using triple decomposition of motion}},
  author  = {Kolář, Václav},
  booktitle={Proceedings of the 15th Australasian Fluid Mechanics Conference},
  editor={Behnia, M. and Lin, W. and McBain, G. D.},
  address={Sydney, Australia},
  publisher={University of Sydney},
  year={2004},
  note={CD-ROM, Paper AFMC00017}
}

@article{Kolar2007VortexID,
  author  = {Kolář, Václav},
  title   = {{Vortex identification: New requirements and limitations}},
  journal = {International Journal of Heat and Fluid Flow},
  year    = {2007},
  volume  = {28},
  number  = {4},
  pages   = {638--652},
}

@inbook{Kolar2011Triple,
  author    = {Kolář, V. and Moses, P. and Šístek, J.},
  title     = {Triple decomposition method for vortex identification in two-dimensional and three-dimensional flows},
  booktitle = {Computational Fluid Dynamics 2010},
  editor    = {Kuzmin, A.},
  publisher = {Springer},
  address   = {Berlin, Heidelberg},
  year      = {2011},
  pages     = {225--231}
}

@article{Schur1909,
  author  = {Schur, I.},
  title   = {{\"U}ber die charakteristischen Wurzeln einer linearen Substitution mit einer Anwendung auf die Theorie der Integralgleichungen},
  journal = {Math.\ Ann.},
  volume  = {66},
  pages   = {488--510},
  year    = {1909}
}

@article{Murnaghan1931,
  author  = {Murnaghan, F. D. and Wintner, A.},
  title   = {A canonical form for real matrices under orthogonal transformations},
  journal = {Proc. Nat. Acad. Sci. U. S. A.},
  volume  = {17},
  pages   = {417--420},
  year    = {1931}
}

@Article{GaoLiu2019,
  author = {Gao, Y. and Liu, C.},
  title =  {Rortex based velocity gradient tensor decomposition},
  year ={2019},
  journal = {Physics of Fluids},
  volume = {31},
  pages ={011704},
}

@Article{LiZhen2014,
  author = {Li, Z. and Zhang, X.-W. and He, F.},
  title =  {Evaluation of vortex criteria by virtue of the quadruple decomposition of velocity gradient tensor},
  year ={2014},
  journal = {Acta Physica Sinica},
  volume = {63},
  number={5},
  pages ={054704},
}

@mastersthesis{LiZhen2010,
   author       = "Li, Z.", 
   title        = "Theoretical {S}tudy on the {D}efinition of {V}ortex", 
   school       = "Tsinghua University", 
   year         = "2010", 
   address      = "Beijing, China",
}

@article{Tian2018,
  author  = {Tian, S. and Gao, Y. and Dong, X. and Liu, C.},
  title   = {A definition of vortex vector and vortex},
  journal = {Journal of Fluid Mechanics},
  volume  = {849},
  pages   = {312--339},
  year    = {2018}
}

@article{Wang2019Liutex,
  author  = {Wang, Y. and Gao, Y. and Liu, J. and Liu, C.},
  title   = {{Explicit formula for the Liutex vector and physical meaning of vorticity based on the Liutex-Shear decomposition}},
  journal = {Journal of Hydrodynamics},
  volume  = {31},
  number  = {3},
  pages   = {464--474},
  year    = {2019},
}

@Article{LiZhen2024,
  author = {Li, Z.},
  title =  {Schur {F}orms and {N}ormal-{N}ilpotent {D}ecompositions},
  year ={2024},
  journal = {Applied Mathematics and Mechanics},
  volume = {45},
  number={9},
  pages ={1-12},
}

@article{Kronborg2023,
  author = {Kronborg, Joel and Hoffman, Johan},
  title = {{The triple decomposition of the velocity gradient tensor as a standardized real Schur form}},
  journal = {Physics of Fluids},
  volume = {35},
  number = {3},
  pages = {031703},
  year = {2023},
}

@article{Keylock2025,
  title = {The role of normal and non-normal contributions to enstrophy production in the near-wall region of a turbulent channel flow},
  author = {Keylock, Christopher J.},
  journal = {Journal of Fluid Mechanics},
  volume = {1006},
  pages = {A3},
  year = {2025},
}

@article{Nagata2020,
  title={Triple decomposition of velocity gradient tensor in homogeneous isotropic turbulence},
  author={Nagata, Ryosuke and Watanabe, Tomoaki and Nagata, Koji and da Silva, Carlos B.},
  journal={Computers and Fluids},
  volume={198},
  pages={104389},
  year={2020},
}

@article{Yu2021,
  author = {Yu, Jia-Lin and Zhao, Zhe and Lu, Xi-Yun},
  title = {Non-normal effect of the velocity gradient tensor and the relevant subgrid-scale model in compressible turbulent boundary layer},
  journal = {Physics of Fluids},
  volume = {33},
  pages = {025103},
  year = {2021}
}

@article{Das2020,
    author = {Rishita Das and Sharath S. Girimaji},
    title = {Revisiting turbulence small-scale behavior using velocity gradient triple decomposition},
    journal = {New Journal of Physics},
    volume = {22},
    pages = {063015},
    year = {2020},
}

@article{Hoffman2021,
  title={Energy stability analysis of turbulent incompressible flow based on the triple decomposition of the velocity gradient tensor},
  author={Hoffman, Johan},
  journal={Physics of Fluids},
  volume={33},
  pages={081707},
  year={2021},
}

@article{Boukharfane2021,
  author = {Boukharfane, R. and Er-Raiy, A. and Parsani, M. and Chakraborty, N.},
  title = {Structure and dynamics of small-scale turbulence in vaporizing two-phase flows},
  journal = {Scientific Reports},
  volume = {11},
  pages = {15242},
  year = {2021}
}

@article{Boukharfane2021s,
  author    = {Boukharfane, Radouan and Er-raiy, Aimad and Alzaben, Lina and Parsani, Matteo},
  title     = {Triple decomposition of velocity gradient tensor in compressible turbulence},
  journal   = {Fluids},
  volume    = {6},
  number    = {3},
  pages     = {98},
  year      = {2021},
}

@article{Arun2024,
    author = {Rahul Arun and Tim Colonius},
    title = {Velocity gradient partitioning in turbulent flows},
    journal = {Journal of Fluid Mechanics},
    volume = {1000},
    pages = {R5},
    year = {2024},
}

@article{Keylock2018,
  title={{The Schur decomposition of the velocity gradient tensor for turbulent flows}},
  author={Keylock, Christopher J.},
  journal={Journal of Fluid Mechanics},
  volume={848},
  pages={876--905},
  year={2018},
}

@article{Moffatt1994,
  title={{Stretched vortices -- the sinews of turbulence; large-Reynolds-number asymptotics}},
  author={Moffatt, H. K. and Kida, S. and Ohkitani, K.},
  journal={Journal of Fluid Mechanics},
  volume={259},
  pages={241--264},
  year={1994},
}

@article{MaoFeng2023,
  title={Vortex dynamics: from formation, structure to evolution},
  author={Mao, F. and Li, Z. and Yao, J. and Wu, J.-Z.},
  journal={Aerodynamic Research \& Experiment},
  volume={1},
  number={2},
  pages={23--46},
  year={2023},
}

@article{YangYue2020,
  title={Theory and applications of the vortex-surface field},
  author={Yang, Yue},
  journal={Chinese Science Bulletin},
  volume={65},
  number={6},
  pages={483--495},
  year={2020},
}

@incollection{LiuCQ2025,
  title={Liutex -- The Past, Current, and Future},
  author={Liu, Chaoqun},
  booktitle={Proceedings of the International Conference on Liutex-Based Vortex Identification Methods},
  editor={Yu, Yifei and Wang, Yiqian and Liu, Chaoqun},
  series={Springer Proceedings in Physics},
  volume={430},
  publisher={Springer},
  address={Singapore},
  year={2025}
}

@article{Jiang2022,
  title={The evolution of coherent vortical structures in increasingly turbulent stratified shear layers},
  author={Jiang, X. and Lefauve, A. and Dalziel, S. B. and Linden, P. F.},
  journal={Journal of Fluid Mechanics},
  volume={947},
  pages={A30},
  year={2022},
}

@article{Helmholtz1858,
  title={{\"U}ber Integrale der hydrodynamischen Gleichungen, welche den Wirbelbewegungen entsprechen},
  author={Helmholtz, H.},
  journal={Journal f{\"u}r die reine und angewandte Mathematik},
  volume={55},
  pages={25--55},
  year={1858},
}

@article{Watanabe2020,
  title={Characteristics of shearing motions in incompressible isotropic turbulence},
  author={Watanabe, T. and Tanaka, K. and Nagata, K.},
  journal={Physical Review Fluids},
  volume={5},
  pages={072601(R)},
  year={2020},
}

@article{Arun2024velocity,
  title={Velocity gradient analysis of a head-on vortex ring collision},
  author={Arun, R. and Colonius, T.},
  journal={Journal of Fluid Mechanics},
  volume={982},
  pages={A16},
  year={2024},
}

@article{Chu2013,
  author    = {You‑Biao Chu and Xi‑Yun Lu},
  title     = {Topological evolution in compressible turbulent boundary layers},
  journal   = {Journal of Fluid Mechanics},
  volume    = {733},
  pages     = {414--438},
  year      = {2013},
}

@article{Zhang2026Bifurcation,
  author    = {Biao Zhang and Tao Chen},
  title     = {Bifurcation phenomena in rotating lid-driven square cavity flow: Hydrodynamic balance and vorticity-dynamical analyses},
  journal   = {Physics of Fluids},
  volume    = {38},
  pages     = {078139},
  year      = {2026},
}

@article{Kronborg2022Computational,
  author    = {Joel Kronborg and Frida Svelander and Samuel Eriksson-Lidbrink and Ludvig Lindström and Carme Homs-Pons and Didier Lucor and Johan Hoffman},
  title     = {Computational analysis of flow structures in turbulent ventricular blood flow associated with mitral valve intervention},
  journal   = {Frontiers in Physiology},
  volume    = {13},
  pages     = {806534},
  year      = {2022},
}

@article{Hayashi2021,
  author    = {Masato Hayashi and Tomoaki Watanabe and Koji Nagata},
  title     = {Characteristics of small-scale shear layers in a temporally evolving turbulent planar jet},
  journal   = {Journal of Fluid Mechanics},
  volume    = {920},
  pages     = {A38},
  year      = {2021},
}

@article{zhou1999mechanisms,
  author    = {Zhou, J. and Adrian, R. J. and Balachandar, S. and Kendall, T. M.},
  title     = {Mechanisms for generating coherent packets of hairpin vortices in channel flow},
  journal   = {Journal of Fluid Mechanics},
  volume    = {387},
  pages     = {353--396},
  year      = {1999},
}

@article{Born1926,
  author    = {Max Born and Werner Heisenberg and Pascual Jordan},
  title     = {Zur Quantenmechanik. {II}},
  journal   = {Zeitschrift f{\"u}r Physik},
  volume    = {35},
  pages     = {557--615},
  year      = {1926},
}

@book{Hodge1941,
  author    = {W. V. D. Hodge},
  title     = {The Theory and Applications of Harmonic Integrals},
  publisher = {Cambridge University Press},
  address   = {Cambridge},
  year      = {1941}
}

@book{Chern2000Lectures,
  author    = {Chern, S. S. and Chen, W. H. and Lam, K. S.},
  title     = {Lectures on Differential Geometry},
  series    = {Series on University Mathematics},
  volume    = {1},
  publisher = {World Scientific},
  address   = {Singapore},
  year      = {2000}
}

@inbook{Lighthill1956Viscosity,
  author    = {M. J. Lighthill},
  title     = {Viscosity effects in sound waves of finite amplitude},
  booktitle = {Surveys in Mechanics},
  editor    = {G. K. Batchelor and R. M. Davies},
  pages     = {250--351},
  year      = {1956},
  publisher = {Cambridge University Press},
  address   = {Cambridge}
}

@article{Chakraborty2005,
  author    = {Pinaki Chakraborty and S. Balachandar and Ronald J. Adrian},
  title     = {On the relationships between local vortex identification schemes},
  journal   = {Journal of Fluid Mechanics},
  volume    = {535},
  pages     = {189--214},
  year      = {2005},
}

@techreport{Hunt1988Eddies,
  author      = {J. C. R. Hunt and A. A. Wray and P. Moin},
  title       = {Eddies, streams, and convergence zones in turbulent flows},
  institution = {Center for Turbulence Research, Stanford University},
  number      = {CTR-S88},
  year        = {1988},
  pages       = {193--208},
  address     = {Stanford, California},
}

@article{Jeong1995Vortex,
  author    = {J. Jeong and F. Hussain},
  title     = {On the identification of a vortex},
  journal   = {Journal of Fluid Mechanics},
  volume    = {285},
  pages     = {69--94},
  year      = {1995},
}

@article{Chen2024boundary,
  title={Boundary sources of velocity gradient tensor and its invariants},
  author={Chen, Tao and Wu, Jie-Zhi and Liu, Tianshu and Salazar, David M.},
  journal={Physics of Fluids},
  volume={36},
  number={11},
  pages={117161},
  year={2024},
}

@article{Ohkitani1995,
    author    = {Ohkitani, K. and Kishiba, S.},
    title     = {Nonlocal nature of vortex stretching in an inviscid fluid},
    journal   = {Physics of Fluids},
    volume    = {7},
    number    = {2},
    pages     = {411--421},
    year      = {1995},
}

@article{Lawson2015,
  title = {On velocity gradient dynamics and turbulent structure},
  author = {Lawson, J. M. and Dawson, J. R.},
  journal = {Journal of Fluid Mechanics},
  volume = {780},
  pages = {60--98},
  year = {2015},
}

@article{Martin1998Dynamics,
  title = {Dynamics of the velocity gradient tensor invariants in isotropic turbulence},
  author = {Martín, Jesús and Ooi, Andrew and Chong, M. S. and Soria, Julio},
  journal = {Physics of Fluids},
  volume = {10},
  number = {9},
  pages = {2336},
  year = {1998},
}

@article{YinW2023,
  title = {Spatial distribution of coherent structures in a self‑similar axisymmetric turbulent wake},
  author = {Yin, Weijun and Tao, Shancong and Nagata, Koji and Ito, Yasumasa and Sakai, Yasuhiko and Zhou, Yi},
  journal = {Physical Review Fluids},
  volume = {8},
  pages = {084603},
  year = {2023},
}

@article{Zhu2021Thermodynamic,
  title = {{Thermodynamic and vortic fine structures of real Schur flows}},
  author = {Zhu, Jian-Zhou},
  journal = {Journal of Mathematical Physics},
  volume = {62},
  pages = {083101},
  year = {2021},
}

@article{Sreenivasan1997,
  title = {The phenomenology of small-scale turbulence},
  author = {Sreenivasan, K. R. and Antonia, R. A.},
  journal = {Annual Review of Fluid Mechanics},
  volume = {29},
  pages = {435--472},
  year = {1997},
}

@article{BilbaoLudena2025,
  author = {Bilbao-Ludena, Juan Carlos and Papadakis, George},
  title = {Non-normality and small-scale statistics in a three-dimensional, separated shear flow},
  journal = {Journal of Fluid Mechanics},
  volume = {1010},
  pages = {A41},
  year = {2025},
}

@article{Xie2020,
  author = {Xie, X. L.},
  title = {Theory of nonholonomic basis of orthogonal coordinate system based on principal directions of surface and related applications in fluid mechanics},
  journal = {Acta Aerodynamica Sinica},
  volume = {38},
  number = {1},
  pages = {171--195},
  year = {2020}
}

@book{Pope2000,
  author = {Pope, S. B.},
  title = {Turbulent Flows},
  publisher = {Cambridge University Press},
  address = {Cambridge, UK},
  year = {2000},
  pages = {771},
}

@article{Oseen1912wirbel,
  author = {Oseen, Carl Wilhelm},
  title = {{\"U}ber die Wirbelbewegung in einer reibenden Fl{\"u}ssigkeit},
  journal = {Arkiv f{\"o}r Matematik, Astronomi och Fysik},
  volume = {7},
  pages = {14--26},
  year = {1912},
  language = {German}
}

@book{Lamb1932hydrodynamics,
  author = {Lamb, Horace},
  title = {Hydrodynamics},
  edition = {6},
  publisher = {Cambridge University Press},
  year = {1932},
  address={Cambridge},
}

@article{Ma2024uniform,
  title={Uniform decomposition of velocity gradient tensor},
  author={Ma, Chenxi and Liu, Chaoqun},
  journal={Journal of Hydrodynamics},
  volume={36},
  number={1},
  pages={24--34},
  year={2024},
}

@article{Chu2025,
  title={{Liutex-shear interaction in Lamb-Oseen vortex and three-dimensional cylinder flow at Re=3900 simulated with SWLBM}},
  author={Chu, Xue-Sen and Jiang, Shang and Zhu, Jian-Song and Wang, Yi-Qian},
  journal={Journal of Hydrodynamics},
  year={2025},
  volume={37},
  number={2},
  pages={230--238},
}

\end{document}